\documentclass[a4paper]{ar}
\usepackage[numbers]{natbib}

\usepackage[ colorlinks = true,
             linkcolor = blue,
             urlcolor  = blue,
             citecolor = blue,
             anchorcolor = green,
]{hyperref}

\usepackage{amsmath}
\usepackage{amssymb}
\usepackage{graphicx}
\usepackage{subcaption}
\usepackage{braket}
\usepackage{slashed}
\usepackage{bm}

\usepackage{algorithm}
\usepackage{algpseudocode}
\usepackage{csquotes}

\usepackage{mathtools}
\mathtoolsset{showonlyrefs}

\newcommand{\mat}[2]{\left[\begin{array}{#1}#2\end{array}\right]}
\newtheorem{theorem}{Theorem}

\newcommand{\bea}{\begin{eqnarray*}}
\newcommand{\eea}{\end{eqnarray*}}

\newcommand{\openone}{1}

\begin{document}

\markboth{Haegeman and Verstraete}{Exact and Computational Methods for Matrix Product Operators}

\title{Diagonalizing transfer matrices and matrix product operators: a medley of exact and computational methods }

\author{Jutho Haegeman,$^1$ and Frank Verstraete$^{1,2}$
\affil{$^1$ Ghent University, Faculty of Physics, Krijgslaan 281, 9000 Gent, Belgium}
\affil{$^2$ Vienna Center for Quantum Technology, University of Vienna, Boltzmanngasse 5, 1090 Wien, Austria}}

\begin{abstract}
Transfer matrices and matrix product operators play an ubiquitous role in the field of many body physics. This paper gives an ideosyncratic overview of applications, exact results and computational aspects of diagonalizing transfer matrices and matrix product operators. The results in this paper are a mixture of classic results, presented from the point of view of tensor networks, and of new results. Topics discussed are exact solutions of transfer matrices in equilibrium and non-equilibrium statistical physics, tensor network states, matrix product operator algebras, and numerical matrix product state methods for finding extremal eigenvectors of matrix product operators.
\end{abstract}

\maketitle

\tableofcontents

\section{Introduction}

The theory of entanglement is providing a novel language by which quantum many body systems can be described. The entanglement features of quantum spin systems are most clearly expressed in the language of tensor networks, in which the wavefunction of a many body system is encoded in a series of tensors, one for each physical spin. As opposed to the usual treatments of many body systems, the tensor network language treats systems in real space, which makes them especially amenable for strongly interacting systems.

The central premise of tensor networks is the fact that an efficient local description of relevant many body wavefunctions can be obtained by modelling the way in which entanglement and correlations are distributed. This means that instead of working with the original degrees of freedom, we want to describe all physics by working on the Hilbert space of entanglement degrees of freedom connecting a bipartition of the system. As those degrees of freedom live on an interface, the spatial dimensionality of this Hilbert space is reduced. Hence, e.g., a one-dimensional system spin chain can be described by an effective zero-dimensional theory, and a two-dimensional theory by an effective one-dimensional Hilbert space. Tensor networks therefore provide an explicit representation for a research program originally laid out by Feynman \cite{Feynman}:

\begin{displayquote}
"Now in field theory, what's going on over here and what's going on over there and all over space is more or less the same. What do we have to keep track in our functional of all things going on over there while we are looking at the things that are going on over here? $\ldots$ It’s really quite insane actually: we are trying to find the energy by taking the expectation of an operator which is located here and we present ourselves with a functional which is dependent on everything all over the map. That’s something wrong. Maybe there is some way to surround the object, or the region where we want to calculate things, by a surface and describe what things are coming in across the surface. It tells us everything that’s going on outside  $\ldots$   I think it should be possible some day to describe field theory in some other way than with wave functions and amplitudes. It might be something like the density matrices where you concentrate on quantities in a given locality and in order to start to talk about it you don't immediately have to talk about what's going on everywhere else."
\end{displayquote}

Tensor networks aim to do exactly that, by defining a new Hilbert space on the surface.  Other powerful methods for simulating strongly correlated systems can also be understood from the point of view of Feynman.  For example, dynamical mean field theory \cite{DMFT} models the "outside" degrees of freedom as a system of free fermions in a self-consistent way; we will later see that such an approximation is exact in e.g. the case of the 2-dimensional classical Ising model. But the message is clear: if we want to describe strongly correlated systems in such a way that we do not encounter an exponential wall, we better try to model the entanglement degrees of freedom.

The theory of quantum information, originally aimed at the study of how to harness the power of the quantum world for doing information theoretic tasks, has in recent years made huge progress in understanding and classifying entanglement \cite{Horodecki4}. A crucial insight has been that ground states of Hamiltonians with local interactions, or equivalently all states whose marginal reduced density matrices are extreme points in the set of all marginals of the set of wavefunctions with certain symmetries, exhibit very few entanglement \cite{faithful}: quantum correlations are essentially restricted to be local, but in such a way that e.g. translational symmetry is not broken, and this implies the so-called area law for the entanglement entropy \cite{arealaw1,hastings2007area,eisert2010colloquium}. The structure of tensor networks is precisely based on that insight \cite{verstraete2004renormalization}: we consider a graph with  an edge between any degrees of freedom which interact with each other through a Hamiltonian term, and then define tensors on the vertices whose indices are contracted according to the underlying graph.

There are essentially three different classes of tensor networks. First of all, there are the ones in which the graph underlying the tensor contractions has no loops, such as matrix product states (MPS) \cite{baxter1978variational,Affleck2004,fannes1992finitely,white1992density,klumper1993matrix,ostlund1995thermodynamic,nishino1995density,vidal2003efficient,verstraete2004density,perez2006matrix,schollwock2005density,schollwock2011density}
and tree tensor networks \cite{shi2006classical,PhysRevB.82.205105}. Those are very well understood due to the fact that a normal form exists for them. Second, there are the tensor networks with loops, such as projected entangled pair states (PEPS) \cite{verstraete2004renormalization,verstraete2008matrix,dukelsky1998equivalence,nishino2001two}. Although such tensor networks could in principle be hard from the computational complexity point of view \cite{verstraete2006criticality,schuch2007computational}, in practice efficient methods have been constructed to deal with them \cite{verstraete2004renormalization,murg2007variational,jordan2008classical,orus2009simulation,bauer2011implementing,corboz2011stripes}. The third class of tensor networks involves the ones with an additional dimension which plays the role of scale in a renormalization group approach; those are called MERA (Multiscale Entanglement Renormalization Ansatz) \cite{vidal2007entanglement,evenbly2009algorithms}. All three of those classes can be shown to arise naturally from a compression of the path integral representation of the quantum state $\ket{\psi}\simeq \lim_{\tau\rightarrow\infty}\exp(-\tau H)\ket{\Omega}$ \cite{evenbly2015tensor,bal2015matrix}.

A central object in all those tensor networks is played by so-called matrix product operators (MPO) \cite{verstraete2004matrix,zwolak2004mixed,pirvu2010matrix,mcculloch2007density}. The aim of this paper is to present a review of the many facets of MPOs. Just as usual operators in quantum mechanics, MPOs come in several flavours. On the one hand, MPOs can represent density operators [such as $\exp(-\beta H)$]. On the other hand, MPOs form a very convenient way of representing a multitude of interesting operators acting on the Hilbert space of interest such as Hamiltonians for quantum spin chains. Similarly, the transfer matrices in a path integral formulation are MPOs, just as cellular automata for non-equilibrium spin systems. Interesting combinatorial problems can also be rephrased as calculating partition functions using the transfer matrix technique. And MPOs also play a crucial role in two-dimensional PEPS, for which the matrix product operators act on the ``virtual'' space and encode how the quantum correlations in the system are distributed.

Matrix product operators have since long been part of the canon of statistical physics, albeit typically under different nomenclatures. The most famous application of an MPO is probably Onsager's solution of the 2-dimensional Ising model \cite{onsager1944crystal}; Onsager managed to diagonalize the corresponding transfer matrix (which is a simple MPO) that was originally introduced by Kramers and Wannier \cite{kramers1941statistics}. The calculation of the entropy of 2-dimensional spin ice, a combinatorial problem, has been achieved by finding the largest eigenvector of an MPO \cite{lieb1967residual}. Similarly, the central ingredient in solving integrable spin systems makes essential use of matrix product operators: the core of the algebraic Bethe ansatz \cite{faddeev1980quantum,korepin1997quantum} is the construction of an algebra of matrix product operators. Matrix product operators can also be used  to construct explicit representations of tensor fusion categories \cite{levin2005string,chen2013symmetry,csahinouglu2014characterizing,williamson2014matrix,bultinck2015anyons}, which makes them central to the study of topologically ordered systems. MPOs have also been studied extensively in the context of non-equilibrium systems \cite{derrida1993exact,blythe2007nonequilibrium}, as one-dimensional cellular automata are MPOs of a special kind \cite{domany1984equivalence}. Finally, MPOs have become a very powerful tool for simulating quantum spin chains numerically, as the MPO formalism forms the backbone of the density matrix renormalization group (DMRG).

During the last years, a lot of progress has been made both on the theoretical description of MPOs and on the computational aspect of diagonalizing them. One of the goals of this article is to discuss how those new developments fit within the many classic applications of MPOs. This article is based on a series of lectures that was given to master students at the university of Vienna in the winter semester '15-'16, which explains the idiosyncratic choice of topics and examples and the fact that plenty of interesting and relevant topics have not been included.

\section{Matrix Product Operators: Definitions and Normal Forms}
In this review, we focus on translational invariant matrix product operators. In particular, we are interested in constructing uniform families of MPOs which act on a tensor product of $N$ $d-$dimensional spins and hence on a Hilbert space of dimension $d^N$. Note that this Hilbert space could equally well represent a quantum many body state
\[|\psi\rangle=\sum_{i_1i_2\ldots i_N}\psi(i_1,i_2,\ldots,i_N)|i_1\rangle|i_2\rangle\ldots|i_N\rangle,\]
normalized with the 2-norm, a classical probability distribution
\[p(i_1,i_2,\ldots,i_N),\]
as used in the case of cellular automata, or just a vector in a $d^N$ dimensional (real) Hilbert space, as occurs in transfer matrices of classical spin systems.

\begin{figure}
\includegraphics[width=\textwidth]{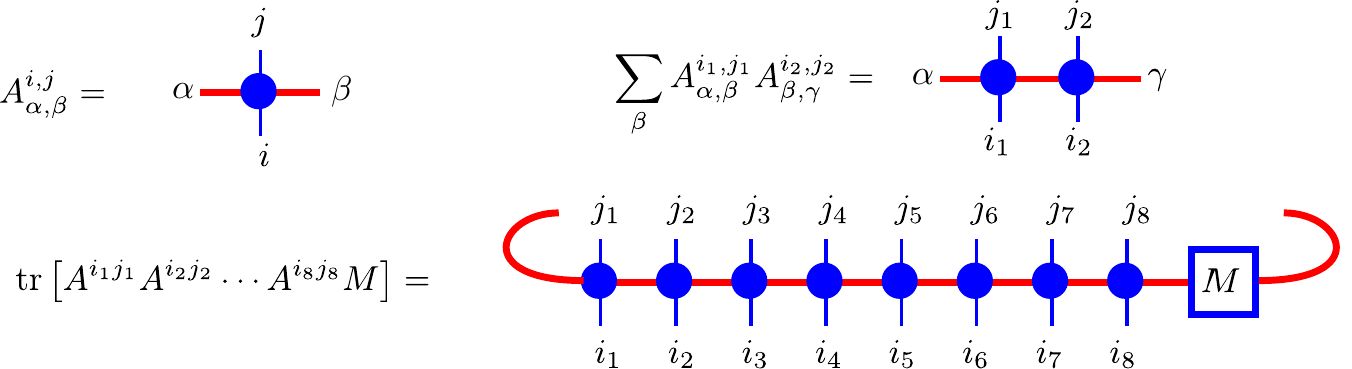}
\caption{\label{fig1} Graphical notation of a tensor contraction and of a matrix product operator.}
\end{figure}

Translational invariant Matrix Product Operators are defined by a single 4-leg tensor and $A^{ij}_{\alpha\beta}$ (see Figure \ref{fig1}) where the upper indices act on the physical space $1\leq i,j\leq d$ while the lower ones are contracted (hereafter called virtual) and run from $1$ to $D$, and by an additional matrix $M_{\alpha\beta}$ which dictates how the MPO has to be closed. Written out in components, we have
\[\hat{O}(A^{ij},M)=\sum_{i_1j_1i_2j_2\ldots}{\rm Tr}\left(A^{i_1j_1}A^{i_2j_2}\ldots A^{i_Nj_N}M\right)|i_1\rangle\langle j_1|\otimes |i_2\rangle\langle j_2| \otimes \ldots\otimes |i_N\rangle\langle j_N|.\]
A sufficient (but not necessary) condition for translational invariance is that $M$ commutes with all $A$: $[M,A^{ij}]=0$. Permutational invariance is only ensured when all the $A^i$ commute with each other (and $M$ can then be arbitrary).

Alternatively, we can of course also parameterize MPOs with 3-leg tensors and a basis of local operators $\{O_i\}$
\[\hat{O}=\sum_{i_1i_2\ldots}{\rm Tr}\left(A^{i_1}A^{i_2}\ldots A^{i_N} M\right)O_{i_1}\otimes O_{i_2} \otimes \ldots O_{i_N}\]
which can be a more useful representation when constructing algebras of MPOs.

This parameterization of MPOs is not unique, as a gauge transformation of the form $A^{ij}\rightarrow XA^{ij}X^{-1}$, $M\rightarrow XMX^{-1}$ leaves the MPO invariant. This is a purely virtual property, so that the discussion of the resulting canonical forms does not distinguish between the case of MPOs (where the matrices are labelled by a double index $A^{ij}$) or MPS (where they are labeled by a single index $A^{i}$). Both terminologies are used interchangeably. If the matrix multiplication algebra of the matrices $A^{ij}$ spans the complete $D^2$  dimensional matrix space, then the corresponding MPO is called injective. If not, then there must exist invariant subspaces $P_1,P_2,\cdots$ such that $\forall i,j,r: A^{ij} P_r =P_r A^{ij} P_r$. We can always perform a \emph{unitary} $U$ gauge transform\footnote{The fact that this can be done in a unitary way can be seen as follows: there must be a basis $X$ such that all $XA^{ij}X^{-1}$ are upper block triangular. Hence the span of all matrices which are in the null space have mutually orthogonal row and column vectors, and we can choose $U$ as the basis for those vectors.} to bring the $P_r$s into block diagonal form, where the unitary is unique up to a direct sum of unitaries on the individual subspaces. Let us for example consider the case with 2 invariant subspaces:
\[UA^{ij}U^\dagger=\mat{cc}{ A^{ij}_{11} & A^{ij}_{12} \\ 0&A^{ij}_{22}}\]
where $A^{ij}_{11}$ and $A^{ij}_{22}$ now represent injective MPOs. If furthermore
\[UMU^\dagger = \mat{cc}{M_{11}&M_{12}\\0&M_{22}}\]
then the matrices $A^{ij}_{12}$ and $M_{12}$ do not affect at all the physical degrees of freedom of the MPO, and we can get rid of them (i.e., setting them equal to zero). In that case, the MPO is nothing but a direct sum of the two other MPOs  $\hat{O}(A^{ij}_{11})$ and $\hat{O}(A^{ij}_{22})$ which can on their turn be injective or not. If not, we can repeat the same procedure. However, if $M$ is not of the form quoted above (i.e. has a block $M_{21}$), then we cannot get rid of the degrees of freedom in $A_{12}^{ij}$; this situation is used to encode e.g., a sum of local operators such as appearing in a Hamiltonian for a system with periodic boundary conditions.

\begin{figure}
\includegraphics[width=12cm]{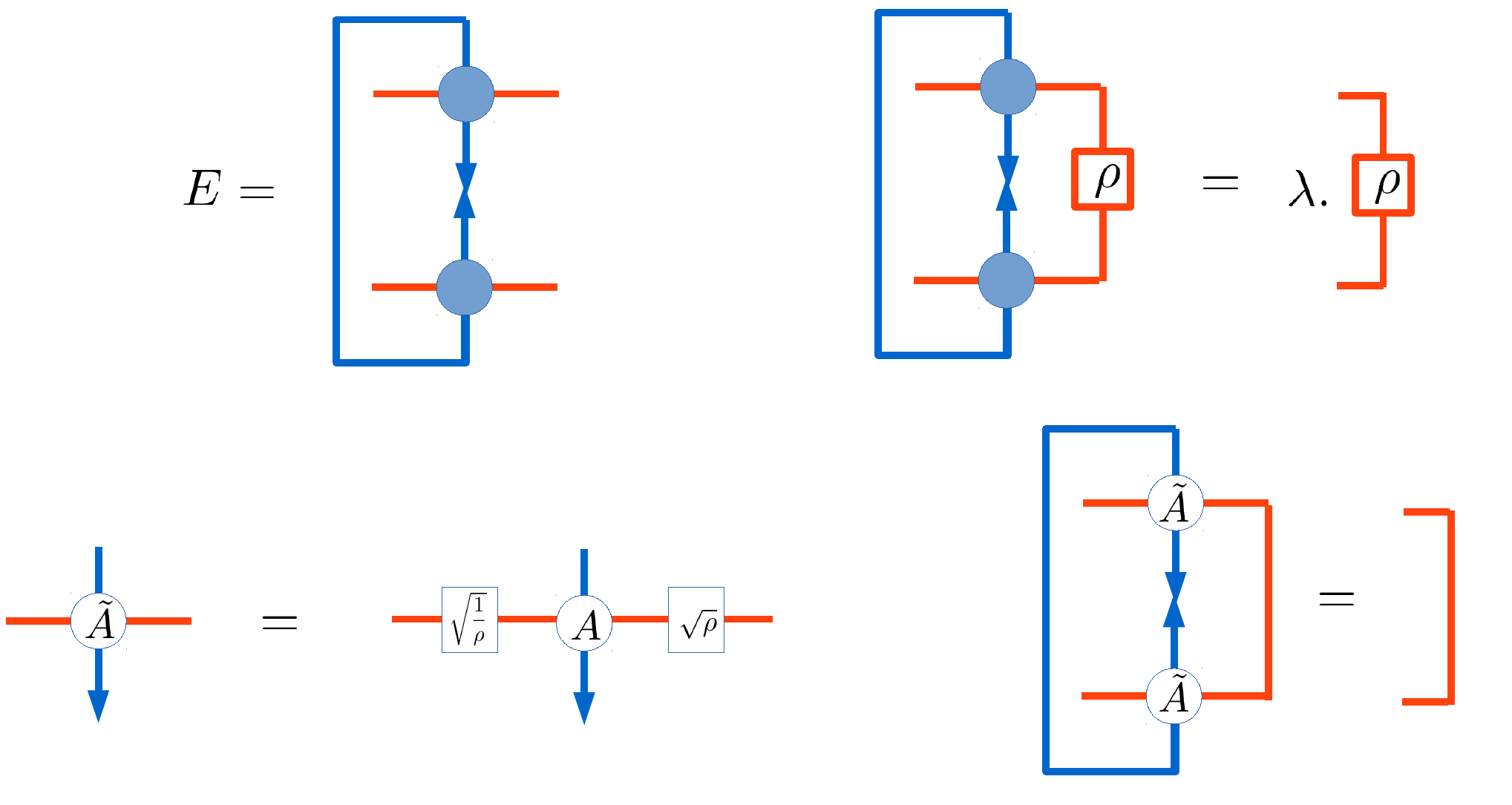}
\caption{\label{fig2} The transfer matrix of a matrix product operator, its leading eigenvector, and a matrix product operator in the right canonical form.}
\end{figure}

In the injective case and when $M=I$ (which is then the only choice that commutes with all $A^{ij}$), it is easy to define a normal form.  A crucial object of interest in that case is the MPS transfer matrix\footnote{Here we switch the discussion to MPS because the transfer matrix originates from interpreting the MPO $\hat{O}(A)$ as a state $\ket{O(A)}$ in a Hilbert space $\mathbb{H}\times\mathbb{H}$ and applying the 2-norm.} $E=\sum_{ij}A^{ij}\otimes \bar{A}^{ij}$ (see Figure \ref{fig2}). This matrix defines a completely positive linear map:
\[E|\rho\rangle\simeq \sum_{ij} A^{ij}\rho A^{ij\dagger}.\]
The quantum version of Perron-Frobenius theory \cite{evans1978spectral,fannes1992finitely} now dictates that injectivity (or ergodicity) implies that the eigenvector $\rho$ corresponding to the largest eigenvalue $\lambda>0$ of $E$ is unique and can be chosen to be strictly positive in the semidefinite sense. Let us now do the gauge transform $X=\rho^{1/2}$ and a rescaling with factor $\lambda^{-1/2}$. The ensuing $\tilde{A}^{ij}=\lambda^{-1/2} \rho^{-1/2}A^{ij}\rho^{1/2}$ then constitute an isometry (see figure \ref{fig2}). We could of course equally well have repeated this procedure starting from the left eigenvector, which would have resulted in an isometry with respect to the different indices. This procedure is called bringing an injective MPS into right or left normal form, which is unique up to an additional unitary gauge transform.

\begin{figure}
\includegraphics[width=12cm]{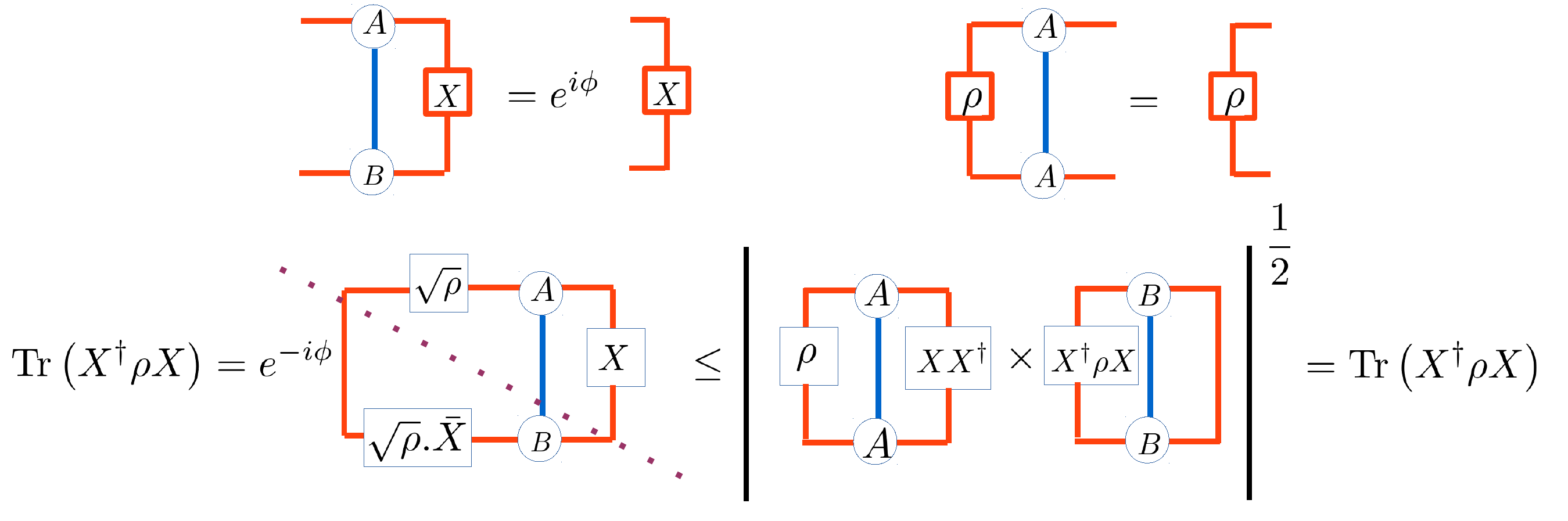}
\caption{\label{fig3} Graphical proof of the fundamental theorem of matrix product states.}
\end{figure}

Let us now prove a very useful theorem, which we call the fundamental theorem of MPS, on the uniqueness of the MPS normal form for translational invariant systems \cite{perez2008string}:
\begin{theorem} Given two injective MPS characterized by $A^{ij}$ and $B^{ij}$, then $\hat{O}(A^{ij})=\hat{O}(B^{ij})$ for $N\rightarrow\infty$ if and only if there exists a gauge transform which transforms $A^{ij}$ into $B^{ij}$: $A^{ij} X = X B^{ij}$.
\end{theorem}
This can readily be proven by using the Cauchy-Schwarz theorem. Without loss of generality, we can assume that both tensors are in right normal form, and we furthermore assume that the largest eigenvalue in magnitude of their respective transfer matrices is equal to 1 (which can always be obtained by rescaling the tensors with a constant). Then $\hat{O}(A^{ij})=\hat{O}(B^{ij})$ iff $|\langle O(A)|O(B)\rangle|=1$, which is possible iff the largest eigenvalue of the mixed transfer matrix $\sum_{ij} A^{ij}\otimes \bar{B}^{ij}$ has magnitude 1. Let us call the associated eigenvector $X$, and then use the Cauchy-Schwarz theorem with respect to the dotted line in Figure~\ref{fig3}. Because the inequality must be an equality, we conclude that $XB^i=A^i X$ which we set out to prove.

We are now in a position to define the normal form of an MPO \cite{perez2006matrix,newmpo2016}: Given a translational invariant  MPO $\{A^{ij},M\}$, then its normal form is obtained by first identifying its invariant subspaces and unitary gauging all $A^{ij}$ in upper triangular block form $A^{ij}_{ab}$. We can also write $M$ in this basis, and write it as $M=M_1+M_2$, where $M_1$ is block upper triangular and $M_2$ only has blocks beneath this block diagonal. The complete MPO is a sum of $O_1\{A^{ij},M_1\}$ and $O_2\{A^{ij},M_2\}$. Two MPOs will be equal to each other iff both of those blocks are equal to each other. For $O_1$, we get rid of the off-diagonal blocks in $A^{ij}$ and $M_1$, and if the remaining diagonal blocks are injective, we are done. Otherwise, we have to find the invariant subspaces within those blocks and repeat the same procedure. Potentially, the same injective MPS can appear multiple times, and if this is the case, we make one block out of it and put the right weights into the corresponding block of $M_1$ by multiplying it with an appropriate scalar. In case 2, translational invariance is difficult to judge, as it could be obtained without all blocks in $A^{ij}$ on which $M_2$ has support to be equal to each other. See \cite{perez2006matrix,newmpo2016} for a proof of uniqueness.

To conclude this section, let us give a few examples.

\begin{itemize}
\item Let us consider the projector on the subspace of all qubit strings with even parity $P=\left(I+Z^{\otimes N}\right)/2$. This can readily be written as an MPO with bond dimension 2 where both physical and virtual indices take value in $\mathbb{Z}_2=\{0,1\}$: $A^{ij}_{\alpha\beta}=\delta_{ij}\delta_{\alpha\beta}\left(-1\right)^{i \cdot \alpha}$ and $M=\mat{cc}{1/2 & 0\\0 & 1/2}$. Equivalently, we can write $A^{ij}_{\alpha\beta}=\left(\mat{cc}{1 & 0\\0 & -1}_{ij}\right)^{\alpha} \cdot \mat{cc}{1 & 0\\0 & 1}_{\alpha\beta}$. Note that this MPO is non-injective and consists of 2 injective blocks with bond dimension 1.
\item A less trivial example, illustrating the existence of so-called trash vectors in the reduction to the canonical form, is encountered in the context of symmetry protected topological (SPT) phases. Let us consider the so-called ``CZX'' \cite{chen2013symmetry} MPO $A^{ij}_{\alpha\beta}=\delta_{\beta i}(-1)^{\alpha \dot i}\mat{cc}{0 & 1\\1 & 0}_{ij}$ for $i,j,\alpha,\beta=0,1$. This MPO is injective, as can easily be seen by considering the algebra generated by the matrices $A^{01}=\mat{cc}{1&1\\0&0}$ and $A^{10}=\mat{cc}{0&0\\1&-1}$. Let us now square this MPO, thereby getting an MPO $B$ with bond dimension 4:
\[B^{00}=\mat{cc}{1 & 1\\0 & 0}\otimes \mat{cc}{0 & 0\\1 & -1}=\mat{cccc}{0 & 0 & 0 & 0\\1 & -1 & 1 & -1\\0 & 0 & 0 & 0\\0 & 0 & 0 & 0} \]
\[B^{11}=\mat{cc}{0 & 0\\1 & -1}\otimes \mat{cc}{1 & 1\\0 & 0}=\mat{cccc}{0 & 0 & 0 & 0\\0 & 0 & 0 & 0\\1 & 1 & -1 & -1\\0 & 0 & 0 & 0}\]
This MPO is clearly not injective, and it has an invariant subspace
\[P=\mat{cccc}{0 & 0 & 0 & 0\\0 & 1 & 0 & 0\\0 & 0 & 1 & 0\\0 & 0 & 0 & 0}\]
We can therefore as well work with the block
\[B^{00}=\mat{cc}{-1 & 1\\0 & 0},\hspace{1cm} B^{11}=\mat{cc}{0 & 0\\1 & -1}\]
But this MPO again has an invariant subspace given by the projector $P=1/2\mat{cc}{1 & -1\\-1 & 1}$. We can rotate the MPO in this basis, get rid of the trash vector, and end up with an MPO with bond dimension 1. In summary, the normal form of the MPO $\hat{O}(A)^2$ is $B^{ij}= (-1)\delta_{ij}$, which globally means $\hat{O}(A)^2=(-1)^N I$.
\item Finally, let us construct the MPO for the operator $\hat{O}=\sum_i X_i$. It can readily be checked that the MPO
\[A^{ij}_{\alpha\beta}=\mat{cc}{1 & 0\\0 & 1}_{ij}\mat{cc}{1 & 0\\0 & 1}_{\alpha\beta}+\mat{cc}{0 & 1\\1 & 0}_{ij}\mat{cc}{0 & 1\\0 & 0}_{\alpha\beta}\]
together with $M=\mat{cc}{0 & 0\\1 & 0}$ does the job. Clearly, $A$ is not injective, but as $M$ is of the form $M_2$, it is not possible to reduce this form further.
\end{itemize}

\section{Matrix product operators and transfer matrices}
Matrix product operators originally popped up as transfer matrices in statistical physics, and more specifically in the study of partition functions of classical spin systems. However, such transfer matrices can also be defined from the path integral representation of quantum spin systems. In a similar fashion, the transfer matrix can be introduced for tensor network representations of quantum states, where it often plays an important role in formulating the associated tensor network algorithms. Finally, we discuss the role of matrix product operators in the context of nonequilibrium statistical physics.

\subsection{Equilibrium stastistical physics and counting problems}
he most famous application of the transfer matrix method in classical statistical physics is for sure Onsager's exact solution of the 2-dimensional Ising model, which was obtained by diagonalizing its transfer matrix, but many other fascinating partition functions have been diagonalized using the more advanced Bethe ansatz. In particular, combinatorial problems on the lattice can often also be formulated as the zero temperature entropy of a partition function and can then be determined using the transfer matrix method.

\subsubsection{Combinatorial problems on the lattice}\label{hardsq}

Counting problems on lattices are notoriously hard and can exhibit very rich behaviour. The prime example of such a problem is calculating the scaling of the number of ways in which a given set of tiles can tile the (infinite) plane. Such problems have a very natural formulation in terms of the leading eigenvalues of a transfer matrix.

Before we move to the interesting case of 2-dimensional lattices, let us warm up with the simple task first undertaken by Fibonacci of counting the number of ways in which we can arrange $N$ bits such that a bit $1$ is never followed by another $1$. We can easily set up a recursion for this: let us call the number of allowed sequences of $n$ bits ending with a $0$ as $x(n)$, and the number of allowed sequences ending with a $1$ as $y(n)$. Then we trivially obtain:
\[\mat{c}{x(n+1)\\y(n+1)}=\underbrace{\mat{cc}{1 & 1\\1 & 0}}_{T}\mat{c}{x(n)\\y(n)}=T^n\mat{c}{x(1)\\y(1)}\]

The relevant (exponential) scaling is governed by the leading eigenvalue of the matrix $T$, which is equal to $(1+\sqrt{5})/2$. The exact result is given by ${\rm Tr}\left(\mat{cc}{1 & 1\\1 & 1}T^N\right)$. For the case of $N$ bits on a ring, the result is just ${\rm Tr}T^N=\left[(1+\sqrt{5})^N+(1-\sqrt{5})^N\right]/2^N$. Let us investigate whether it is possible to write the projector on those allowed sequences as a matrix product operator. The MPO $\hat{O}$ defined by the following tensor does exactly that:
\[A^{ij}_{\alpha\beta}=\delta_{i\alpha} \delta_{ij} \mat{cc}{1 & 1\\1 & 0}_{\alpha\beta}\]
We indeed recover the counting result by taking the trace of $\hat{O}$. The situation is much more interesting however in two dimensions. Let us stick to this Fibonacci example but try to count the ways in which I can put bits on a square lattice with $N\cdot M$ sites in such a way that a bit $1$ is solely surrounded by $0$s. This MPO can be constructed by constructing an MPO with periodic boundary conditions on $N$ sites :
\[A^{ij}_{\alpha\beta}=\delta_{i\alpha} \mat{cc}{1 & 1\\1 & 0}_{ij} \mat{cc}{1 & 1\\1 & 0}_{\alpha\beta}\]
The result is then obtained by calculating ${\rm Tr}\left(\hat{O_N}^M\right)$. In the limit of large $M$, this amounts to calculating the leading eigenvalue of the MPO $\hat{O}$.  We will see in later sections how such problems can be solved to a very good precision using tensor network techniques.

A slight modification of this problem is obtained by adjusting the weight of the occupied sites (i.e. $1$s):
\[A^{ij}_{\alpha\beta}=\mat{cc}{1 & 0\\0 & p}_{i\alpha}\mat{cc}{1 & 1\\1 & 0}_{ij}\mat{cc}{1& 1\\1 & 0}_{\alpha\beta}\]
By varying $p$ from $0$ to $\infty$, we can adjust for the density of occupied sites from $0$ to $1/2$, and can count the number of possible solutions with the corresponding density: in the thermodynamic limit, the law of large numbers dictates that the weight of the partition function will be very peaked around solutions equal to the average density $\rho$, and hence the total number of solutions can be obtained by calculating $Z/p^{\rho N M}$ or exponential scaling $\frac{1}{N M}\log(Z)-\rho \log(p)$. Note that $\rho$ itself can either be calculated as an expectation value in the tensor network, or as a derivative of the partition function with respect to $p$.

This model turns out to exhibit a nontrivial phase transition at a finite density, and such a phase transition is called a jamming transition. The history of such transitions goes back a long time to the birth of the famous Metropolis sampling algorithm, in which Rosenbluth et al. tried to calculate the entropy of hard disks \cite{metropolis1953equation}. The MPO discussed here yields a discretized version of this hard disk problem, in which hard disks are replaced by occupied sites which cannot be adjacent to other occupied sites. If we would have considered a triangular lattice as opposed to a square one, this problem is equivalent to solving the hard hexagon\footnote{Note that the hexagonal lattice is the dual lattice of the triangular lattice} problem, which has been solved exactly by Baxter using an ingenious combination of the Bethe ansatz and corner transfer matrix ideas \cite{baxter1980hard}.

Given a square lattice, another classic problem is to count the number of ways in which we can fill the lattice with dimer configurations. This problem is equivalent to counting the number of ways in which we can label the edges of a square lattice with a bit $0$ or $1$ in such a way that around any vertex there is precisely one edge equal to 1 and three of them are equal to zero. Let us therefore consider the tensor $A^{ij}_{\alpha\beta}$ equal to one iff one of its indices is one and equal to zero otherwise. If we now construct the partition function out of this tensor by putting it on any vertex of the square lattice and contract this two-dimensional tensor network, we will count the number of possible configurations. From the point of view of MPOs, this partition function is equal to $Z=\lim_{N\rightarrow\infty}{\rm Tr}\left(\hat{O}_N\right)^N$. The relevant quantity is $\log(Z)/N^2$, as this yields the exponential scaling of the number of configurations per lattice site.

It turns out that this problem is exactly solvable \cite{kasteleyn1963dimer}. Within the computer science community, the preferred method for finding an exact solution is in terms of Pfaffians. Physicists  prefer to tackle it using the very related method of reinterpreting the transfer matrix as a thermal state of free fermions, which can be diagonalized exactly \cite{lieb1967solution}. This will be discussed in Section~4.

Another classic counting problem was originally formulated by Pauling \cite{pauling1935structure}, where he was able to explain the experimentally determined zero entropy of ice at absolute zero by counting the possible ways in which two of the four vertices of tetrahedra tiling the hexagonal wurtzite crystal are occupied. Solving this  3-dimensional problem would in principle  involve finding the largest eigenvalue of a projected entangled pair operator (PEPO), which is the two-dimensional generalization of the one-dimensional MPO. The two-dimensional version of the problem was formulated by Lieb \cite{lieb1967residual}, and is as follows: given a square lattice, put degrees of freedom $s_{ij}\in \{\uparrow,\downarrow\}$ on the edges. Count in how many ways we can label the edges such that around every vertex there are exactly two spins up and two spins down.

The transfer matrix MPO for this problem can readily be written down:
\begin{equation}
 A^{ij}_{\alpha\beta} = \begin{cases}
1,& i+j+\alpha+\beta = 2\\
0,& i+j+\alpha+\beta \neq 2
\end{cases}
\end{equation}
For obvious reasons, this model is also called the six vertex model, and it turns out that this transfer matrix can be diagonalized exactly using the Bethe ansatz. The Bethe ansatz itself can be formulated in an algebraic way and involves a very nice application of matrix product operator algebras \cite{korepin1997quantum}.

\subsubsection{Equilibrium Statistical physics in 2D}
The central quantity of interest in the context of statistical physics is the partition function $Z$: given a system of $N$ classical spins $\{s_i\}$ and an energy functional $H(s_1,s_2,\cdots)$, then the partition function is given by
\[Z=e^{-\beta F}=\sum_{s_1s_2\cdots}e^{-\beta H(s_1,s_2,\cdots)}.\]
where $F=E-T\ S$ is the free energy which is an extensive thermodynamic quantity. For infinite systems and Hamiltonians with local interactions (such as already encountered in the counting problems), this free energy is proportional to the logarithm of the  leading eigenvalue of a transfer matrix / MPO with bond dimension equal to the local spin dimension. Indeed, let us illustrate this for the case of a translational invariant 2-body nearest neighbour Hamiltonian acting on $d-$level systems. Define the matrix $G$
\[G_{ij}=\exp\left(-H(s_i,s_j)/T\right)\]
then we can parameterize the transfer matrix by the MPO
\[A^{ij}_{\alpha\beta}=\delta_{i\alpha}G_{ij}G_{\alpha\beta}\]
For the case of the ferromagnetic Ising model, $G$ is given by $G=\mat{cc}{e^\beta & e^{-\beta}\\e^{-\beta} & e^\beta}$.

The problem of diagonalizing such transfer matrices is equivalent to the problem of diagonalizing Hamiltonians of quantum spin systems, which is notoriously hard. However, beginning with the density matrix renormalization group (DMRG) \cite{white1992density,baxter1978variational,nishino1995density} a wealth of methods has recently been developed for tackling that problem, which effectively allows one to find the leading eigenvalues and eigenvectors. These methods are effectively variational methods within the manifold of matrix product states, and we will discuss this framework in Section~5. These methods allow one to simulate local classical spin systems in two spatial dimensions to almost machine precision.

The Ising model was originally introduced by Lenz as a toy model for explaining ferromagnetism and its related phase transition from a disordered phase to an ordered one. The first breakthrough happened when Kramers and Wannier wrote down the partition function of the two-dimensional Ising model in terms of the transfer matrix \cite{kramers1941statistics}. They realized that the Ising model is self dual. This means that, by a change of variables, the partition function at high temperature can be mapped to the one at low temperature. This way they could exactly identify the critical temperature, under the condition that there is a phase transition going on.

Onsager's exact solution of this model in two dimensions is without a doubt one of the highlights of $20^{\mathrm{th}}$ century statistical physics. The exact solution forced the whole community to rethink the whole basic formalism of second order phase transitions,  and this was the starting point for studying critical phenomena and culminated in the formulation of the renormalization group. The reason for that upheaval is the fact that the exact critical exponents did not coincide at all with the mean field ones as were predicted. Onsager's solution was obtained by diagonalizing the transfer matrix introduced above; it turns out that it can both be diagonalized using the Bethe ansatz or using the Jordan Wigner transformation as discussed in the case of dimers.

\subsection{Path Integrals in 1+1 D}
An extremely powerful way of dealing with quantum many body systems is given by the path integral approach. It also forms the basis of quantum Monte Carlo methods \cite{foulkes2001quantum,evertz2003loop}. The objective of the path integral formalism is to calculate local expectation values with respect to thermal states of interacting Hamiltonians. This is obtained by  mapping a 1D quantum problem to an effective 2D classical-like partition function. The main idea is to introduce resolutions of the identity within the Gibbs state $\exp\left(-\beta H\right)$ as
\[e^{-\beta H} = \sum_{x_1}\sum_{x_2}\cdots\sum_{x_N}|x_1\rangle\langle x_1|e^{-\frac{\beta}{N}H}|x_2\rangle\langle x_2| e^{-\frac{\beta}{N}.H} |x_3\rangle \cdots \langle x_{N-1}|e^{-\frac{\beta}{N}.H}|x_N\rangle\langle x_N|\]

If the resolutions of the identity have a tensor product structure ($|x\rangle\equiv|x(1)\rangle\otimes |x(2)\rangle \otimes |x(3)\rangle \cdots$), then  $\hat{T}=\langle x_i|e^{-\frac{\beta}{N}H}|x_j\rangle$ is in the form of a matrix product operator, albeit possibly with a very large bond dimension. The ground state of the Hamiltonian $H$ is then nothing else than the leading eigenvector of this transfer matrix.

In principle, is is very hard to deal with this transfer matrix  $T=\langle x_i|e^{-\frac{\beta}{N}H}|x_j\rangle$ as $H$ consists of non-commuting operators. However, for $N\gg 1$, a very good approximation of $\hat{T}=\langle x_i|e^{-\frac{\beta}{N}H}|x_j\rangle$ can readily be found using the Trotter-Suzuki formula.

Let us e.g. consider the Heisenberg antiferromagnetic spin chain with Hamiltonian

\[H_{heis}=\sum_{i} X_iX_{i+1}+Y_iY_{i+1}+Z_iZ_{i+1} \]

where $X,Y,Z$ represent the spin $1/2$ Pauli matrices. Then a natural choice for the Trotter-Suzuki formula is to split the Hamiltonian into the terms that act on the even-odd sites and the odd-even ones:

\[\exp(-\epsilon H)\simeq \prod_i \underbrace{\exp\left(-\epsilon \hat{H}_{2i,2i+1}\right)}_{\exp(-\epsilon)\left(I+\left(exp(4\epsilon)-1\right)|\psi^-\rangle\langle \psi^-| \right)} \prod_i\exp\left(-\epsilon \hat{H}_{2i-1,2i}\right)\]

The leading eigenvector of this MPO will be very close to the ground state of the Heisenberg model. For this particular example, we can rotate every second spin with the unitary operator $Z$, rotating every $\psi^-$ into $\psi^+$. The ensuing matrix product operator only has positive elements, and we can hence interpret is as a classical statistical mechanical model with only positive Boltzmann weights. This makes it possible to use classical Monte Carlo sampling techniques to simulate this model, and this is called quantum Monte Carlo. However, the fact that we could make all weights in the MPO positive is certainly an accident, and it is unknown how to achieve this for many of the most interesting problems; this is precisely the infamous sign problem \cite{troyer2005computational}.

Note that there are many ways in which we can split our Hamiltonian within the Trotter-Suzuki splitting. The one just discussed breaks translational invariance, as the even and odd sites were treated on an unequal basis. It is also possible to keep translational invariance at the cost of breaking the $SU(2)$ invariance by splitting the Hamiltonian into three parts, containing all $X$, $Y$ and $Z$ terms. As shown in \cite{pirvu2010matrix}, an exact representation of $\exp\left(-x\sum_{i}Z_iZ_{i+1}\right)$ can be obtained in terms of a MPO with bond dimension 2:
\[ \exp\left(\alpha\sum_i Z_iZ_{i+1}\right)=\sum_{k_1k_2\cdots}{\rm Tr}\left(C^{k_1}C^{k_2}\cdots\right)Z_1^{k_1}\otimes Z_2^{k_2}\cdots\]
where
\bea C^0&=&\mat{cc}{\cosh(\alpha) & 0 \\ 0 & \sinh(\alpha)}\\
C^1 &=& \mat{cc}{ 0 & \sqrt{\sinh(\alpha)\cosh(\alpha)}\\ \sqrt{\sinh(\alpha)\cosh(\alpha)} & 0}
\eea
The formulas for the Pauli $X$ and $Y$ terms are of course equivalent.

\begin{figure}
\includegraphics[width=6cm]{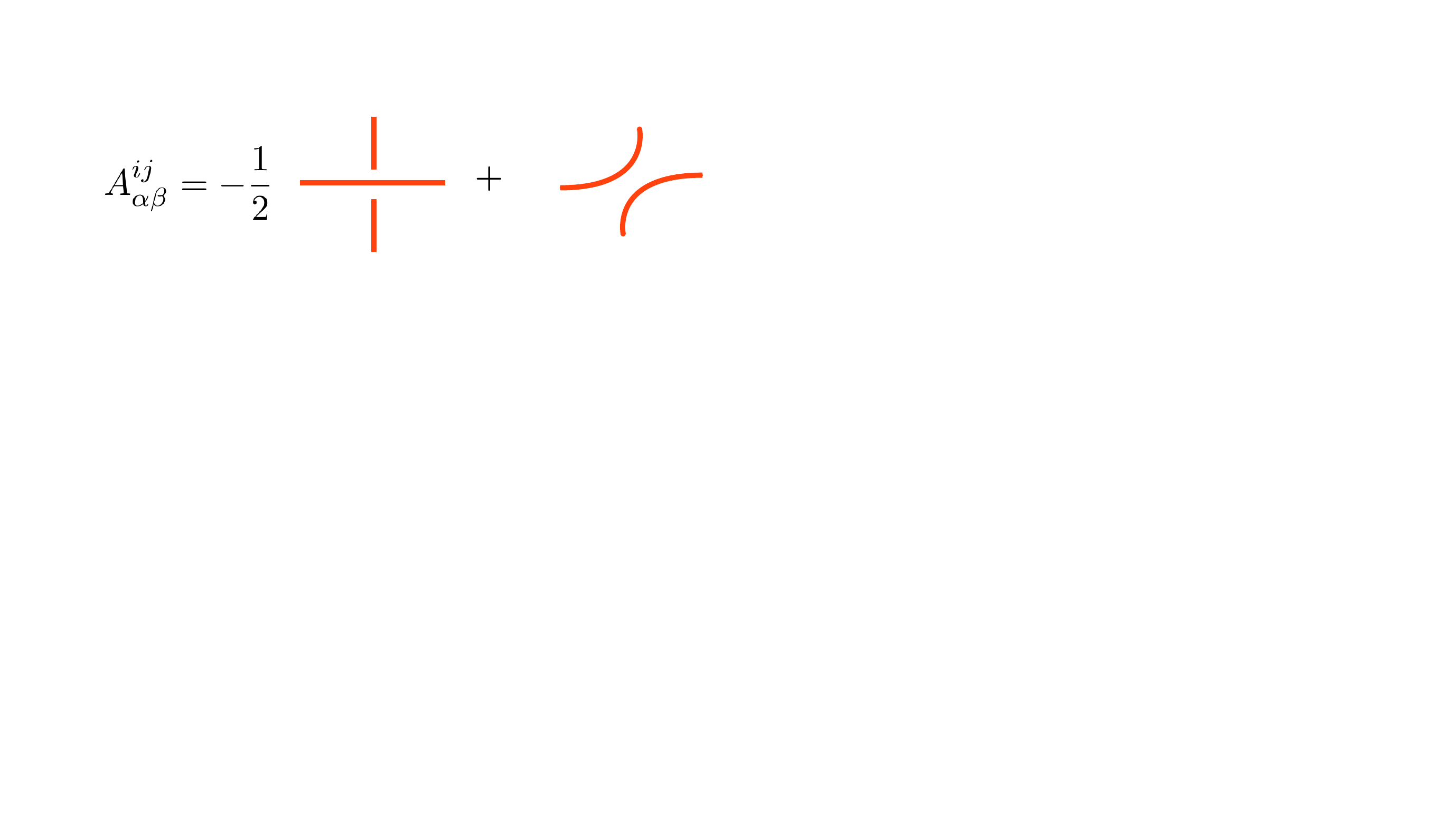}
\caption{ \label{fig4}
 The tensor $A^{ij}_{\alpha\beta}$ for which the leading eigenvector of the corresponding matrix product operator is the ground state of the Heisenberg Hamiltonian.}
\end{figure}

For the particular case of the ground state of the  one dimensional Heisenberg model, the exact ground state can actually be found as the leading eigenvector of the transfer matrix associated to a 6-vertex model. This MPO has bond dimension 2 and is equal to $A^{ij}_{\alpha\beta}=\delta_{ij}\delta_{\alpha\beta}-\frac{1}{2}\delta_{i\alpha}\delta_{j\beta}$, as shown in figure \ref{fig4}.

\subsection{Matrix product operators in PEPS: physics at the edge}
\begin{figure}
\includegraphics[width=12cm]{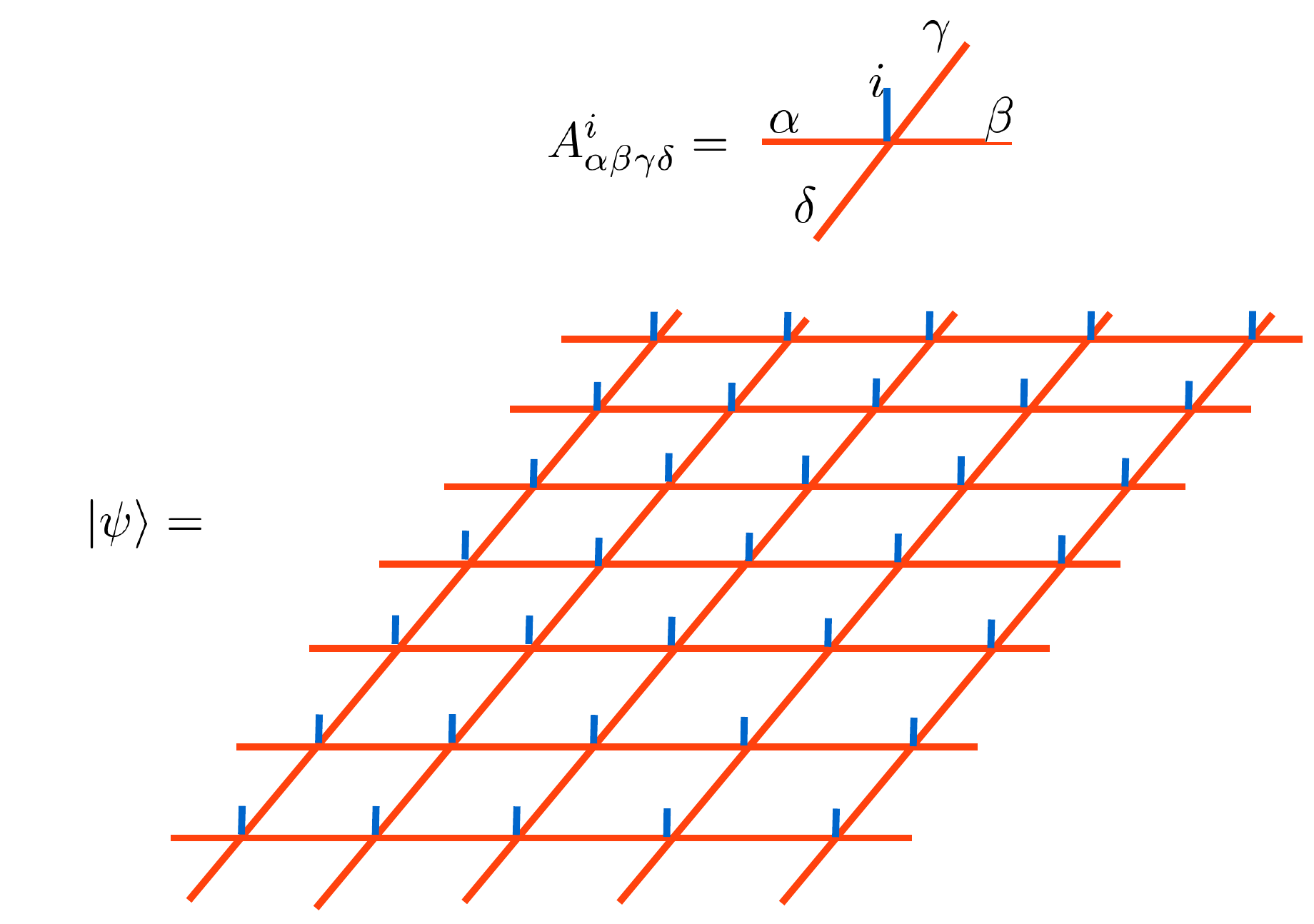}
\caption{\label{fig5}
 Definition of a projected entangled pair state parmaterized by the tensor $A^{i}_{\alpha\beta\gamma\delta}$.}
\end{figure}

Projected entangled pair states (PEPS) \cite{verstraete2004renormalization,verstraete2008matrix,dukelsky1998equivalence,nishino2001two} provide a systematic way of parameterizing ground state wavefunctions of strongly correlated systems on two-dimensional lattices. PEPS form the natural generalization of matrix product states to higher dimensions. The archetypical example of a PEPS is the wavefunction of Affleck-Kennedy-Tasaki-Lieb on the square lattice \cite{Affleck2004}, which is a spin-2 antiferromagnet obtained by projecting halves of four ``virtual'' singlet pairs onto the spin-2 subspace. More generally, a translation invariant PEPS on a square lattice is parameterized by a tensor $A^i_{\alpha\beta\gamma\delta}$, with $i$ the physical dimension and $ \alpha,\beta,\gamma,\delta$ virtual indices having dimension $D$; see figure \ref{fig5}.

\subsubsection{PEPS transfer matrix}
\begin{figure}
\includegraphics[width=0.8\textwidth]{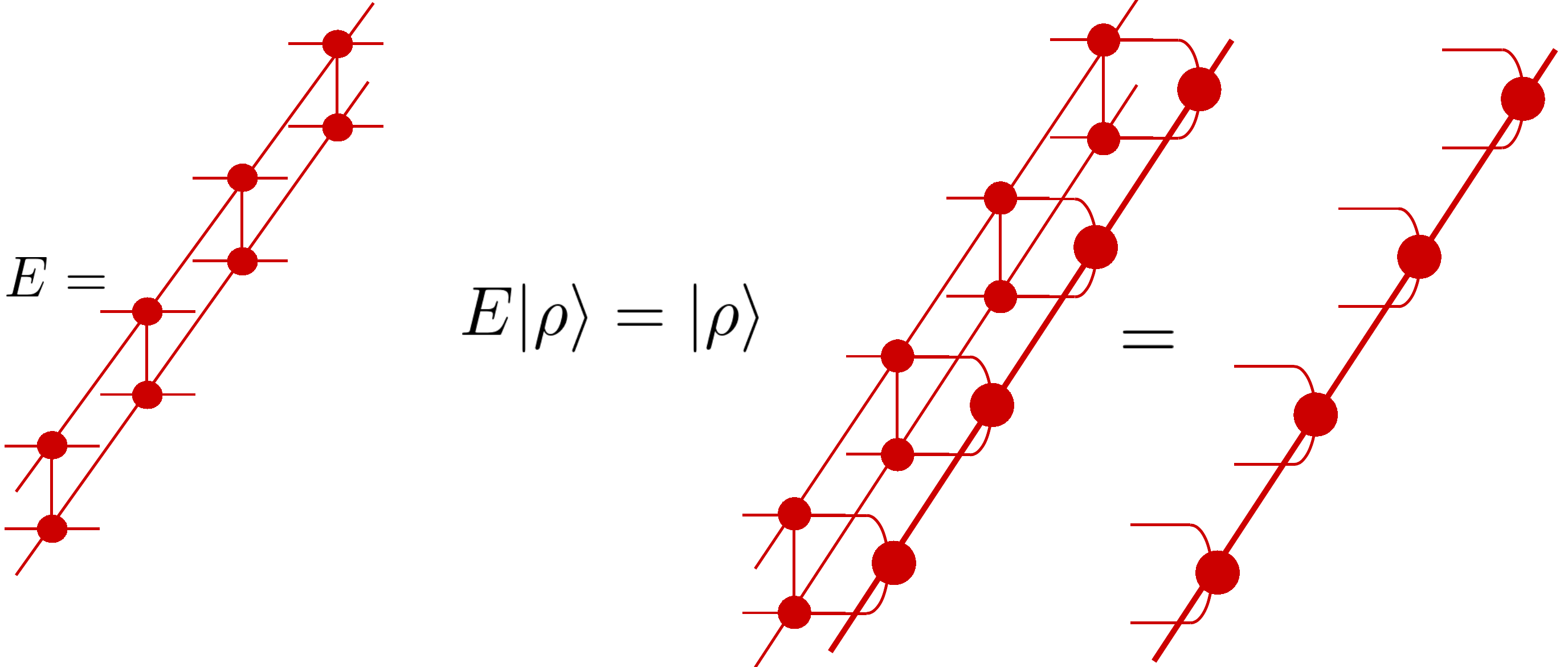}
\caption{\label{fig6} Definition of a projected entangled pair state parameterized by the tensor $A^{i}_{\alpha\beta\gamma\delta}$.}
\end{figure}

The complexity of calculating expectation values of observables is equal to the complexity of calculating expectation values of classical spin systems in 2D. Indeed, the quantity $\langle \psi|O|\psi\rangle$ involves the contraction of a tensor network in which the bra and the ket are connected to each other. Just as the central interest in 2D classical systems is contained in the eigenstructure of the transfer matrix, here the essential ingredients are encoded into the leading eigenvector of the double layer transfer matrix (see Figure \ref{fig6}), which is a vector in a Hilbert space of dimension $D^{2N}$. Due to the symmetric double layer structure of the transfer matrix, it makes however more sense to interpret this transfer matrix as a completely positive map on operators of dimension $D^N\times D^N$. The theory of completely positive maps (quantum Frobenius theorem \cite{evans1978spectral}) then dictates that the leading eigenvector can be chosen to be a positive operator $\rho$. The quantum analogue of ergodicity is called injectivity, and injectivity ensures the uniqueness of this fixed point \cite{perez2008peps}.

From the computational point of view, the class of PEPS seems to be a very rich variational class by which it should be possible to describe the physics of many gapped phases of matter. The central difficulty and bottleneck in implementing the variational optimization is exactly the step of finding the leading eigenvector of the transfer matrix, and the motivation of coming up with better algorithms for finding fixed points of MPOs stems to a large extent from this problem.

\subsubsection{Entanglement Hamiltonians}
As the leading eigenvector of the transfer matrix can be interpreted as a hermitian operator itself, it has itself eigenvalues and eigenvectors. From the point of view of the theory of entanglement, those eigenvalues are of great interest as they represent the so-called entanglement spectrum \cite{li2008entanglement} and their entropy is the entanglement entropy which exhibits the famous area law \cite{eisert2010colloquium}. Much of the recent progress in understanding exotic and strongly correlated phases of matter stems from studying the structure of those eigenvalues.

The framework of tensor networks is very well suited for studying this entanglement spectrum in more detail \cite{cirac2011entanglement}. As opposed to other methods, tensor networks also give access to the eigenvectors corresponding to the entanglement spectrum, which is especially interesting from the point of view of topological phases where one might study how those eigenvectors transform under certain symmetries.

As shown in Section~5, it is possible to find the leading eigenvector of the double-layer transfer matrix of the PEPS in terms of an MPS of a certain bond dimension $\chi$, and as just discussed, this MPS can effectively be interpreted as an MPO of the same bond dimension. As the double layer transfer matrix is typically not hermitian, the left and right eigenvectors are typically different. Let us call the corresponding MPOs $\rho_L$ and $\rho_R$. Without loss of generality, we can always rescale $\rho_L,\rho_R$ such that they are hermitian and positive. In analogy to the way in which the entanglement spectrum is determined in 1D MPS, the entanglement spectrum is obtained by calculating the singular value decomposition of the operator $\sqrt{\rho_L}\sqrt{\rho_R}$, or equivalently the eigenvalue decomposition of $\sqrt{\rho_L}\rho_R\sqrt{\rho_L}$, the logarithm of which we call the entanglement Hamiltonian. Note that this operator has the same eigenvalues as the operator $\rho_L\rho_R$, which is the product of two MPOs and hence an MPO itself. The entanglement spectrum can be obtained by finding the eigenvectors and eigenvalues of this MPO. We will show that numerical techniques based on MPS allow to do exactly that. 

\subsubsection{Matrix product operators and topological order}
PEPS turn out to be also very useful for characterizing systems with topological quantum order. In that case, PEPS exhibit  nontrivial symmetries on the virtual level \cite{schuch2010peps,buerschaper2014twisted,csahinouglu2014characterizing,bultinck2015anyons}. Such systems with topological order are currently under intense study as they present a natural framework for constructing quantum error correcting codes which allow to build fault tolerant quantum memories and quantum computers.

The simplest of those systems is the so-called toric code \cite{kitaev2003fault}, originally introduced by Kitaev to exemplify the connection between quantum error correcting codes and systems exhibiting topological quantum order and anyonic excitations. By blocking 4 sites into 1, one can parameterize one of the ground states by a translational invariant PEPS with bond dimension 2 \cite{verstraete2006criticality,schuch2010peps}:
\[ A^{ijkl}_{\alpha\beta\gamma\delta}=\delta_{i\oplus j,\alpha}\delta_{j \oplus k,\beta}\delta_{k\oplus l,\gamma}\delta_{l\oplus i,\delta} \]
with $\oplus$ addition modulo $2$. It can easily be seen that this PEPS is not injective, i.e.\ if we look at the tensor as a matrix from the physical level to the virtual one, then the matrix is not full rank. The full subspace on  which the virtual system has support is in this case parameterized by the projector
\[ P_4 = \frac{1}{2}\left(I+Z^{\otimes 4}\right)\]
Here $Z$ is the Pauli $Z$ spin $1/2$ matrix. Note that this is a very simple MPO with bond dimension 2 which we encountered in the examples of Section~2: $A^{ij}_{\alpha\beta}=\delta_{\alpha\beta} (Z_{ij})^{\alpha}$.  Note also that the same MPO characterizes the virtual subspace that a larger block of spins has support on.

\begin{figure}
\includegraphics[width=12cm]{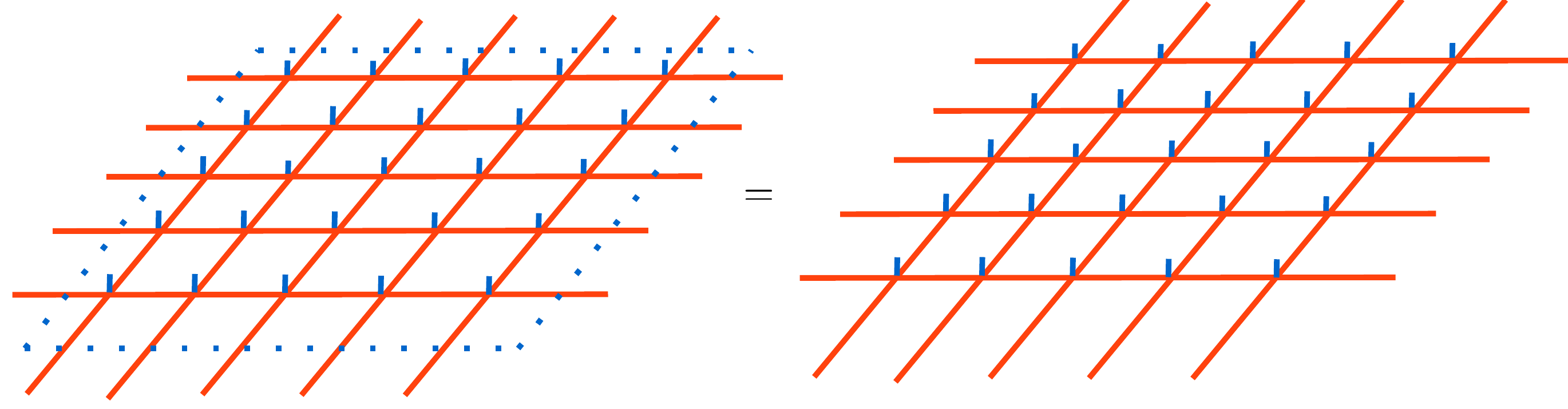}
\caption{\label{fig7}
For a topologically ordered projected entangled-pair state, the symmetry on the virtual level is characterized by a set of matrix product operators (dotted).}
\end{figure}

Let us indeed consider a contiguous block of spins on the lattice, and consider the entanglement entropy of this block. A well established property of topological systems is the fact that this entanglement entropy scales like the length $L$ of the boundary of this block (so-called ``area law''), and that there is a constant correction $\gamma$ to this area law which characterizes the anyonic content of it: $S(\rho_L)=c L-\gamma$ \cite{kitaev2006topological,levin2006detecting}. This $\gamma$ turns out to be exactly related to the the non-injectivity of the PEPS: it arises as a consequence of  the hidden symmetry on the virtual level of the PEPS. This hidden symmetry is characterized by a set of MPOs (see figure \ref{fig7}), and the full richness of the topological theory (including e.g. anyon statistics and braiding properties) can be deduced from studying the algebra generated by those MPOs \cite{csahinouglu2014characterizing,bultinck2015anyons}. The so-called quantum double models and string nets \cite{levin2005string} provide systems for which those MPOs can be constructed analytically. In the case of the toric code discussed before, those MPOs are obtained by bringing the MPO defined before into normal form, which is trivial in this case as it is the sum of 2 MPOs $\hat{O}_i$ of bond dimension 1: $\hat{O}_0$ is the identity matrix and $\hat{O}_1$ the tensor product of $Z$s. The ensuing algebra is in this case given by $\hat{O}_i\hat{O}_j=\hat{O}_{i\oplus j}$.

\begin{figure}
\includegraphics[width=12cm]{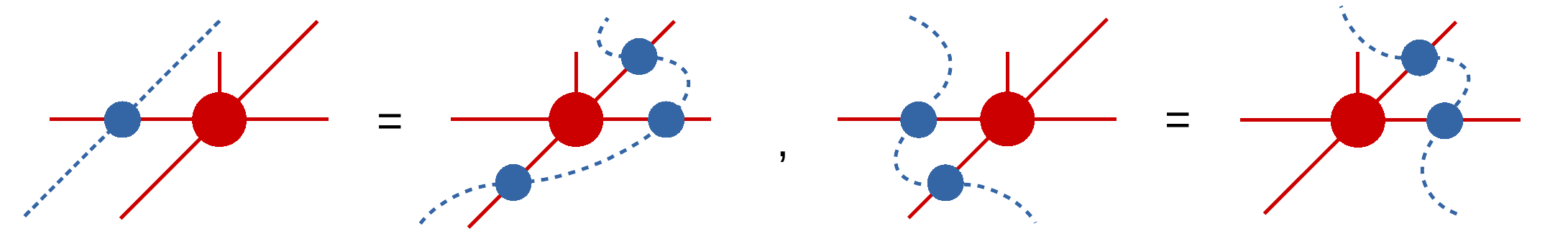}
\caption{\label{fig8}
 The local ``pulling through'' condition in topological projected entangled-pair states.}
\end{figure}

The way in which those symmetries are encoded in the PEPS tensors can be most easily understood in terms of the ``pulling through'' condition, pictorially given in figure \ref{fig8}; those MPOs can be interpreted as Wilson loops acting on the virtual level, but opposed to Wilson loops, they remain exact symmetries even in the presence of a finite correlation length. A consequence of this pulling through condition is the fact that the different MPOs commute with the double layer transfer matrix:
\[\left[\hat{O}_i\otimes \hat{O}_j,T\right]=0\]
Those symmetries are of course reflected in its leading eigenvectors, and the way in which those symmetries are broken characterize the different ways by which the different anyons in the theory can condense \cite{schuch2013topological,haegeman2014shadows}.

Conversely, it is possible to try to characterize all possible sets of MPOs which form a closed algebra. Given such MPOs, it is possible to explicitly construct PEPS wavefunctions for which those MPOs represent their symmetries. The resulting classes of topological systems turn out to be precisely those proposed by Kitaev (quantum doubles \cite{kitaev2003fault}) and Levin and Wen (stringnets \cite{levin2005string}) in the case of spin systems, but certainly allow for more general systems such as the ones consisting of fermions \cite{gu2014symmetry} and/or symmetry protected and symmetry enhanced topological phases \cite{barkeshli2014symmetry}.

\subsection{Nonequilibrium statistical physics and cellular automata in 1D}

Transfer matrices also naturally occur in the context of cellular automata and non-equilibrium physics of one-dimensional systems [see e.g.\ \cite{domany1984equivalence}]. In this context the MPO parameterizes a stochastic matrix which maps a probability distribution defined on a 1-dimensional line of $d$-level systems and hence of dimension $d^N$ to another probability distribution. If this stochastic matrix is ergodic, then Perron-Frobenius theory imposes that it has a unique eigenvector with eigenvalue 1, and this fixed point probability distribution plays the role of the stationary distribution under the stochastic updates.

\subsubsection{Directed Percolation}
As archetypical example, let us consider the problem of directed percolation: imagine an infinite square lattice, and erase any edge randomly with probability $p$; what is the probability that a connected path will exist from one side to the other side? This problem is an idealization of asking the question whether water can percolate through a rock with a certain density of cracks. The so-called Domany-Kinzel cellular automaton \cite{domany1984equivalence} can be considered as the Ising model of this class of percolation problems. Their cellular automaton acts on $N$ bits $x_i$, with $x_i=1$ if there is a particle present and $0$ otherwise. It acts successively on all even and then on all odd sites:

For even times:
\bea x_{2k}(t+1=0& &\textrm{ if }  x_{2k-1}(t)+x_{2k+1}(t)=0\\
 x_{2k}(t+1)=1& \textrm{ with probality } p_1 &\textrm{ if } x_{2k-1}(t)+x_{2k+1}(t)=1 \textrm{ and otherwise } x_{2k}(t+1)=0\\
 x_{2k}(t+1)=1& \textrm{ with probality } p_2 &\textrm{ if } x_{2k-1}(t)+x_{2k+1}(t)=2 \textrm{ and otherwise } x_{2k}(2t+1)=0\\
\eea
For odd times:
\bea x_{2k+1}(t+1)=0& &\textrm{ if }  x_{2k}(t)+x_{2k+2}(t)=0\\
 x_{2k+1}(t+1)=1& \textrm{ with probality } p_1 &\textrm{ if } x_{2k}(t)+x_{2k+2}(t)=1 \textrm{ and otherwise } x_{2k+1}(t+1)=0\\
 x_{2k+1}(t+1)=1& \textrm{ with probality } p_2 &\textrm{ if } x_{2k}(t)+x_{2k+2}(t)=2 \textrm{ and otherwise } x_{2k+1}(t+1)=0\\
\eea

\begin{figure}
\includegraphics[width=12cm]{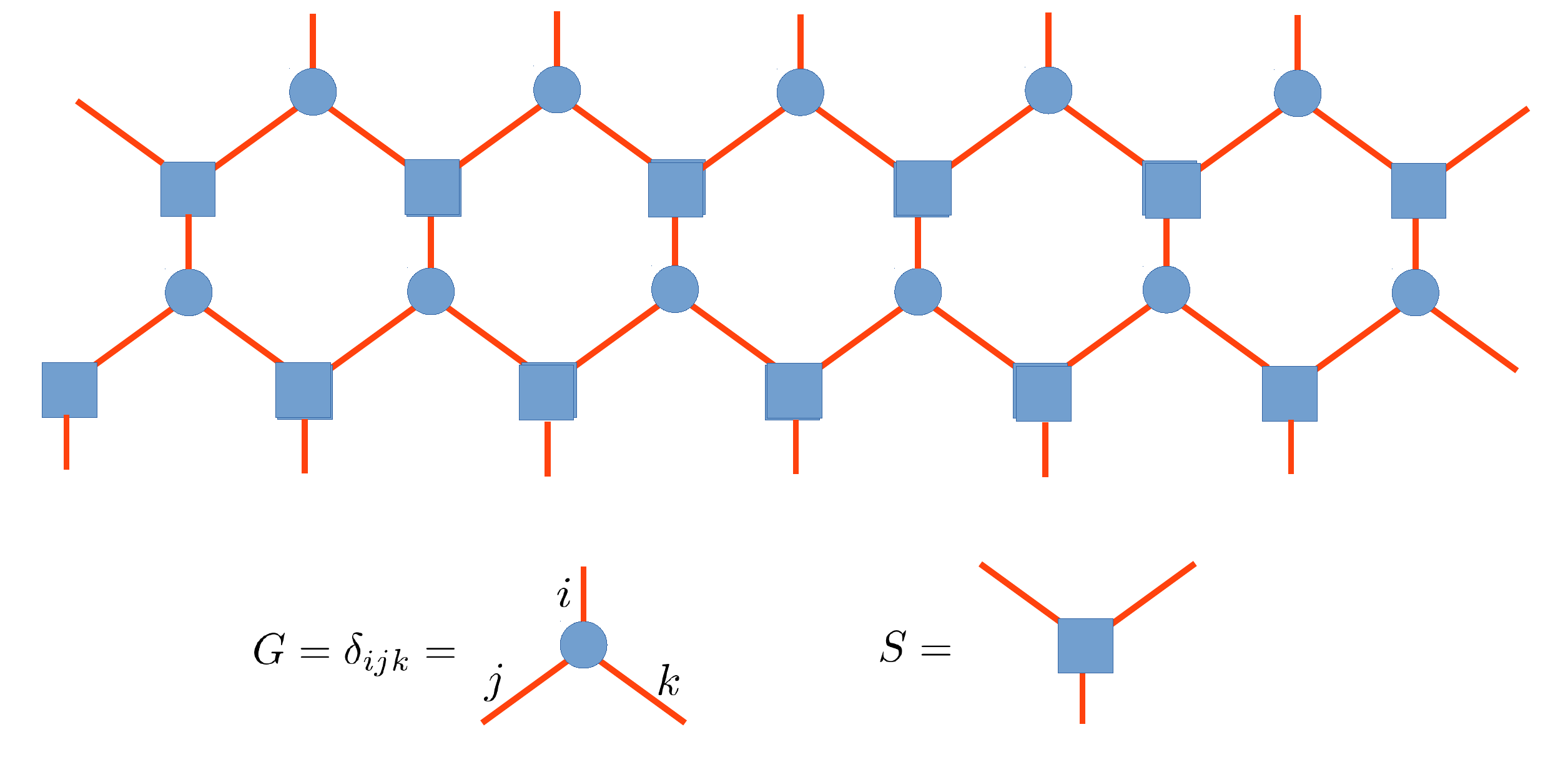}
\caption{\label{fig9}
Matrix product operator representation of the cellular automaton for directed percolation.}
\end{figure}

The physics is contained in the eigenstructure of the transfer matrix MPO for $2$ time steps. It  has bond dimension $4$ and is most easily depicted as in figure \ref{fig9}. Here the different tensors are $G=|0\rangle\langle 00|+|1\rangle\langle 11|$ and
\[S=\mat{cccc}{1 & 1-p_1 & 1-p_1 & 1-p2\\0&p_1&p_1&p_2}\]
The whole MPO is a stochastic matrix. The model exhibits an interesting phase diagram exhibiting second  order phase transitions for a 1-parameter family $p_1(c),p_2(c)$. Those phase transitions are of the class of directed percolation. However, this stochastic matrix is not ergodic, which means that it has several fixed points. Although matrix product state methods have been succesfully used for finding the leading eigenvectors of the continuous time version of this problem (where the MPO becomes a Liouvillian which is a sum of local terms), diagonalizing the above translational MPO in the thermodynamic limit using MPS methods remains a challenge, mainly due to the fact that a trivial (zero-entanglement) fixed point exists. This should be related to the complications which arise in the description of the universal behaviour of the percolation phase transition in terms of non-unitary conformal field theories.

\subsubsection{Asymmetric Exclusion Processes}

A different setting where stochastic cellular automata have been used to a great success is in the field of transport phenomena. Let us consider a line of $N$ bits, and model traffic as a hopping process of particles  from one site to the one to the right of it: this is called the asymmetric simple exclusion process (ASEP) \cite{derrida1998exactly}. More specifically, at every time step, a particle has probability $p$ of moving to the right on the condition that there is no particle to the right of it. Furthermore, there is a constant inflow of particles on site $1$ with probability $p_{in}$ and outflow of particles at site $N$ with probability $p_{out}$. This model exhibits an interesting phase transition as a function of these probabilities which many people experience on a daily basis: if the inflow becomes too large, the traffic gets stuck! In the literature, this model is called the ASEP with fully parallel updates.

\begin{figure}
\includegraphics[width=12cm]{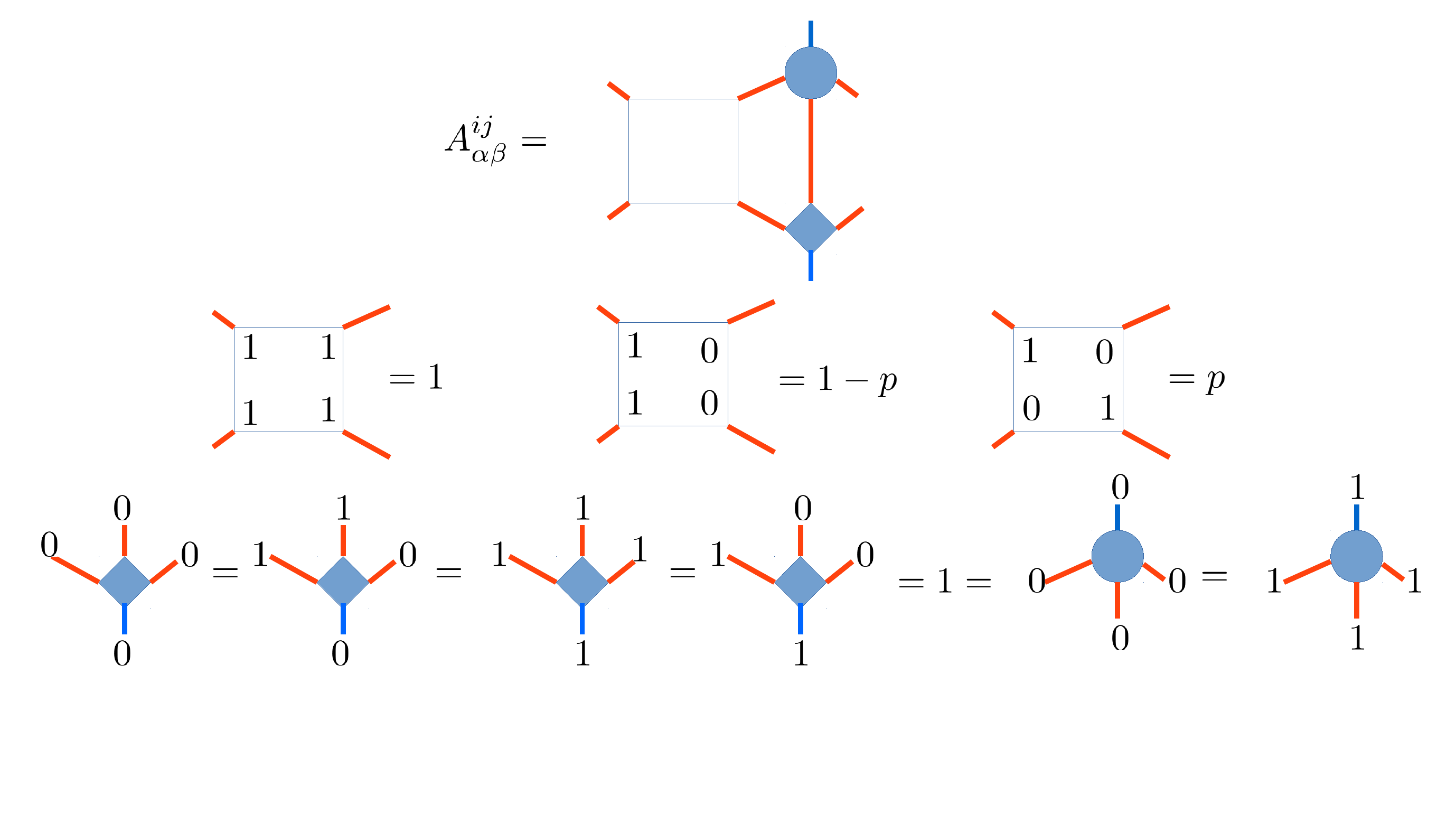}
\caption{\label{fig20}
Matrix product operator representation of the cellular automaton governing the asymmetric simple exclusion process with fully parallel updates.}
\end{figure}

The MPO of this process can readily be constructed, see figure \ref{fig20}. Remarkably, an analytical form can be found for the leading eigenvector of this matrix product operator in terms of matrix product states \cite{evans1999exact,de1999exact}. The essential ingredient of this construction is an explicit construction of a matrix representation of an algebra of operators \cite{derrida1993exact,blythe2007nonequilibrium}. This will be discussed in the section on exact solutions.

\section{Exact methods for diagonalizing matrix product operators}
There exists a vast literature in mathematical physics on exact solutions of eigenvalues and eigenvectors of transfer matrices and MPOs, and most of those exact solutions concern equilibrium and non-equilibrium statistical physics.  In this article, we will only be able to discuss a few remarkable results.

This section is divided in 4 parts. The first part discusses the mapping of a class of transfer matrices to  Gibbs states of  quadratic Hamiltonians of free fermions. This method allows for a huge simplification into diagonalizing the transfer matrix of the classical Ising model in 2D, as originally done by Onsager. As this proof is very well known, we present here the exact solution of another fascinating problem, namely the dimer problem on the square lattice. The partition function of that model was originally solved using Pfaffian methods by Kasteleyn \cite{kasteleyn1963dimer}, but here we follow Lieb \cite{lieb1967solution} and show how a mapping to free fermions allows to diagonalize the transfer matrix exactly.

In a second part, we discuss the algebraic Bethe ansatz, as originally formulated by Faddeev and collaborators \cite{faddeev1980quantum,korepin1997quantum}. A central role in integrable models solvable by the Bethe ansatz is played by an algebra of commuting matrix product operators. We show how an application of the fundamental theorem of MPOs to the associativity of this algebra immediately leads to the so-called Yang Baxter equation, and hence allows to reproduce classic results in the field of the algebraic Bethe ansatz in a very concise way. As an application of that formalism, we show how to diagonalize the transfer matrix of the 6 vertex model, and show how those eigenvectors are also the eigenvectors of the Heisenberg XXZ model. We also show how the coordinate Bethe ansatz can be formulated in terms of MPS, and how that representation is related to the one of the algebraic Bethe ansatz by a gauge transformation \cite{katsura2010derivation}.

In a third part, we apply the logic as the one presented in the part on the algebraic Bethe ansatz, but now on a discrete finite set of MPOs fulfulling a nontrivial closed algebra. This allows us to characterize a large class of topological phases in 2D on the hand of matrix product operators.

In the last part, we discuss the case of finding the leading eigenvectors of  matrix product operators corresponding to nonequilibrium systems. This work was pioneered by B. Derrida and collaborators \cite{derrida1993exact}, and the central tool is to reformulate exact eigenstates in terms of matrix product states with infinite virtual bond dimension. It is then shown that the associated matrices form a very simple algebra, and the matrices satisfying this algebra have a very peculiar form. A central role is again played by the fundamental theorem of matrix product states.

\subsection{Mapping to free fermions}

The most famous method for diagonalizing MPOs involves the mapping of the transfer matrix to a Gibbs state of free fermions; this happens to be possible for quite some interesting models. We illustrate this method by diagonalizing the MPO corresponding to the dimer problem on the square lattice. Recall that the MPO $\hat{O}$ was given by the tensor $A^{ij}_{\alpha\beta}$, which was equal to 1 iff one of its indices is one and equal to zero otherwise. From a conceptual point of view, the only thing we do is a basis change such that $\hat{O}$ becomes a tensor product of $4\times 4$ matrices. We follow Lieb \cite{lieb1967solution} in the following steps:

\begin{itemize}
\item It can readily be checked that the MPO can be rewritten as
\[\hat{O}=\left(\bigotimes_{n=1}^N X_n\right) \exp\left(\sum_{n=1}^{N} \sigma^-_n\sigma^-_{n+1}\right)\]
where $X=\mat{cc}{0 & 1\\1 & 0}$ and $\sigma^-=\mat{cc}{0 & 1\\0 & 0}$. Note that this works because $\sigma^-$ is nilpotent and that all terms $\sigma^-_n\sigma^-_{n+1}$ commute.

\item We now take the square of this MPO, which obviously has the eigenvalues squared:
\bea \left(\hat{O}\right)^2&=&\bigotimes_{n=1}^N X_n \exp\left(\sum_{n=1}^{N} \sigma^-_n\sigma^-_{n+1}\right)\bigotimes_{n=1}^N X_n \exp\left(\sum_{n=1}^{N} \sigma^-_n\sigma^-_{n+1}\right)\\
&=&\exp\left(\sum_{n=1}^{N} \sigma^+_n\sigma^+_{n+1}\right)\exp\left(\sum_{n=1}^{N} \sigma^-_n\sigma^-_{n+1}\right)
\eea

\item Next, we define operators which are fermionic using the Jordan Wigner prescription:
\[\psi_n=(\bigotimes_{m=1}^{n-1}Z_m) \sigma^-_n\]
Those operators and their hermitian conjugates obey the algebra of fermionic annihilation operators $\{\psi_m^\dagger,\psi_n\}_+=\delta_{mn}$ and $\{\psi_m,\psi_n\}_+=0$. The important point is that this algebra is invariant under canonical transformations. One can check that $\hat{O}^2$ can be rewritten in terms of quadratic forms:
\[\hat{O}^2=\exp\left(\sum_n \psi^\dagger_n\psi^\dagger_{n+1}\right)\exp\left(-\sum_n \psi_n\psi_{n+1}\right)\]

\item We then perform a canonical transformation of the fermionic operators  to momentum space
\[\psi_n=\frac{1}{\sqrt{N}}\sum_k \exp(ikn)\chi_k\]
where $k$ ranges from $-\pi+2\pi/N$ to $\pi$.
We now have
\bea \sum_n\psi_n\psi_{n+1}&=&\frac{1}{N}\sum_{n,k,k'}e^{ikn+ik'(n+1)}\chi_k\chi_{k'}=\sum_k e^{-ik}\chi_k\chi_{-k}\\
 \sum_n\psi^\dagger_n\psi^\dagger_{n+1}&=&\frac{1}{N}\sum_{n,k,k'}e^{-ikn-ik'(n+1)}\chi^\dagger_k\chi^\dagger_{k'}=\sum_k e^{ik}\chi^\dagger_k\chi^\dagger_{-k}
\eea
Due to  $e^{-ik}\chi_k\chi_{-k}+e^{ik}\chi_{-k}\chi_{k}=2i\sin(k)\chi_{-k}\chi_{k}$ and the fact that pairs of distinct fermionic operators commmute, we get
\[\hat{O}^2=\bigotimes_{k=2\pi/N}^{\pi(1-2/N)}\underbrace{\exp\left(2i\sin(k)\chi^\dagger_{k}\chi^\dagger_{-k}\right)\exp\left(2i\sin(k)\chi_{k}\chi_{-k}\right)}_{M_k}\] All matrices $M_k$ commute with each other, so the largest eigenvalue of our MPO is just the product of the largest eigenvalue of all of them.
\item $M_k$ can be written out in spin components (inverse Jordan Wigner on 2 modes):
\[M_k=\mat{cccc}{1 & 0 & 0 & -2i\sin(k)\\0&1&0&0\\0&0&1&0\\2i\sin(k)&0&0&1+4\sin^2(k)}\]
The largest eigenvalue is readily obtained as
\[\lambda_{\max}(k)=\left(\sin(k)+\sqrt{1+\sin^2(k)}\right)^2\]
The leading scaling term in the partition function is therefore (where we include an extra factor $1/2$ as we calculated the eigenvalues of $\hat{O}^2$):
\[\begin{split}
\frac{1}{2}\frac{1}{N}\log Z&=\frac{1}{2N}\sum_{k=2\pi/N}^{\pi(1-2/N)}\log\lambda_{\max}(k)\\
&\simeq \frac{1}{2\pi}\int_{0}^{\pi}\mathrm{d}k\,\log\left(\sin(k)+\sqrt{1+\sin^2(k)}\right)\simeq 0.29156\ldots
\end{split}\]
So the number of dimer configurations grows as $(1.33851\ldots)^{N M}$. This classic result was originally proven by Kasteleyn using Pfaffians. Note that this method allows for finding all eigenvectors of the transfer matrix.
\end{itemize}

This diagonalization of the transfer matrix of the dimer model automatically leads to an MPO description of all eigenstates: following \cite{verstraete2009quantum}, any canonical transformation of free fermions can be implemented exactly using a quantum circuit on spin $1/2$s with only nearest neighbour interactions and of depth $N^2$ with $N$ the number of spins involved. This automatically yields an MPO description, as any quantum circuit with nearest neigbour interactions is of course an MPO of a very special kind. Alternatively, we can use the matrix product operator representation of the Bethe ansatz to describe wavefunctions with fermionic statistics.

The exact solution of the classical Ising model can be obtained along  similar lines \cite{schultz1964two}. There is however an interesting complication in the symmetry broken regime, which manifests itself as a degeneracy of the leading eigenvector. To determine the order parameter in this symmetry broken state, one has to do perturbation theory with respect to an external magnetic field within this twofold degenerate subspace, and this leads to the famous magnetization formula of Onsager and Yang $M\sim |T_c-T|^{1/8}$ \cite{yang1952spontaneous}.

\subsection{The Bethe Ansatz}
We first discuss the algebraic Bethe ansatz, as originally formulated by Faddeev and collaborators \cite{faddeev1980quantum,korepin1997quantum}, in terms of an algebra of commuting MPOs. As an application, we diagonalize the transfer matrix of the 6 vertex model, and show how those eigenvectors are also the eigenvectors of the Heisenberg XXZ model. Finally, we show how the coordinate Bethe ansatz, originally formulated by Hans Bethe \cite{bethe1935statistical}, can be formulated in terms of MPS.
\subsubsection{The algebraic Bethe ansatz}\label{aba}
\begin{figure}
\includegraphics[width=8cm]{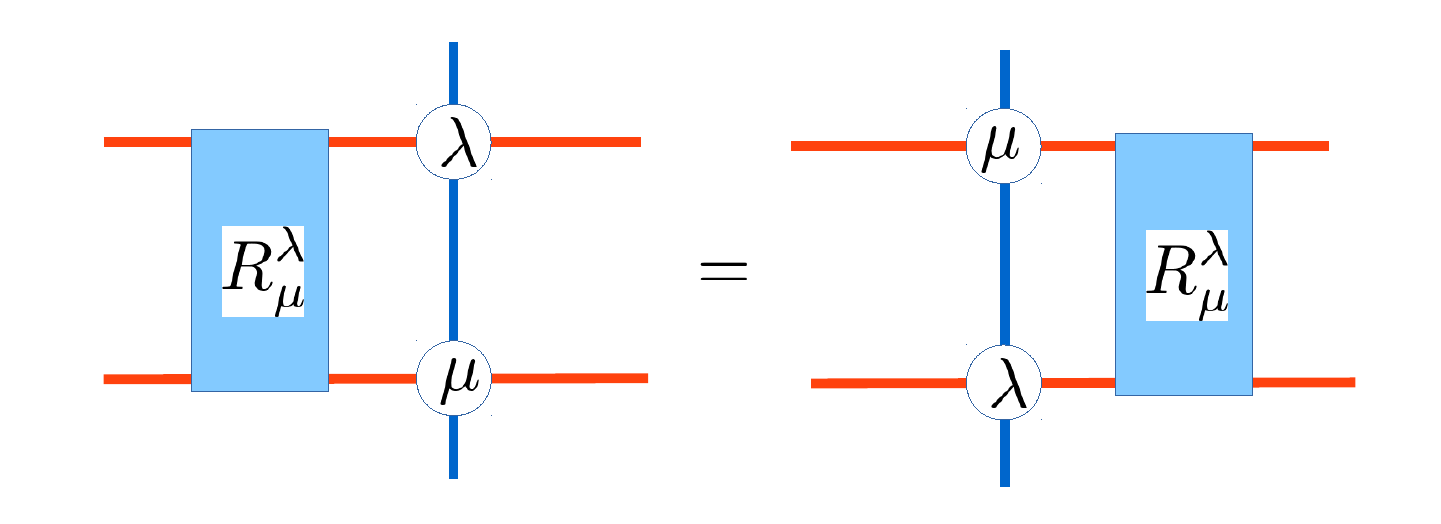}
\caption{\label{fig10}
$R$-matrix as the gauge transformation switching $O(\lambda)$ and $O(\mu)$. }
\end{figure}

Matrix product operators are also at center stage in the algebraic Bethe ansatz, which has allowed for the solution of a multitude of models in statistical physics. The starting point is a continuous one-parameter family of MPOs with periodic boundary conditions, with the special property that any two of those MPOs commute with each other:
\[\forall \lambda,\mu: \left[O(\lambda),O(\mu)\right]=0.\]
The fundamental theorem of MPOs dictates that two (injective) MPOs (in this case $O(\lambda) O(\mu)$ and $O(\mu)O(\lambda)$) are equal to each other if and only if there exists a gauge transform which transforms one MPO into the other one. In other words, there must exist a tensor $R^{\alpha\beta}_{\alpha'\beta'}(\lambda,\mu)$ which acts on the virtual indices which switches the local tensors of $O(\lambda)$ with $O(\mu)$ (see Figure~\ref{fig10}).

Indeed, if there exists a tensor $R$ with that property, we can put $I=R^{-1}R$ somewhere on the virtual level, pull through the $R$ matrix all along the MPO with periodic boundary conditions, and then let it annihilate again as $R R^{-1}=I$. This way of identifying two different MPOs by pulling through a different tensor $R$ is also the way in which the non-equilibrium problems will be solved, and also plays a crucial role in the numerical treatment of diagonalizing MPOs.

\begin{figure}
\includegraphics[width=12cm]{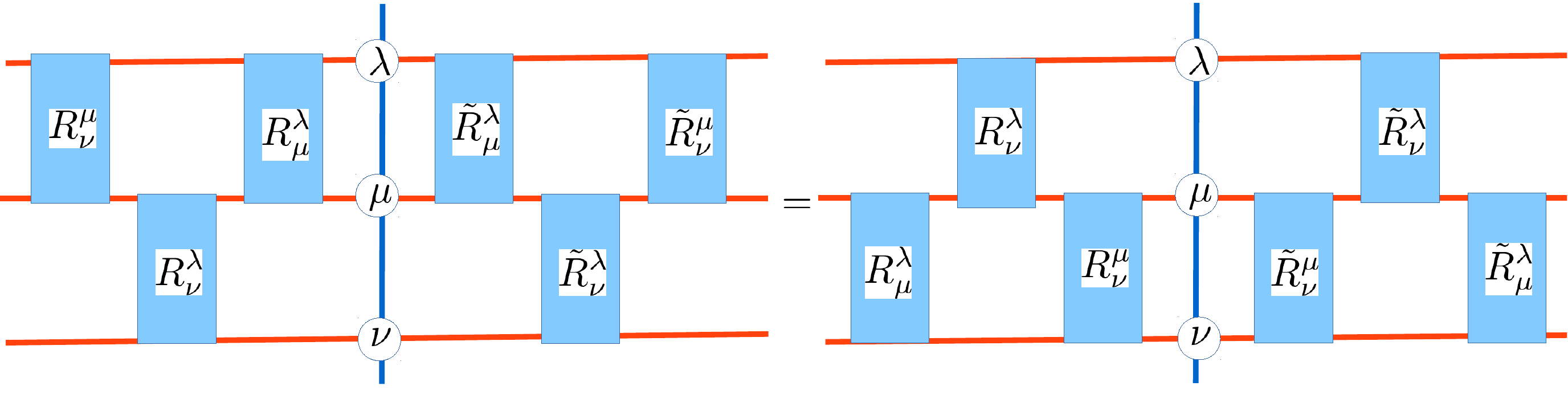}
\caption{\label{fig12}
Yang-Baxter equations as associativity conditions for the $R$-matrices in the algebraic Bethe ansatz. We use the notation $\tilde{R}$ for the inverse of $R$.}
\end{figure}

The essence of the algebraic Bethe ansatz now consists of realizing that those $R$-matrices have to satisfy an associativity condition, and that the corresponding algebraic equations yield a very rigid framework for possible solutions.  The associativity property of the $R$-matrices follows from the fact that there are two different ways in which we can reorder three commuting MPOs in a local way (see Figure~\ref{fig12}). The two triples of R-matrix contractions hence have to be equal to each other, i.e.
\[R^{\alpha\beta}_{\alpha'\beta'}(\mu,\nu)R^{\beta'\gamma}_{\beta"\gamma'}(\lambda,\nu)R^{\alpha'\beta"}_{\alpha"\beta'''}(\lambda,\nu)=R^{\beta\gamma}_{\beta'\gamma"}(\lambda,\mu)R^{\alpha\beta'}_{\alpha'\beta"}(\lambda,\nu)R^{\beta"\gamma"}_{\beta'''\gamma'}(\mu,\nu)\]
where Einstein summation convention is assumed. These equations are called the Yang-Baxter equations \cite{yang1968s,baxter1971eight}.

The solutions of these Yang-Baxter equations are very restrictive, and effectively all integrable models in statistical physics can be constructed from those solutions. It turns out that the defining property of the $R$-matrix, as illustrated in Figure~\ref{fig10}, is of the exact same form as this Yang-Baxter equation. Given such a set of solutions of the Yang-Baxter equations $\{R(\lambda,\mu)\}$, it is now easy to construct a specific set of MPOs $\{O(\lambda)\}$ satisfying the required commutativity relations. This is the so-called fundamental representation which is obtained by choosing $A^{ij}_{\alpha\beta}(\lambda)=R^{\alpha i}_{j\beta}(\lambda,0)$.


Let us now illustrate how to proceed for finding the eigenvectors of this set of commuting MPOs for the simplest nontrivial model, the XXX model. We will follow the steps as explained in the book of Korepin, Bogoliubov and Izergin \cite{korepin1997quantum}:

\begin{enumerate}
\item The simplest solution of the Yang-Baxter is given by the $R$-matrices
\[R(\lambda,\mu)=\mat{cccc}{f(\mu,\lambda) & 0 & 0 & 0\\0 & g(\mu,\lambda) & 1 & 0\\0 & 1 & g(\mu,\lambda) & 0\\0 & 0 & 0 & f(\mu,\lambda)}\]
with $g(\mu,\lambda)=i/(\mu-\lambda)$ and $f(\mu,\lambda)=1+g(\mu,\lambda)$ for the specific XXX model. The corresponding fundamental representation is of the form $A^{ij}_{\alpha\beta}(\lambda)=i\lambda\delta_{ij}\delta_{\alpha\beta}+\delta_{i\alpha}\delta_{j\beta}$ (see Figure~\ref{fig11}) and hereafter we have constructed all MPOs with those elements.

\begin{figure}
\includegraphics[width=6cm]{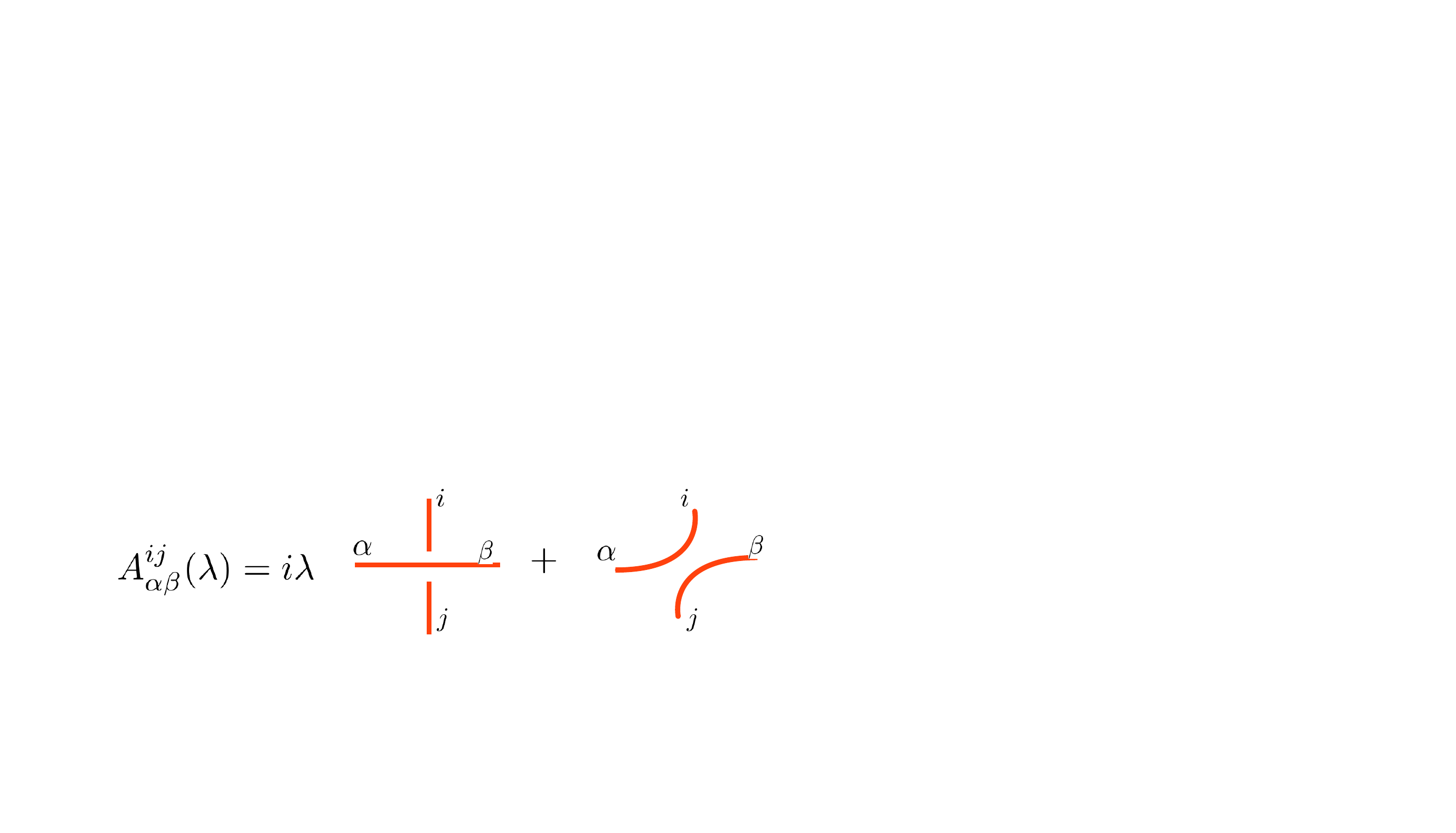}
\caption{\label{fig11}
Local tensors of the matrix product operator in the XXX model.}
\end{figure}

\item It is now possible to construct 4 different families of MPOs by varying the  boundary conditions $M$:

\[\mat{cc}{1 & 0\\0 & 0}\hspace{1cm} \mat{cc}{0 & 0\\1 & 0}\hspace{1cm}
\mat{cc}{0 & 1\\0 & 0}\hspace{1cm}
\mat{cc}{0 & 0\\0 & 1}\]
We will call the 4 corresponding MPOs $A(\lambda)$, $B(\lambda)$, $C(\lambda)$, $D(\lambda)$, where $T=A+D$ corresponds to closing with the identity matrix $M=1$. As can easily be checked, the state $|\Omega\rangle=|000\cdots 0\rangle$ is an eigenstate of $A(\lambda)$, $C(\lambda)$, $D(\lambda)$ with eigenvalues  $a(\lambda)=(1+i\lambda)^N$, $0$, $d(\lambda)=(i\lambda)^N$, respectively.

\item The Yang-Baxter relations impose the following commutation relations on those MPOs:
\bea \left[A(\lambda),A(\mu)\right]&=&0=\left[B(\lambda),B(\mu)\right]=\left[D(\lambda),D(\mu)\right],\\
 A(\mu)B(\lambda)&=&f(\mu,\lambda)B(\lambda)A(\mu)-g(\mu,\lambda)B(\mu)A(\lambda),\\
 D(\mu)B(\lambda)&=&f(\lambda,\mu)B(\lambda)D(\mu)-g(\lambda,\mu)B(\mu)D(\lambda).
\eea
We can now check that
\bea A(\mu)\prod_i B(\lambda_i)|\Omega\rangle=\Lambda\prod_i B(\lambda_i)A(\mu)|\Omega\rangle+\sum_n \Lambda_n B(\mu)\prod_{j\neq n}B(\lambda_j)A(\lambda_n)|\Omega\rangle,\\
D(\mu)\prod_i B(\lambda_i)|\Omega\rangle=\tilde{\Lambda}\prod_i B(\lambda_i)D(\mu)|\Omega\rangle+\sum_n\tilde{\Lambda}_n B(\mu)\prod_{j\neq n}B(\lambda_j)D(\lambda_n)|\Omega\rangle,
\eea
The coefficients for $\Lambda$ and $\tilde{\Lambda}$ are easily obtained, as there is only one way by which $A(\mu)$ and $D(\mu)$ were commuted to the very right:
\bea \Lambda &= &\prod_i f(\mu,\lambda_i),\\
\tilde{\Lambda}&=&\prod_i f(\lambda_i,\mu)
\eea
The equations for $\Lambda_n$ look much more complicated, as there are in principle exponentially many ways ($2^{n-1}$) by which $A(\lambda_n)$ could be commuted to the very right. Note that there is only one way by which the term with $A(\lambda_1)$ can be obtained, and the coefficient in front of the corresponding term is thus easily obtained. Without loss of generality, we could however have reordered all commuting $B(\lambda_i)$, and hence could have obtained all other terms in the same way. This leads to the following equations:
\bea
\Lambda_n&=&g(\lambda_n,\mu)\prod_{j\neq n}f(\lambda_n,\lambda_j),\\
\tilde{\Lambda}_n&=&g(\mu,\lambda_n)\prod_{j\neq n}f(\lambda_j,\lambda_n).
\eea
\item If we can now find solutions $\{\lambda_i\}$ for which $a(\lambda_n)\Lambda_n+d(\lambda_n)\tilde{\Lambda}_n$ vanishes for all $n$, then we clearly have found an eigenstate of $T(\mu)=A(\mu)+D(\mu)$. It turns out that the equation $a(\lambda_n)\Lambda_n+d(\lambda_n)\tilde{\Lambda}_n=0$ exactly corresponds to the well known Bethe equations which also appear in the coordinate Bethe ansatz. For many cases, and in particular for the case considered here, it is known how to obtain a complete set of solutions of those nonlinear equations, and every different solution corresponds to an orthogonal eigenstate $\prod_i B(\lambda_i)|\Omega\rangle$ with eigenvalue $a(\mu)\Lambda+d(\mu)\tilde{\Lambda}$. $B(\lambda_i)$ can hence be interpreted as a creation operator, creating a particle with (quasi-)momentum $\lambda_i$ and energy $f(\mu,\lambda_i)$.
\end{enumerate}

It is interesting to note that the algebraic Bethe ansatz yields a matrix product ansatz for all eigenstates, albeit one where the dimension scales exponentially with the number of particles and/or applications of $B$ operators \cite{murg2012algebraic}. Indeed, e.g.\ the spin zero eigenstates on a chain with $N$ sites are obtained by multiplying $N/2$ MPOs $B(\lambda)$ with bond dimension $2$ with the vacuum, yielding an MPS with bond dimension $2^{N/2}$.

By constructing an algebra of MPOs, we have hence shown that is is possible to diagonalize a set of commuting MPOs $T(\mu)$. Most of the integrable models in statistical physics can be solved along those lines, and there is a very rich mathematical literature dedicated to working out further intriguing algebraic properties of those equations.

\begin{figure}
\includegraphics[width=12cm]{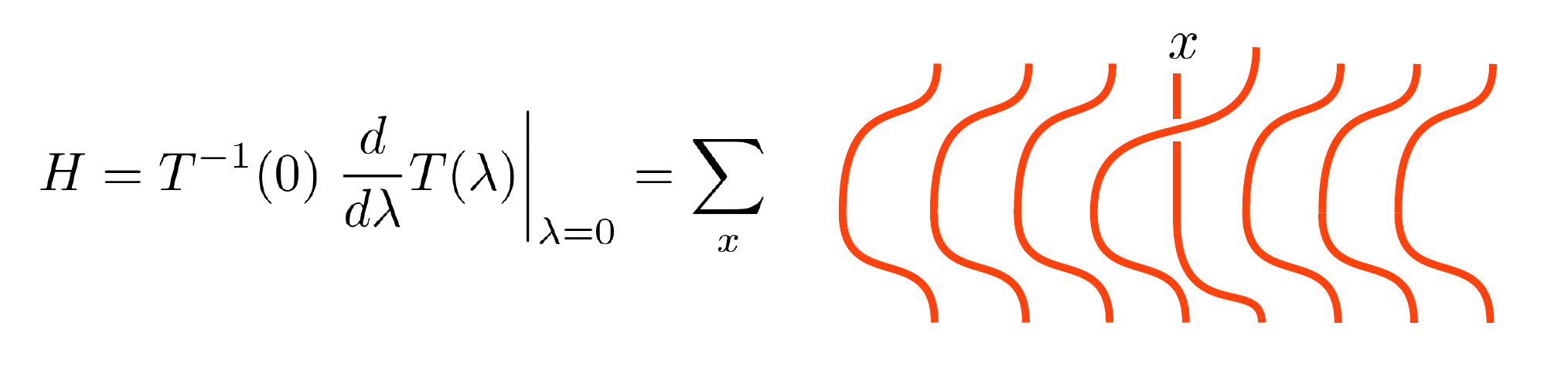}
\caption{\label{fig13}
The logarithmic derivative of the 6-vertex matrix product operator yields the Heisenberg spin 1/2 Hamiltonian.}
\end{figure}

It turns out that there is a  beautiful  connection between the above XXX classical statistical model and the Heisenberg antiferromagnet. Figure~\ref{fig13} illustrates the fact that the logarithmic derivative of the MPO $T(\lambda)$ at $\lambda=0$ $H=T^{-1}\frac{\mathrm{d}\ }{\mathrm{d}\lambda} T(\lambda)|_{\lambda=0}$ is precisely the Heisenberg model. As all $T(\lambda)$ commute, this automatically implies that they share the same eigenstates. The leading eigenvector of $T(\lambda)$ corresponds to the ground state of the Heisenberg antiferromagnet for $-1\leq \lambda\leq 0$. From the point of view of numerics, the optimal choice seems to be $\lambda=-1/2$, as this value maximizes the gap of the transfer matrix.

A similar construction can in principle be done for the one-dimensional Hubbard model, which was also shown to be integrable by Lieb and Wu \cite{lieb1968absence}. Numerical MPS method can readily be applied to this model, which allows to calculate correlation and spectral functions.

\subsubsection{The coordinate Bethe ansatz}
Historically, the Bethe ansatz was first formulated in terms of first quantization \cite{bethe1935statistical} and called the coordinate Bethe ansatz. We will show that this formulation can also immediately be converted into an MPS description with exponential bond dimension. It will turn out that this MPS description is equivalent to the MPS description emerging from the algebraic approach up to a gauge transform \cite{katsura2010derivation}.

Let us consider the Heisenberg spin $1/2$ ferromagnet for simplicity, although similar results can easily be derived for more complicated integrable models. Bethe assumed that the empty vacuum state is all spins down, and wrote down an ansatz for all eigenstates of a given magnetization $M$ with spins up at positions $n_1,n_2,\cdots n_M$:
\[ \psi\left(n_1,n_2,\cdots,n_M\right)=\sum_{\mathcal{P}}\exp\left(i\sum_{j=1}^M n_j\xi_{\mathcal{P}_j}\right)\exp\left(i\frac{1}{2}\sum_{j<k} \theta\left(\xi_{\mathcal{P}_j},\xi_{\mathcal{P}_k}\right)\right)\]
Here $\mathcal{P}$ is the set of all permutation operators, the $\xi_i$ are the spectral parameters, $\xi_{\mathcal{P}_j}$ is the $j^{\mathrm{th}}$ spectral parameter after the permutation, and $\theta\left(\xi_i,\xi_j\right)$ is the antisymmetric logarithm of the scattering matrix which can be determined from the two-particle problem. The spectral parameters are obtained by imposing that this ansatz is an actual eigenstate, leading again to the Bethe equations.

It is remarkably simple to write an MPS in the second quantized form which is equivalent to $\psi$. Let us first consider the case of $1$ particle, which has the form
\[|\psi\rangle = \frac{1}{\sqrt{N}}\sum_{j=1}^N e^{i\xi j}\psi^\dagger_j|\Omega\rangle\]
and which can readily be identified with the MPS $A^0(\xi)=D(\xi)$, $A^1=J$, $M=V$:
\[ D(\xi)=\mat{cc}{e^{i\xi} & 0\\0 & 1}\hspace{1cm} J=\mat{cc}{0 & e^{i\xi}\\0 & 0}\hspace{1cm} V=\mat{cc}{0 & 0\\1 & 0}\]

The $2$-particle case with spectral parameters $\xi_1,\xi_2$ can then be obtained by introducing $A^0=\bigotimes_i D\left(\xi_i\right)$, $M=\bigotimes_i V$ and $A^1=B_1+B_2$ with
\bea B_1&=& \mat{cc}{e^{i\left(\xi_2+\frac{1}{2}\theta(\xi_1,\xi_2)\right)} & 0\\0 & e^{i\frac{1}{2}\theta(\xi_2,\xi_1)}}\otimes \mat{cc}{0 & e^{i\xi_1}\\0 & 0},\\
B_2&=& \mat{cc}{0 & e^{i\xi_2}\\0 & 0}\otimes \mat{cc}{e^{i\xi_1} & 0\\0 & 1}.
\eea
The important property of those matrices is the fact that $B_i \left(A^0\right)^k B_j\cdots B_i =0$, and the boundary conditions then imply that both of them have to appear exactly once in the MPS wavefunction. If $B_1$ appears to the left of $B_2$, then the scattering phase $\theta\left(\xi_1,\xi_2\right)/2$ is introduced, and in the other case the minus $\theta\left(\xi_2,\xi_1\right)/2$ is introduced. This is clearly a MPS of bond dimension $4$, and we can readily proceed to the $M$-particle case:
\bea A^0&=&\bigotimes_j \mat{cc}{e^{i\xi_j} & 0\\0 & 1}\hspace{1cm} A^1=\sum_{j=1}^M B_j\hspace{1cm} M=\bigotimes_j \mat{cc}{0 & 0\\1 & 0}\\
B_j&=&\bigotimes_{k=j+1}^M  \mat{cc}{e^{i\left(\xi_k+\frac{1}{2}\theta\left(\xi_j,\xi_k\right)\right)} & 0\\0 & e^{i\frac{1}{2}\theta\left(\xi_k,\xi_j\right)}}\otimes\mat{cc}{0 & e^{i\xi_j}\\0 & 0}\bigotimes_{k=1}^{j-1}\mat{cc}{e^{i\xi_k} & 0\\0 & 1}
\eea
We can indeed check that this ordering reproduces the right phases. The resulting  MPS has bond dimension $2^M$. This proves that an MPS is able to encode the permutation properties of Bethe ansatz wavefunctions very easily. Note also that the case of free fermions can readily be obtained by imposing $\theta(\lambda_i,\lambda_j)=\pi$, which makes the wavefunction antisymmetric. The matrices involved in the MPS satisfy the following commutation relations, which are related to the Zamalodchikov algebra of creation operators in a 2-dimensional quantum field theory \cite{zamolodchikov1979z}:
\bea B_jA^0 & = & e^{i\xi_j}A^0 B_j\\
\left(B_j\right)^2&=&0\\
B_j B_k&=& e^{i\theta\left(\xi_j,\xi_k\right)}B_k B_j
\eea

Those relations were first discovered by Alcaraz and Lazo \cite{alcaraz2003bethe,alcaraz2006generalization}, whose goal was to find a matrix product ansatz for eigenstates of integrable systems. They showed that the condition of an MPS to be an eigenstate of the Heisenberg Hamiltonian is equivalent to those relations, and henceforth managed to find representations of this algebra. At first sight, this MPS solution looks distinct from the one obtained from the algebraic Bethe ansatz. However, Katsura and Maruyama constructed a gauge transformation \cite{katsura2010derivation} which transforms both into each other: The fundamental theorem is at work again and allows going form a first quantized description to a second quantized one! We illustrate this utilizing notation used in the previous section. It can readily be checked that $A^0$ in the MPS description in the algebraic Bethe is upper diagonal. An upper diagonal matrix can be diagonalized by a similarity transformation with an upper triangular matrix $Q$ with $1$s on the diagonal, and this leads to
\[QA^0Q^{-1}=\bigotimes_{j}\mat{cc}{i\lambda_j+1 & 0\\0 & i\lambda_j}\]
Applying the same similarity transform on $A^1$ leads to $QA^1Q^{-1}=\sum_i B_i$ with
\[B_j=\bigotimes_{k>j}\mat{cc}{(1+i\lambda_k)f(\lambda_k,\lambda_j) & 0\\0 & i\lambda_k f(\lambda_j,\lambda_k)}\bigotimes\mat{cc}{0 & 0\\1  & 0}\bigotimes_{k<j}\mat{cc}{i\lambda_k+1 & 0\\0 & i\lambda_k}\]
We can now identify the corresponding terms as a function of $\left\{\lambda_i\right\}$ with the terms obtained in the coordinate Bethe ansatz as a function of $\left\{\xi_i\right\}$.

As an example, let's go back to the 2-dimensional spin ice problem of Lieb, for which the entropy was given by the leading eigenvalue of the MPO
\begin{equation}
 A^{ij}_{\alpha\beta} = \begin{cases}
1,& i+j+\alpha+\beta = 2\\
0,& i+j+\alpha+\beta \neq 2
\end{cases}
\end{equation}
By multiplying the vertical legs of the MPO by  $\sigma_x$ Pauli operators, this MPO tensor is of the form
\[A\equiv \mat{cccc}{1 & 0 & 0 & 0\\0 &  1 & 1 & 0\\0 & 1 & 1 & 0\\0 & 0 & 0 & 1}\]
which is precisely of the same form as the R-matrices introduced in section \ref{aba}. We can hence imbed this MPO in a 1-parameter family of Bethe ansatz integrable models, and find the exact eigenvectors either by means of the algebraic or coordinate Bethe ansatz. The resulting eigenvalue per site is given by Lieb's square ice constant $8\sqrt{3}/9$.

\subsection{Discrete MPO algebras and tensor fusion categories}

The section on the algebraic Bethe ansatz already demonstrated that MPOs can exhibit a very rich algebraic structure. The central ingredient of the algebraic Bethe ansatz arose from the fundamental theorem of MPOs: the global commutativity of an algebra of MPOs leads to the existence of a local $R$-matrix with nontrivial algebraic conditions which follow from associativity. In the case of the Bethe ansatz, the set of MPOs was characterized by a continuous spectral parameter $\lambda$. What about algebras of MPOs with a discrete label? Such algebraic constructions exactly lead to representations of tensor fusion categories \cite{etingof2015tensor}, which form the mathematical basis for describing topological order and theories exhibiting anyons in two spatial dimensions. As found out by Drinfield and Jimbo, such constructions are very much related to the Yang Baxter equations, and the discrete structure is obtained by taking the limiting case of the spectral parameters being equal to infinity. Their work gave rise to the field of quantum groups.

\begin{figure}
\includegraphics[width=8cm]{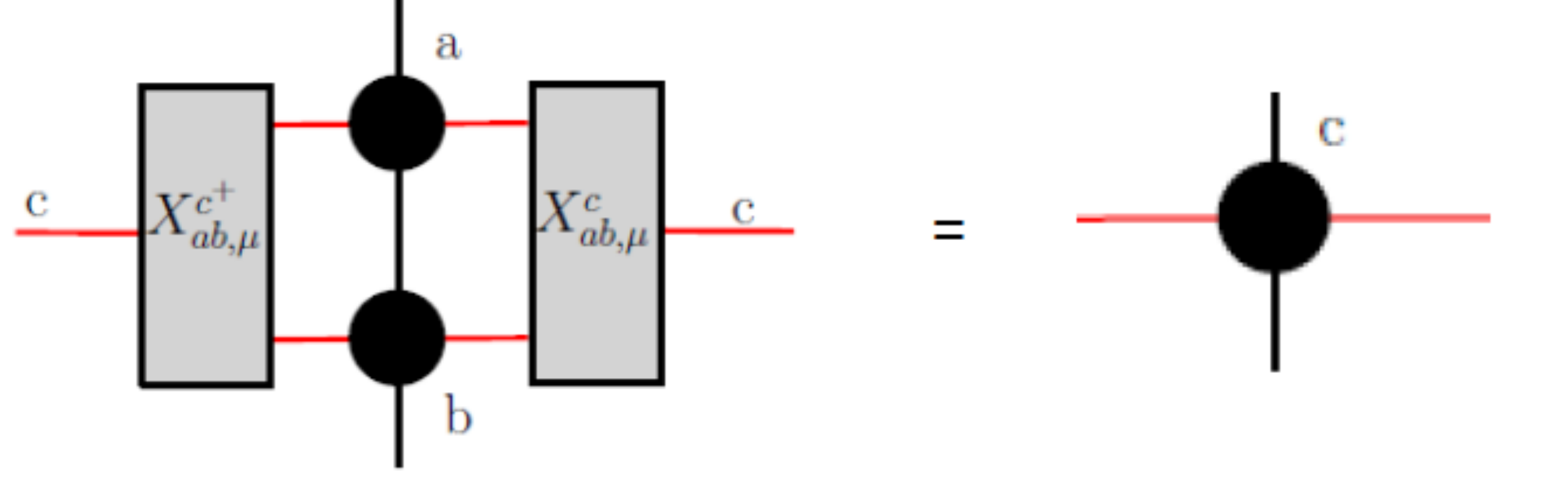}
\caption{\label{fig17}
 The fundamental theorem at work on a closed algebra of matrix product operators.}
\end{figure}

The logic to find solutions parallels the logic followed in the previous section: solutions of the associativity conditions of the gauge transforms will allow us to construct fundamental representations. First of all, we want to construct a discrete set of injective MPOs $\{O_a\}$ which form a closed algebra of MPOs with structure factors independent of the size of the MPOs:
\[O_a O_b=\sum_c N^c_{ab}O_c\]
The tensor $N^c_{ab}$ consists of integers and encodes the so-called ``fusion rules'', i.e.\ the different ways in which the MPOs can fuse into other ones. The fundamental theorem of MPOs then implies that there must be a gauge transform $X^c_{ab\mu}$ (where $\mu$ stands for a possible degeneracy) which decomposes the joint MPO $O_aO_b$ into a direct sum of blocks (see Figure~\ref{fig17}).

\begin{figure}
\includegraphics[width=\textwidth]{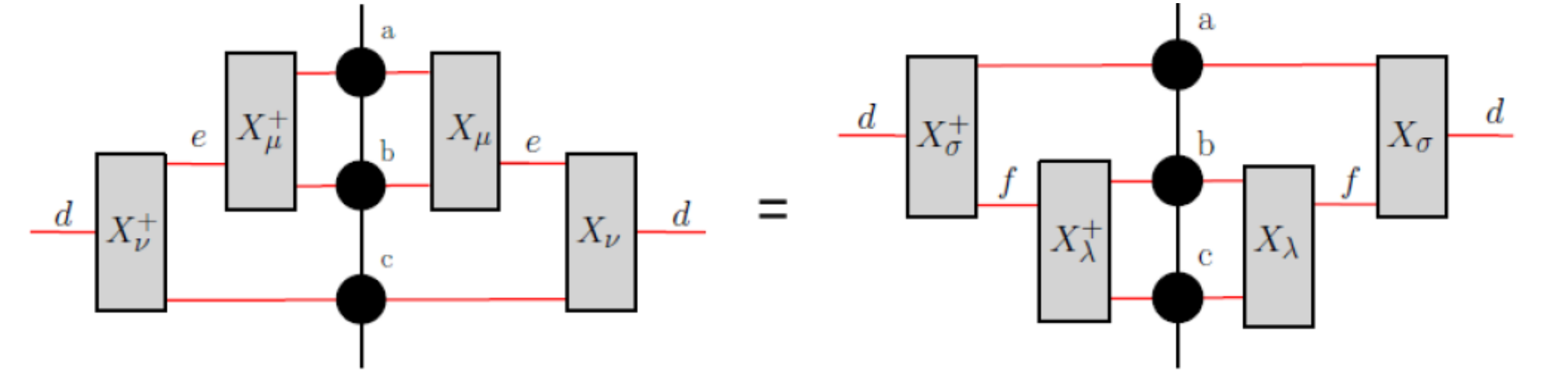}
\caption{\label{fig16}
Associativity conditions for the fusion tensors $X$.}
\end{figure}

\begin{figure}
\includegraphics[width=\textwidth]{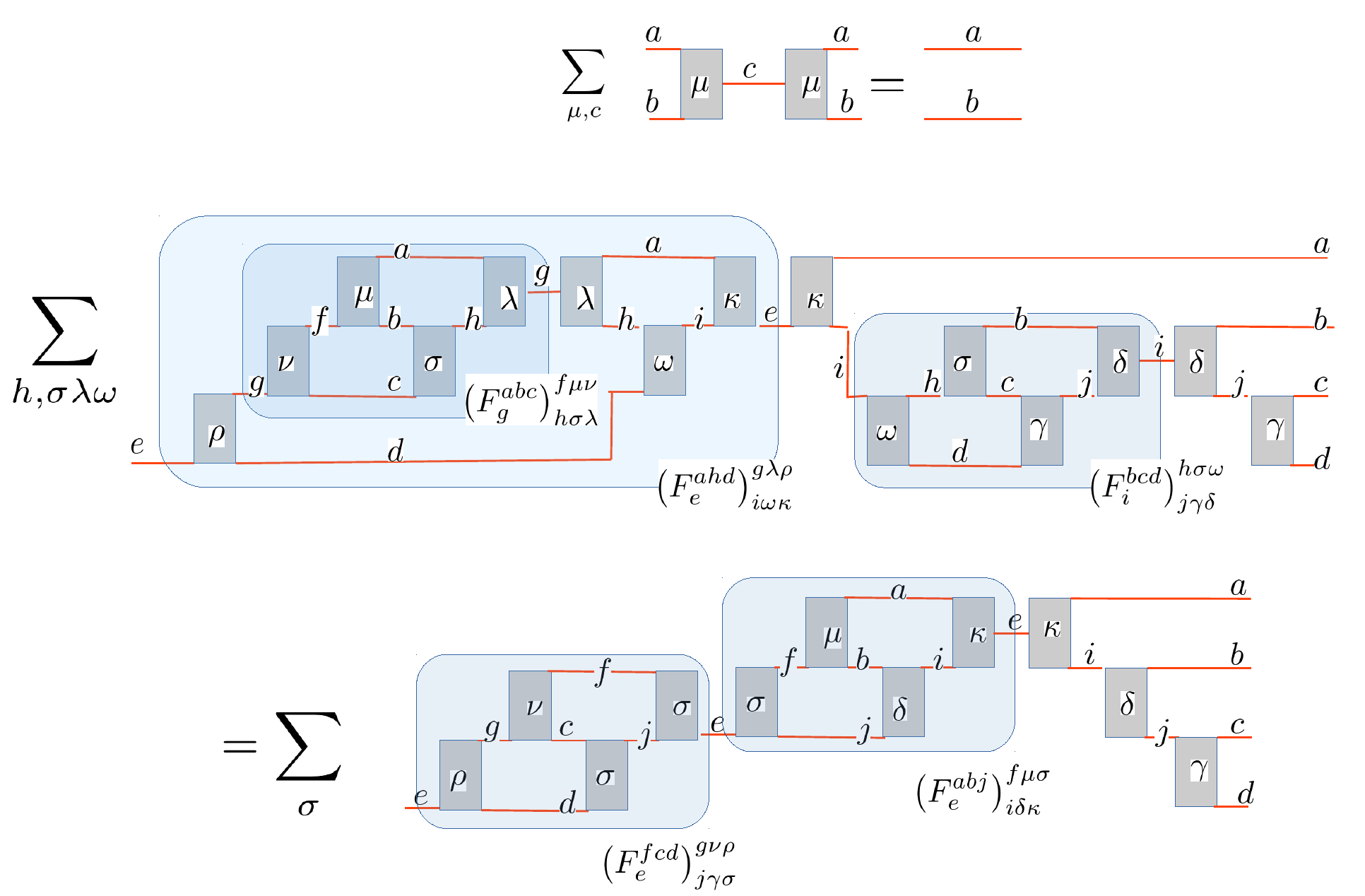}
\caption{\label{fig22}
Pentagon equation for the $F$-symbols obtained by rearranging the order in which tensors are fused.}
\end{figure}

The $X$-tensors have to satisfy the  associativity condition depicted in Figure~\ref{fig16}. Due to the fundamental theorem and the injectivity of the MPOs, this implies the existence of a new tensor, $\left(F^{abc}_d\right)^{f\lambda\sigma}_{e\mu\nu}$, which satisfies the following equation (see Figure~\ref{fig22}):
\[\left(X^e_{ab,\mu}\otimes I \right)X^d_{ec,\nu}=\sum_{f=1}^N\sum_{\lambda=1}^{N^f_{bc}}\sum_{\sigma=1}^{N_{af}^d}\left(F^{abc}_d\right)^{f\lambda\sigma}_{e\mu\nu}\left(I\otimes X^f_{bc,\lambda}\right)X^d_{af,\sigma}\]
This object itself has to satisfy another associativity condition, obtained by rearranging the product of 4 MPO tensors in 2 different ways:
\[\sum_{h,\sigma\lambda\omega}\left(F^{abc}_g\right)^{f\mu\nu}_{h\sigma\lambda}\left(F^{ahd}_e\right)^{g\lambda\rho}_{i\omega\kappa}\left(F^{bcd}_i\right)^{h\sigma\omega}_{j\gamma\delta}=
\sum_\sigma\left(F^{fcd}_e\right)^{g\nu\rho}_{j\gamma\sigma}\left(F^{abj}_e\right)^{f\mu\sigma}_{i\delta\kappa}\]
This equation is the celebrated pentagon equation, and has been studied very extensively recently due to its relevance in the field of topological quantum computation \cite{nayak2008non,wang2010topological}. It is known that for a given set of fusion rules $N^c_{ab}$, there are only a finite set of essentially different solutions.

Just as in the case of the Yang Baxter equation in Bethe ansatz, solutions of the pentagon equation allow us to construct  fusion tensors and MPOs, as well as the related PEPS, by defining all those ``fundamental'' tensors in terms of those $F$-symbols. This construction allows us to define string nets \cite{levin2005string}  and quantum doubles \cite{kitaev2003fault} on arbitrary lattices. Furthermore, this construction decomposes the physical Hilbert space into a direct sum of anyonic sectors: the MPO formalism gives an explicit expression for the elementary excitations (anyons) and their braiding properties in terms of central idempotents of the algebra generated by the ensuing MPOs \cite{bultinck2015anyons}. It is truly remarkable that such a rich algebraic structure emerges out of a very innocuous looking set of MPOs.

\subsection{Non-equilibrium steady states as matrix product states}\label{ss:exact:asep}
Exact fixed points of non-equilibrium stochastic process were found using a formalism introduced by Derrida, Evans, Hakim and Pasquier \cite{derrida1993exact}. Their idea was that such solutions can be found for translation invariant master equations with open boundary conditions using a translation invariant MPS with open boundary conditions. They reduced the problem to an algebraic set of equations for the different matrices $A^i$, and quite a lot of nontrivial systems could be solved that way. Although the matrix product ansatz version of the algebraic Bethe ansatz by Alcaraz and Lazo \cite{alcaraz2003bethe} was actually inspired by the approach of Derrida et al., the solutions are quite different, as the last ones  give rise to  matrices with bond dimensions scaling linearly in the systems size as opposed to exponentially.

\begin{figure}
\includegraphics[width=12cm]{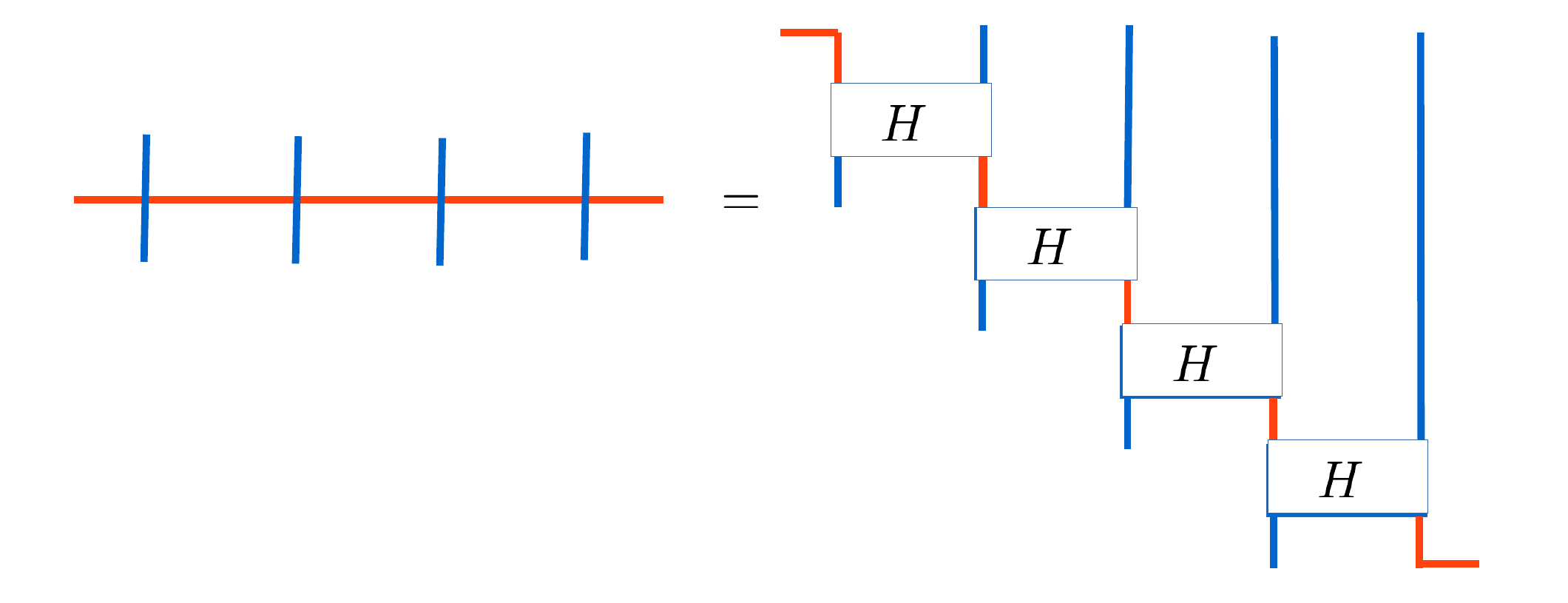}
\caption{\label{fig14}
Rewriting a matrix product operator (MPO) in terms of a staircase. Note that the virtual dimension (red) in the staircase might be larger than the one in the original MPO.}
\end{figure}

In line with this article, let us illustrate their approach for obtaining exact solutions for steady states of stochastic cellular automata. More specifically, let us consider the cellular automaton for the asymmetric exclusion process (ASEP) with fully parallel updates, which is supposedly a good model for traffic and has been solved exactly using the MPS formalism \cite{evans1999exact,de1999exact}. There are 3 parameters, namely the probability that a particle comes in at site 1 ($\alpha$), the probability ($p$) that a particle hope one side forward, conditioned on the fact that the next site is not occupied, and the probability of a particle hopping out at the last site ($\beta$). The goal is to find a class of MPS $\langle L|,A^i,|R\rangle$ with $i\in\{0,1\}$ which depends on $\alpha,\beta,p$ parameterizing the exact fixed points of this cellular automaton:
\[|p\rangle=\sum_{i_1i_2\cdots i_N}\langle L|A^{i_1}A^{i_2}\cdots A^{i_N}|R\rangle |i_1\rangle|i_2\rangle\cdots|i_N\rangle\]
It is easy to rewrite the cellular automaton in terms of a staircase (see figure \ref{fig14}) where $H$ is now a $3\times 2 \times 2\times 3$ tensor with nonzero elements
\[ H=\mat{cc|cc|cc}{1 & 0 &   0 & 0 & 0 & 0\\
                    0 & 1-p & 0 & 0 & p & 0\\
                    0 & 1 & 0 & 0 & 0 & 0\\
                    \hline
                    0 & 0 & 1 & 0 & 0 & 0\\
                    0 & 0 & 0 & 1 & 0 & 0\\
                    0 & 0 & 0 & 1 & 0 & 0}
                    \]
and the boundary stochastic matrices are given by
\[ S_L=\mat{ccc}{1-\alpha & 0 & \alpha\\0 & 1 & 0}\hspace{1cm}
S_R=\mat{cc}{1 & 0\\\beta & 1-\beta\\0 & 1}
\]

\begin{figure}
\includegraphics[width=8cm]{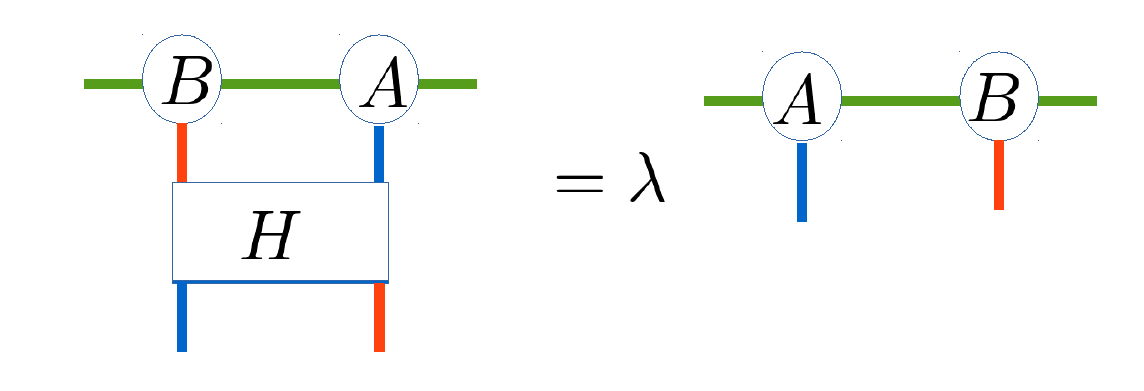}
\caption{\label{fig15}
The zipper condition as a sufficient condition for having an exact eigenvector.}
\end{figure}

A sufficient condition for an MPS $A^i,|L\rangle,|R\rangle$ to be an exact fixed point of the MPO with eigenvalue $1$ is the existence of a tensor  $B^i$, $i\in\{0,1,2\}$ satisfying the ``zipper'' conditions illustrated in figure \ref{fig15}, with $\lambda = 1$. Written out in algebraic terms, we have to find matrices $A^0,A^1,B^0,B^1,B^2$ satisfying
\bea
A^1 B^0&=&(1-p)B^1 A^0 + B^2 A^0\\
A^0 B^2&=&p B^1 A^0\\
A^0 B^0 &=& B^0 A^0\\
A^0 B^1&=& B^0 A^1\\
A^1 B^1&=& B^1 A^1 + B^2 A^1
\eea
and vectors $|L\rangle,|R\rangle$ satisfying
\bea
\langle L|\left(B^2-\alpha A^0\right)&=&0\\
\langle L|\left(B^0-(1-\alpha) A^0\right)&=&0\\
\langle L|\left(B^1- A^1\right)&=&0\\
\left(A^1-B^2-(1-\beta)B^1\right)|R\rangle&=&0\\
\left(A^0-B^0-\beta B^1\right)|R\rangle&=&0\\
\eea
The nature of the game is to make a guess of infinite dimensional matrices exactly satisfying those equations. As exact solutions for the $A^i$ were already obtained in \cite{evans1999exact,de1999exact}, it was straightforward to find solutions for the matrices $B^i$ satisfying the above equations. See appendix A for the explicit representations.

As shown in \cite{evans1999exact,de1999exact}, the exact MPS solution $A^i$ enables one to calculate any property of the steady state such as current, current-current correlations, critical exponents, etc. It is truly remarkable that a translational invariant MPS solution exists, as the physics is completely determined by the non-translational invariant boundary terms. The given solution holds for any number of sites $N$. The recurring theme in all similar solutions of non-equilibrium steady state is the fact that the matrices $A^i$ are band diagonal. Given the fact that the boundary vectors impose the fact that the virtual index starts and halts at the left above corner, the virtual variable effectively undergoes a random walk governed by 1-site hopping terms. This implies that, for a finite system with $N$ sites, the virtual variable can only explore the first $N/2$ indices of the matrices $K_p$ and $K_m$. This implies the very peculiar property that the steady state solution has exponentially fewer correlations than a random steady state solution could have: for a finite system with $N$ sites, the steady state is an exact MPS with bond dimension $N$. In other words, if we would cut the system in two halves, and calculate the singular values of the matrix
\[p^{i_1i_2\cdots i_{N/2}}_{i_{N/2+1}i_{N/2+2}\cdots i_N}\]
then the rank of that matrix would be $N$ as opposed to $2^{N/2}$. The MPS hence yields an exponential compression of the fixed point tensor, and this is precisely the reason why MPS methods are so useful in practice. Unlike in the case of quantum spin chains however, no clear operational meaning has been given to the corresponding Schmidt values; see however \cite{temme2010stochastic} for an attempt to give an operational meaning to the virtual correlations in classical stochastic MPS.

It turns out that it is not a coincidence that the ASEP model has an exact solution in terms of a MPS. In the master equation (continuous time) formulation of the same problem, a similarity transformation in the tensor product form  $\otimes_n D_n$ transforms this master equation into the XXZ Hamiltonian (but with nonhermitian boundary conditions) \cite{essler1996representations}. The steady state problem is then solvable by the algebraic Bethe ansatz, and furthermore all excited states can now also be determined \cite{de2005bethe}. Note however that this leads to an MPS with exponential bond dimension. The boundary conditions must then be such that they only access an exponentially small subspace of the space spanned by the virtual qubits, and just like in the case of the 6-vertex model there should be a gauge transform which makes the MPS block-diagonal.

\section{Numerical methods for diagonalizing matrix product operators}
The previous section illustrated a case where the  leading eigenvector of a stochastic MPO was given exactly by a MPS, whose virtual dimension only scales linearly rather than exponentially in the system size. For quantum spin chain Hamiltonians, the set of MPS has been well established as a powerful variational class thanks to efficient numerical methods such as the DMRG algorithm\cite{white1992density} and later the time-evolving block decimation (TEBD) method\cite{vidal2004efficient,vidal2007classical}. For Hamiltonians with local interactions and an energy gap, the existence of an efficient MPS approximation of the ground state can also rigorously be proven \cite{hastings2007area}.

For leading eigenvectors of MPOs, no such rigorous bounds are available. Nevertheless, efficient numerical methods based on MPS and related ideas can still be formulated, both for finite systems and in the thermodynamic limit. The first of such algorithms is the corner transfer matrix method of Baxter \cite{baxter1978variational}, which was formulated as a variational approach for reflection invariant MPOs (in both directions). Based on White's DMRG method \cite{white1992density}, various generalizations and extension have been formulated \cite{nishino1996corner,bursill1996density,wang1997transfer,shibata1997thermodynamics,sirker2005real,murg2005efficient,orus2009simulation}, including for nonhermitian problems where the eigenvalue problem can no longer be formulated variationally and for non-equilibrium problems \cite{carlon1999density,carlon2001critical}. Here, we will formulate the eigenvalue problem using the differential geometric properties of MPS \cite{haegeman2014geometry}, which also arise in the context of the Dirac-Frenkel time-dependent variational principle \cite{haegeman2011time}.

Our first interest goes towards an extremal eigenvalue, typically the one of largest magnitude. The corresponding eigenvector then represents the fixed point of the MPO, i.e.\ the state obtained after successively applying the MPO (infinitely) many times. Then, we also consider excitations around this fixed point. Throughout the remainder of this section, we focus on translation-invariant MPOs in the thermodynamic limit $N\to\infty$. We furthermore assume that the dominant eigenvalue scales as the exponential of an extensive quantity, i.e.\ $\lambda^N = \exp(-f N)$, where we colloquially refer to $f = -\log \lambda$ as the (not necessarily real-valued) free energy (per site). In addition, we assume that translation invariance is not broken for the dominant eigenvector. Generalisations of the presented algorithms for MPS with a $p$-site unit cell (in case of explicit or spontaneous breaking of translation invariance) or for finite lattices can readily be formulated. We first discuss the properties of the MPS manifold. We then present numerical algorithms for bringing translation invariant MPS into a normal form. Then, we discuss how to formulate the eigenvalue problem for the MPO fixed point in the manifold of MPS. We conclude by presenting a strategy for targeting the excited eigenstates of the MPO around the fixed point.

\subsection{Manifold of uniform matrix product states and its tangent bundle}
As argued, we will restrict our discussion to the set of uniform MPS. Their definition is analogous to that of MPOs, and they are specified by a single (3-index) tensor $A$ and a matrix $M$ as
\begin{equation}
\ket{\Psi(A)} = \sum_{\{i_n\}} \mathrm{tr}\left[ A^{i_{1}}A^{i_{2}}\cdots A^{i_{N}} M\right] \ket{\cdots i_{1} i_{2}\cdots i_{N}}
\end{equation}
where $M$ is assumed to commute with $A^i$, $i=1,\ldots,d$ to ensure translation invariance. The discussion of gauge invariance and normal forms is identical to the MPO case. However, for approximating eigenvectors, we can restrict ourselves to injective MPS tensors $A$, which are also the ones encountered in numerical simulations with unit probability. In that case, expectation values in the bulk become independent of $M$ in the thermodynamic limit. In fact, as the algebra of the matrices $A^i$ and their products span the full $D\times D$ matrix algebra, the only matrix $M$ that commutes with $A^i$ is proportional to the identity.

\begin{figure}
\includegraphics[width=\textwidth]{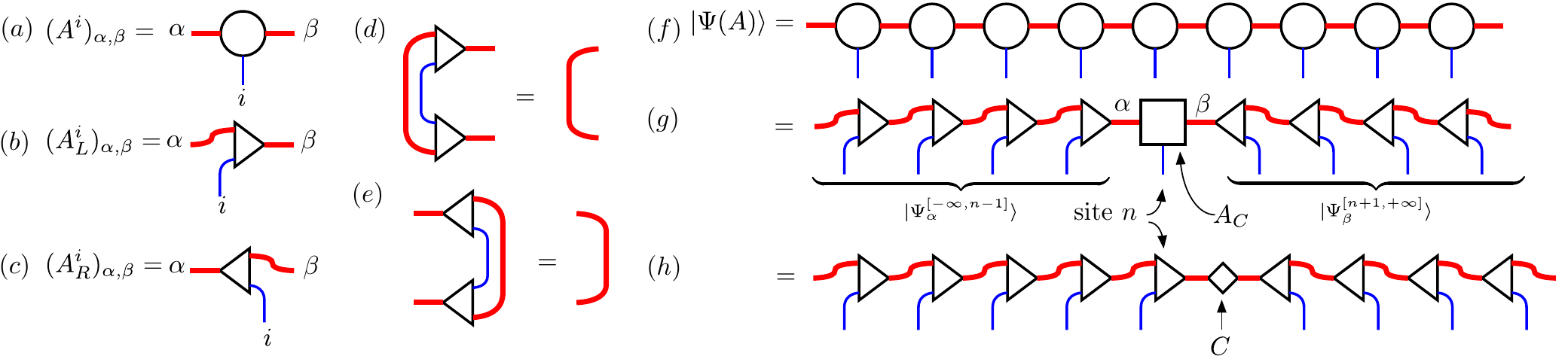}
\caption{ \label{fig:mpsgraphs} The matrix product state (MPS) tensor (a) and its left (b) and right (c) normal forms as defined by the isometric constraints shown in panels (d) and (e). This allows expressing the uniform MPS (f) in the form (g) or (h) corresponding to Eq.~\protect\eqref{eq:mps:ACform} or Eq.~\protect\eqref{eq:mps:Cform}, respectively.}
\end{figure}

Using a suitable gauge transformation (and normalization), the injective MPS tensor $A$ can be brought into a left or right orthonormal form, both of which are defined by the isometric constraints
\begin{align}
	\sum_{i=1}^{d} (A_L^i)^\dagger A_L^i &= \openone &&\text{or} & 	\sum_{i=1}^{d} A_R^i (A_R^i)^\dagger  &= \openone.\label{eq:mps:defALR}
\end{align}
In the numerical algorithms presented below, we will use both normal forms. The gauge transformation that relates both isometries is denoted with $C$ such that
\begin{align}
	A_L^i C = C A_R^i,\quad \text{for $i=1,\ldots,d$}.\label{eq:mps:defC}
\end{align}
With these definitions, we can write the MPS $\ket{\Psi}$ in the thermodynamic limit as
\begin{align}
\ket{\Psi} = \sum_{\alpha,\beta=1}^{D} C_{\alpha,\beta} \ket{\Psi_\alpha^{[-\infty,n]}} \otimes \ket{\Psi_\beta^{[n+1,+\infty]}}\label{eq:mps:Cform}
\end{align}
where the states $\ket{\Psi_\alpha^{[-\infty,n]}}$ ($\ket{\Psi_\beta^{[n+1,+\infty]}}$) constitute an orthonormal basis for support of the state on the left (right) half of the system, and are given as
\begin{align}
 \ket{\Psi_{\alpha}^{[-\infty,n]}} &= \sum_{\{s_{k}\}} (\bra{L}\cdots A_L^{s_{n-1}} A_L^{s_{n}})_{\alpha} \ket{\cdots s_{n-1} s_{n}},\\
 \ket{\Psi_{\alpha}^{[n+1,+\infty]}} &= \sum_{\{s_{k}\}} (A_R^{s_{n+1}} A_R^{s_{n+2}}\cdots\ket{R})_{\beta} \ket{s_{n+1} s_{n+2}\cdots}.
\end{align}
with $\bra{L}$ and $\ket{R}$ irrelevant boundary vectors at $\mp \infty$. That  Eq.~\eqref{eq:mps:Cform} is independent of $n$ follows from the definition in Eq.~\eqref{eq:mps:defC}. By further computing the singular value decomposition $C=U S V^\dagger$ and transforming $A_L$ and $A_R$ with $U$ and $V$ respectively, we obtain the Schmidt decomposition of the state $\ket{\Psi}$ and the left (right) canonical form for $A_L$ ($A_R$), which is unique up to a diagonal unitary matrix. For the remainder, we do not require $C$ to be diagonal. Normalization of the physical state amounts to $\lvert C \rvert_{2}^2 =\mathrm{tr}( C C^\dagger ) =  1$. For further reference, we also define the quantity
\begin{align}
A_C^i = A_L^i C = C A_R^i\label{eq:mps:defAC}
\end{align}
which is known as the center site tensor and allows us to write the state $\ket{\Psi}$ as
\begin{align}
\ket{\Psi} = \sum_{\alpha,\beta=1}^{D}\sum_{i_n=1}^{d} (A_C^{i_n})_{\alpha,\beta} \ket{\Psi_\alpha^{[-\infty,n-1]}} \otimes \ket{i_n} \otimes \ket{\Psi_\beta^{[n+1,+\infty]}}
\label{eq:mps:ACform}.
\end{align}
These definitions are graphically represented in Figure~\ref{fig:mpsgraphs}

The set of (uniform) MPS with a given bond dimension does not constitute a linear subspace of the full Hilbert space $\mathbb{H}$. Rather, the set of (injective) MPS can be shown to form a smooth (complex) submanifold $\mathcal{M}$ of $\mathbb{H}$, with additional properties that make it into a K\"{a}hler manifold \cite{haegeman2014geometry}. At any point $\ket{\Psi(A)}$ of the manifold, we can define a (holomorphic) tangent space $T_{\ket{\Psi(A)}}\mathcal{M}$, which is a linear subspace of $\mathbb{H}$. It is given by the set of states of the form
\begin{align}
\ket{\Phi(B;A)} = \sum_{n=-\infty}^{+\infty} B^{s_n}_{\alpha,\beta} \ket{\Psi_{\alpha}^{[-\infty,n-1]}}\otimes \ket{s_n}\otimes \ket{\Psi_{\beta}^{[n+1,+\infty]}}.\label{eq:mps:deftangent}
\end{align}
This definition is invariant under the substitution $B^s \to B^s + A_L^s X - X A_R^s$ for any $X\in\mathbb{C}^{D\times D}$, which indicates that the parameterisation of tangent vectors is not unique. This is a consequence of the gauge invariance (non-uniqueness) present in the original MPS map $\ket{\Psi(A)}$. A convenient expression for the projector onto $T_{\ket{\Psi(A)}}\mathcal{M}$ was obtained for generic finite MPS in Ref.~\cite{haegeman2014unifying}. In the thermodynamic limit, it readily generalises to uniform MPS \footnote{There is a subtlety which is obscured in the thermodynamic limit, which is that there are $N$ projector terms with a $+$ sign (one for every site) and only $N-1$ terms with a $-$ sign (one for every link between the sites, where the system is supposed to have open boundary conditions). Since every individual projector term $P$ satisfies $P \ket{\Psi(A)} = \ket{\Psi(A)}$, we do indeed recover $P_{T_{\ket{\Psi(A)}}\mathcal{M}}\ket{\Psi(A)} = \ket{\Psi(A)}$, i.e.\ the MPS is itself contained in the tangent space}, i.e.\
\begin{equation}
P_{T_{\ket{\Psi(A)}\mathcal{M}}} = \sum_{n=-\infty}^{+\infty} P^{[-\infty,n-1]} \otimes \openone_{n} \otimes P^{[n+1,+\infty]} - \sum_{n=-\infty}^{+\infty} P^{[-\infty,n]} \otimes \otimes P^{[n+1,+\infty]}.
\end{equation}
Here, the projectors $P^{[-\infty,n]}$ and $P^{[n,+\infty]}$ are given by
\begin{align}
P^{[-\infty,n]} &= \sum_{\alpha=1}^{D} \ket{\Psi^{[-\infty,n]}_\alpha}\bra{\Psi^{[-\infty,n]}_\alpha},&P^{[n,+\infty]} &= \sum_{\alpha=1}^{D} \ket{\Psi^{[n,+\infty]}_\alpha}\bra{\Psi^{[n,+\infty]}_\alpha}
\end{align}
and are clearly independent of the unitary gauge freedom in the definition of $A_L$ and $A_R$ (which amounts to a unitary transformation on the label $\alpha$). The projectors $P^{[-\infty,n]}$ and $P^{[n,+\infty]}$ project onto the support of the reduced density matrix of the state $\ket{\Psi(A)}$ in the respective half-infinite regions of the lattice.

\subsection{Algorithms for obtaining normal forms of uniform matrix product states}
The typical approach [see e.g.\ Ref.~\cite{orus2008infinite}] to compute e.g. the left orthonormal form $A_L$ from a tensor $A$ for the uniform MPS $\ket{\Psi(A)}$ is to first compute the fixed point (leading eigenvector) $\rho_L$ of the completely positive map $\mathcal{E}$ given by
\begin{equation}
\mathcal{E}:\rho \to \sum_{s} (A^i)^\dagger \rho A^i
\end{equation}
using an iterative eigensolver such as the (restarted) Arnoldi method \cite{saad2011numerical,lehoucq1998arpack}. The fixed point $\rho_L$ can be chosen to be hermitian and positive, and $C$ is found as any matrix satisfying $\rho_L = C^\dagger C$, e.g.\ via a Cholesky decomposition. Then, $A_L^i$ is computed by performing the transformation $A_L^i = C A^i C^{-1}$. From a numerical point of view, this approach has some downsides. When $\ket{\Psi}$ provides a good approximation to a physical ground state, it will have several small Schmidt coefficients. When e.g. $A=A_R$ (the right orthonormal form), this is equivalent to small singular values in $C$, which correspond to the square roots of the eigenvalues of the matrix $\rho_L$. Given a machine precision $\epsilon$, the smallest singular values of $C$ determined using this method will only be accurate up to order $\sqrt{\epsilon}$. In fact, $\rho_L$ will typically have some eigenvalues of $\mathcal{O}(\epsilon)$ which can be negative and have to be made positive by hand. Furthermore, as $C$ will also have large singular values, it has a large condition number, such that the use of its inverse in the computation $A_L^i = C A_R^i C^{-1}$ represents a second loss of accuracy. As a consequence, the isometric constraint for $A_L$ obtained in this way can be violated well above the machine precision $\epsilon$.

Here we present an algorithm that is closer in spirit to what is used in MPS algorithms for finite lattices. Given e.g.\ $A_R$, we start from a guess $C^{[0]}$ (e.g. the identity matrix or a better guess obtained from a previous step of the encompassing algorithm). We compute the quantity $A_C^{[n]i} = C^{[n]} A_R^{i}$ and perform a QR decomposition to obtain $A_L^{[n+1]i} C^{[n+1]} = A_C^{[n] i}$. Here, we have reinterpreted the 3-index tensors as rectangular matrices by combining the physical index $s$ with the first virtual index. Using the uniqueness of the QR decomposition (when the diagonal elements of the upper triangular matrix are fixed to be positive), this approach can be shown to converge to an isometry $A_L=\lim_{n\to\infty} A_L^{[n]}$ (exact up to machine precision) and an upper triangular matrix $C=\lim_{n\to\infty} C^{[n]}$. However, the convergence rate of this approach is the same as a that of a simple power method for finding the left fixed point $\rho_L$ of $\mathcal{E}$, which is typically insufficient when $\mathcal{E}$ has a small gap (corresponding to a long correlation lengths in the system). Therefore, after having obtained an updated guess $A_L^{[n+1]i}$, we can further improve the corresponding guess $C^{[n+1]}$ by replacing it with the fixed point $\tilde{C}^{[n+1]}$ of the map $X \to \sum_{s} (A_L^{[n+1]i})^\dagger X A_R^i$, which can be obtained using the Arnoldi method. Although it has the same computational scaling as the aforementioned approach, this algorithm seems slower at first sight because the Arnoldi routine is applied in every iteration step $n\to n+1$. However, as we do not want to obtain the fixed point $\tilde{C}^{[n+1]}$ to machine precision when $A_L^{[n+1]}$ itself is not converged, a single Arnoldi run for some small Krylov subspace dimension is typically sufficient in every iteration step. When carefully implemented, this algorithm (see Algorithm~\ref{alg:leftorth}), can perform equally fast as the standard approach but at a higher accuracy.

\begin{algorithm}
\caption{Gauge transform a uniform MPS $A$ into left orthonormal form\label{alg:leftorth}}\label{euclid}
\begin{algorithmic}[1]
\Procedure{LeftOrthonormalize}{$A,C=\openone, \eta$}\Comment{Initial guess $C$ and a tolerance $\eta$}
\State $\sim,C \gets $ \Call{QRPos}{C} \Comment{Select upper triangular part of $C$, discard unitary part}
\State $C \gets C/\lVert C\rVert$ \Comment{Normalize C}
\State $A_C^i\gets C A^i$
\State $C_{\text{old}} \gets C$
\State $A_L,C \gets $ \Call{QRPos}{$A_C$} \Comment{QR decomposition of matrix $(A_C)_{(\alpha,i),\beta} = (A_C^i)_{\alpha,\beta}$}
\State $\lambda \gets \lVert C\rVert$, $C \gets \lambda^{-1} C$ \Comment{Normalize new $C$ and save norm change}
\State $\delta \gets \lVert C - C_{\text{old}}\rVert$ \Comment{Compute measure of convergence}
\While{$\delta < \eta$} \Comment{Repeat until converged to specified tolerance}
\State $\sim~,C\gets $\Call{Arnoldi}{$X\to \sum_{s} (A_L^i)^\dagger X A^i$, $C$, $\delta/10$}
\State \Comment{Compute fixed point using initial guess $C$ , up to a tolerance depending on $\delta$}
\State $\sim,C \gets $ \Call{QRPos}{C}
\State $C \gets C/\lVert C\rVert$
\State $A_C^i\gets C A^i$
\State $C_{\text{old}} \gets C$
\State $A_L,C \gets $ \Call{QRPos}{$A_C$}
\State $\lambda \gets \lVert C\rVert$, $C \gets \lambda^{-1} C$
\State $\delta \gets \lVert C - C_{\text{old}}\rVert$
\EndWhile
\State \textbf{return} $A_L,C,\lambda$
\EndProcedure
\end{algorithmic}
\end{algorithm}

This algorithm depends on the auxiliary routines \texttt{QRPos}, which computes the QR-decomposition of a matrix with guaranteed positive diagonal elements of $R$ (to ensure uniqueness) and \texttt{Arnoldi}, which computes the leading eigenvector of a linear map up to a specified tolerance using some iterative Arnoldi-based method. It returns the eigenvalue, which we don't use (hence the $\sim$) and the corresponding eigenvector. Note that we should supply the current guess for $C$ as starting vector and that we only want to find an improved guess converged up to a tolerance dependent on the current error $\delta$, e.g. something like $\delta/10$. Typically then, even for a small Krylov subspace dimension, no restarts of the Arnoldi routine within the \texttt{for}-loop should be necessary to reach the specified tolerance. In fact, the outer loop can be interpreted as the restart loop, but where the linear map changes in between the restarts (because $A_L$ is redefined). While this excludes the possibility of using the implicit restart scheme of Sorensen \cite{lehoucq1998arpack}, other tick restart schemes could in principle be used \cite{morgan1996restarting}.

Note that Algorithm~\ref{alg:leftorth} does not depend on the input tensor $A$ being in a right orthonormal form, and does indeed work for any tensor $A$. For a generic tensor, the relation to be found is $C A^i = \lambda A_L^i C$. The positive scalar $\lambda$ represents the norm per site of the state $\ket{\Psi(A)}$, i.e.\ $\lVert \ket{\Psi(A)}\rVert \to  \lambda^N$ in the thermodynamic limit $N\to\infty$. For $A=A_R$, we automatically have $\lambda = 1$ at any step in the algorithm. An algorithm for bringing $A$ into the right orthonormal form can be formulated analogously, or simply be implemented by transposing the matrices $A^i$ and then transposing the output matrices $A_L^i$ and $C$ of Algorithm~\ref{alg:leftorth}.


\subsection{Approximating the MPO eigenvalue problem in the MPS manifold}

In many applications where matrix product operators appear, the primary interest is in the dominant (i.e.\ largest magnitude) eigenvalue and corresponding leading eigenvector of the matrix product operator. The reason is that the MPO appears in a network where it is successively applied $M$ times, and the dominant eigenvalue and corresponding eigenspace projector is what remains in the limit $M\to \infty$, irrespective of the boundary conditions (provided that there is a gap in the spectrum).

Let us first discuss some standard terminology associated with the literature of large scale numerical methods for computing (leading) eigenvectors of (nonhermitian) matrices [see many good books such as Ref.~\cite{saad2011numerical} on this topic]. Given an approximate eigenpair $(\theta,v)$ of a matrix $A$ with $\lVert v \rVert = 1$, the residual is defined as $r = Av - \theta v$. Because of the nonhermiticity, it is important to remember that the left and right eigenvectors corresponding to a given eigenvalue are different and not simply related by (hermitian) conjugation. Typical computational problems that are considered have a dimension $n$ such that the full $n\times n$ matrix $A$ can probably not be stored in memory, and can certainly not be fully diagonalized. However, the corresponding $n$-dimensional vector $v$ can be stored in memory (sometimes using symmetries) and the matrix vector product $A v$ can efficiently be computed. Directly implementing the idea of applying the operator a large number of times to a random initial vector (i.e. computing $\lim_{M\to\infty} A^M v$) is known as the power method. While this approach provably converges if the spectrum of the operator is gapped, the convergence speed is often too slow to be practical. Rather, most methods (so-called projection methods) are based on constructing approximate eigenvectors $v$ in a certain subspace $V$ of dimension $m \ll n$. The condition for being an approximate eigenvector is the so-called Ritz-Galerkin condition, namely the residual $r$ is orthogonal to another $m$-dimensional subspace $W$. The choice $W=V$ gives rise to so-called orthogonal projection methods, and requires $\theta = v^\dagger A v$, which is referred to as the Ritz value. For nonhermitian problems, however, there is no a priori reason for choosing $W$ equal to $V$, in which case an oblique projection method is obtained which hopefully captures the left eigenvector in the subspace $W$. The most well-known methods are the Krylov subspace methods where the subspaces $V$ (and $W$) are obtained by successively applications of $A$. This yields the Arnoldi algorithm in the orthogonal case ($V=W$) and the nonhermitian Lanczos algorithm in the oblique case. For hermitian problems, both methods unify into the famous Lanczos algorithm.

MPS provide an efficient parameterisation of a set of vectors with a dimension $n=d^N$, which is typically too large for the vectors to be stored explicitly. Here, in fact, we consider the limit $N\to\infty$. For a given MPO $\hat{O}(T)$ with tensor $T^{i,j}_{\alpha,\beta}$ with virtual dimension $D'$, the power method can be implemented \cite{orus2008infinite} owing to the fact that applying the MPO $\hat{O}(T)$ to a given MPS $\ket{\Psi(A)}$ with virtual dimension $D$ gives rise to an MPS $\ket{\Psi(\tilde{A})}$ with $\tilde{A}^i = \sum_{j=1}^{d} T^{i,j} \otimes A^{j}$, which thus has virtual dimension $\tilde{D} = D' \times D$. A standard MPS truncation step can then be used to reduce the virtual dimension to a smaller value. If the virtual MPO dimension $D'$ is big (e.g.\ in the case of a PEPS transfer matrix, $D'$ is the square of the PEPS bond dimension), this increase can already be prohibitive. Furthermore, this approach suffers from the same slowness as the generic power method when the gap around the leading eigenvalue is small. Inspired by the idea of Krylov subspaces, one could think of combining $\ket{\Psi(A)}$, $\ket{\Psi(\tilde{A})}$ and MPS obtained by further applications of the MPO into a subspace. As the set of MPS does itself not constitute a subspace, the linear combination of (injective) MPS can only be written as a non-injective MPS with larger bond dimension. But there is a more fundamental obstruction to this approach. In the thermodynamic limit $N\to\infty$, any two injective MPS are either equal or orthogonal\footnote{This is in fact not an MPS specific problem but a generic feature in many body systems known as the (infrared) orthogonality catastrophe \cite{anderson1967infrared}.}. As such, no improvement can be obtained by constructing a linear combination of injective MPS.

We thus try to find the best approximation to the leading eigenvector of the MPO as an injective MPS by generalising the idea of projection methods from subspaces to general manifolds. In order for $\ket{\Psi(A)}$ to be an approximate eigenvector of the operator $O$, we require as a Galerkin condition that the residual $\ket{r} = O \ket{\Psi(A)} - \theta \ket{\Psi(A)}$ is orthogonal to the tangent space $T_{\ket{\Psi(A)}} \mathcal{M}$. Geometrically, when interpreting $\ket{r}$ as a vector emanating from the point $\ket{\Psi(A)}$, it implies that $\ket{r}$ is orthogonal to the manifold $\mathcal{M}$ itself. Note that, since $\ket{\Psi(A)}$ is itself in $T_{\ket{\Psi(A)}} \mathcal{M}$ (by choosing $B\sim A_C$ in Eq.~\eqref{eq:mps:deftangent}), we obtain the Ritz value expression $\theta = \braket{\Psi(A) | O | \Psi(A)}$, provided $\ket{\Psi(A)}$ is normalized to one. In principle, we could also try an oblique projection method by expressing that the residual has to be orthogonal to a different subspace, such as the tangent space $T_{\ket{\Psi(\tilde{A})}}\mathcal{M}$ at the point $\ket{\Psi(\tilde{A})}$ which provides an approximation of the left fixed point. However, we find the orthogonal projection method more stable in practice, which has also been observed in the simpler case of subspace methods.

\begin{figure}
\includegraphics[width=\textwidth]{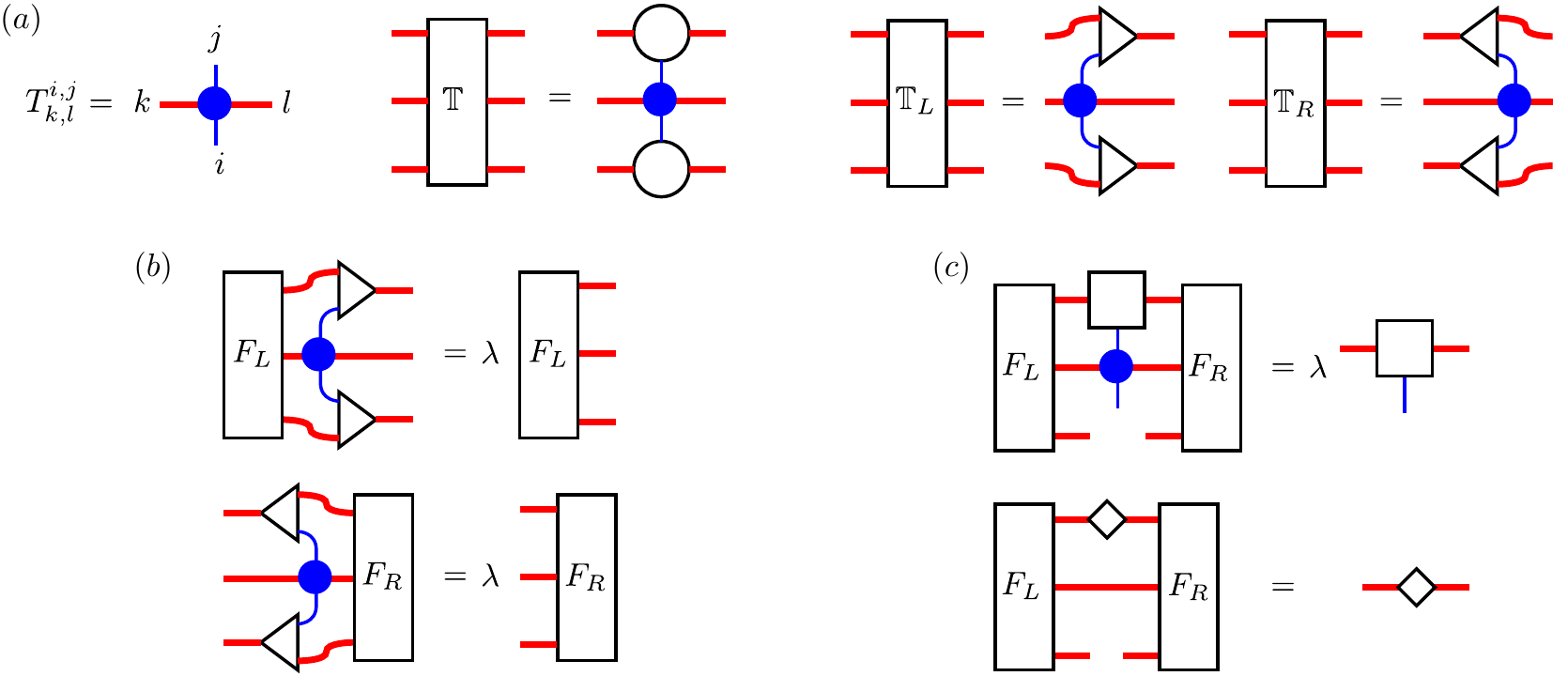}
\caption{ \label{fig:mpographs}
The MPO tensor and corresponding (transfer) matrices $\mathbb{T}$, $\mathbb{T}_L$ and $\mathbb{T}_R$. At a stationary point for the MPS approximation of an MPO fixed point, the four eigenvalue problems in (b) and (c) are satisfied. The second eigenvalue problem in (c) follows from the first one in combination with the solutions of (b).}
\end{figure}

We have yet to discuss how to obtain an MPS that satisfies the Galerkin condition on the residual. Let us therefore consider the problem in greater detail. The Ritz value $\theta = \braket{\Psi(A)|\hat{O}(T)|\Psi(A)}$ converges to $\lambda^N$ in the thermodynamic limit, where $\lambda$ is the dominant eigenvalue of the $D^2D' \times D^2 D'$ matrix
\begin{equation}
\mathbb{T} = \sum_{i,j} \bar{A}^{i} \otimes T^{i,j} \otimes A^{j},
\end{equation}
provided that it is non-degenerate and without Jordan block structure. The eigenvalue $\lambda$ is independent of the gauge choice of $A$, unlike the corresponding eigenvectors. If $\mathbb{T}_L$ is defined using the tensor $A_L$ and $\mathbb{T}_R$ using the tensor $A_R$, then we denote the left eigenvalue of $\mathbb{T}_L$ as $\bra{F_L}$ and the right eigenvector of $\mathbb{T}_R$ as $\ket{F_R}$. Equivalently, $F_L$ and $F_R$ can be interpreted as 3-index tensors $(F_L^k)_{\alpha,\beta}$ and $(F_R^l)_{\alpha,\beta}$ satisfying the eigenvalue equation
\begin{align}
\sum_{i,j,k} T^{i,j}_{k,l} (A_L^i)^\dagger F_L^k A_L^j &= \lambda F_L^l,\label{eq:mps:eigFL}\\
\sum_{i,j,l} T^{i,j}_{k,l} A_R^j F_R^l (A_R^i)^\dagger &= \lambda F_R^k.\label{eq:mps:eigFR}	
\end{align}
Normally, the proper normalization for a left and right eigenvector of a given operator is such that their mutual overlap is one. But since $F_L$ and $F_R$ correspond to eigenvectors of two distinct operators that differ by a similarity transform with $C \otimes \openone\otimes \bar{C}$, we arrive at the normalization condition
\begin{align}
\sum_{k,l} \delta_{k,l} \mathrm{tr}\left[ F_L^k C F_R^l C^\dagger \right] = 1.
\end{align}
With these definitions at hand, we can now evaluate the Galerkin condition
\begin{equation}
P_{T_{\ket{\Psi(A)}\mathcal{M}}} (\hat{O}(T) - \theta) \ket{\Psi(A)} = 0.
\end{equation}
as
\begin{equation}
\begin{split}
\sum_{n} (\lambda^{N-1} \tilde{A}_C^{i_n}- \theta A_C^{i_n})_{\alpha,\beta} \ket{\Psi^{[-\infty,n-1]}_\alpha} \otimes \ket{i_n} \otimes \ket{\Psi^{[n+1,+\infty]}_\beta}\\
- \sum_{n} (\lambda^N \tilde{C}- \theta C)_{\alpha,\beta}\ket{\Psi^{[-\infty,n]}_\alpha} \otimes \ket{\Psi^{[n+1,+\infty]}_\beta} =0,\label{eq:mps:galerkin}
\end{split}
\end{equation}
with
\begin{align}
\tilde{A}_C^{i} &= \sum_{j,k,l} T^{i,j}_{k,l} F_L^k A_C^j F_R^l,\label{eq:mps:defACtilde}\\
\tilde{C} &= \sum_{k,l} \delta_{k,l} F_L^k C F_R^l.\label{eq:mps:defCtilde}
\end{align}
When partially projecting the Galerkin condition onto the states $\ket{\Psi^{[-\infty,n-1]}_{\alpha}}$, the $k$th term in the first sum cancels with the $k$th term in the second sum for all $k\leq n-1$. Further projecting onto $\ket{\Psi^{[n+1,+\infty]}_{\beta}}$, the $k$th term in the first sum cancels with the term $k-1$ in the second sum for all $k \geq n+1$. As such, only the $n$th term in the first sum remains and gives rise to an eigenvalue equation for $A_C$
\begin{equation}
\tilde{A}_C^{i} = \sum_{j,k,l} T^{i,j}_{k,l} F_L^k A_C^j F_R^l = \lambda A_C^i
\label{eq:mps:eigAC}
\end{equation}
where we have also used $\theta = \lambda^N$. Repeating this projection with $\ket{\Psi^{[-\infty,n]}_{\alpha}}\otimes \ket{\Psi^{[n+1,+\infty]}_{\beta}}$ instead, the $n$th term of the second sum is canceled twice, or thus, is added with a plus sign and gives rise to an eigenvalue equation for $C$
\begin{equation}
\tilde{C} =  \sum_{k} F_L^k C F_R^k = C.\label{eq:mps:eigC}
\end{equation}
When equation \eqref{eq:mps:eigAC} and \eqref{eq:mps:eigC} are satisfied, the resulting uniform MPS $\ket{\Psi(A)}$ is a stationary solution for the MPO eigenvalue problem within the MPS manifold, i.e. no further decrease of the residual is possible without leaving the MPS manifold with given bond dimension. Note that \eqref{eq:mps:eigC} is not independent but follows from Eq.~\eqref{eq:mps:eigAC} by projecting onto $A_L$ or $A_R$ and using Eq.~\eqref{eq:mps:eigFL} or \eqref{eq:mps:eigFR}. These conditions are graphically represented in Figure~\ref{fig:mpographs}. Away from such a stationary point, $A_C$ and $C$ will not be eigenvectors of the respective linear maps $A_C\to \tilde{A}_C$ and $C\to\tilde{C}$. Furthermore, the dominant eigenvalues that one would get from these maps do not necessarily correspond to $\lambda$ (defined as the largest magnitude eigenvalue of $E$) in the case of $A_C\to\tilde{A}_C$, or to $1$ in the case of $C\to\tilde{C}$.

\subsection{Algorithms for the leading eigenvector of a matrix product operator}
Typically, algorithms for contracting two-dimensional tensor networks (either two-dimensional classical partition functions or PEPS expectation values) are divided into three categories:
\begin{enumerate}
	\item via the leading eigenvector of the MPO: typically using the power method, which is often referred to as imaginary time-TEBD \cite{orus2008infinite};
	\item via the corner transfer matrix (CTM) method \cite{baxter1978variational,nishino1996corner,nishino1997corner,orus2009simulation,orus2012exploring}
	\item via one of the many variations of the tensor network renormalization group \cite{levin2007tensor,gu2008tensor,Evenbly:2015aa,yang2015loop}.
\end{enumerate}

As our central object of study is the MPO representing the (linear) transfer matrix, we set out to follow the first strategy. However, by providing a more efficient alternative than the power method, we obtain a method that can be compared to (specific formulations of) the CTM method. Finally, we also relate the resulting algorithm to the exact algebra of matrices used to solve e.g.\ the ASEP as discussed in Section~\ref{ss:exact:asep}. Given that the stationary point is characterized by the coupled eigenvalue problems in Figure~\ref{fig:mpographs}, we can follow the DMRG spirit of iteratively optimizing every single eigenvalue problem. Suppose we currently have an MPS $\ket{\Psi(A)}$ where the tensor $A$ satisfies
\begin{align}
C_1 A^i &= A^i_L C_1,& A^i C_2 &= C_2 A^i_R &\Rightarrow C &= C_1 C_2,& A_C^i &= C_1 A^i C_2.
\end{align}
$C_1$ and $C_2$ can be obtained from the algorithms \texttt{LeftOrthonormalize} and \texttt{RightOrthonormalize}. As the second eigenvalue problem in Figure~\ref{fig:mpographs}(c) is not independent, we only set out to solve the first one. We can transform this into an eigenvalue problem for $A$ by multiplying it with $C_1^{-1}$ and $C_2^{-1}$. We can also compute the norm of the residual projected onto the tangent space. This norm scales as the number of sites times $\lambda^{N}$ times $\varepsilon$, where the error quantity $\varepsilon$ is given by
\begin{align}
\varepsilon=\left(\sum_{i}\lVert (\lambda^{-1} \tilde{A}^{i}_C - A^{i}_C) -  A^{i}_L(\tilde{C}-C)\rVert)\right)^{1/2} = \left(\sum_{i}\lVert \lambda^{-1} \tilde{A}^{i}_C - A^{i}_L \tilde{C}\rVert\right)^{1/2},
\end{align}
and can thus be used as a measure of convergence. This gives rise to Algorithm~\ref{alg:mpofixedpoint1}.
\begin{algorithm}
\caption{Find the optimal MPS approximation for the fixed point of the MPO $\hat{O}(T)$ \label{alg:mpofixedpoint1}}
\begin{algorithmic}[1]
\Procedure{MPOFixedPoint1}{$T,A,\eta$}\Comment{Initial guess $A$ and a tolerance $\eta$}
\State $A_L,C_1 \gets$ \Call{LeftOrthonormalize}{$A$}
\State $A_R,C_2 \gets$ \Call{RightOrthonormalize}{$A$}
\State $\lambda, F_L \gets $ \Call{Arnoldi}{$F^l \to \sum_{i,j,k} T^{i,j}_{k,l} (A_L^i)^\dagger F^k A_L^j$}
\State $\sim, F_R \gets $ \Call{Arnoldi}{$F^k \to \sum_{i,j,l} T^{i,j}_{k,l} A_R^j F^l (A_R^i)^\dagger$} \Comment{Eigenvalue is also $\lambda$}
\State $C \gets C_1 C_2$
\State $F_L \gets F_L / (\sum_{k} \mathrm{tr}[F_L^k C F_R^k C^\dagger])$ \Comment{Proper normalization}
\State $A_C^i \gets C_1 A^i C_2$
\State $\varepsilon \gets $\Call{Norm}{$\lambda^{-1} \sum_{j,k,l} T^{i,j}_{k,l} F_L^k A_C^j F_R^l - A_L^i \sum_{k} F_L^k C F_R^k$}\Comment{Convergence measure}
\While{$\varepsilon > \eta$}
\State $\lambda',A \gets $ \Call{Arnoldi}{$A^i \to  C_1^{-1}(\sum_{j,k,l} T^{i,j}_{k,l} F_L^k C_1 A^j C_2 F_R^l) C_2^{-1}$, $A$, $\varepsilon/10$}
\State $A_L,C_1 \gets$ \Call{LeftOrthonormalize}{$A$, $C_1$, $\varepsilon/10$}
\State $A_R,C_2 \gets$ \Call{RightOrthonormalize}{$A$, $C_1$, $\varepsilon/10$}
\State $\lambda, F_L \gets $ \Call{Arnoldi}{$F^l \to \sum_{i,j,k} T^{i,j}_{k,l} (A_L^i)^\dagger F^k A_L^j$, $F_L$, $\varepsilon/10$}
\State $\sim, F_R \gets $ \Call{Arnoldi}{$F^k \to \sum_{i,j,l} T^{i,j}_{k,l} A_R^j F^l (A_R^i)^\dagger$, $F_L$, $\varepsilon/10$}
\State $C \gets C_1 C_2$
\State $F_L \gets F_L / (\sum_{k} \mathrm{tr}[F_L^k C F_R^k C^\dagger])$
\State $A_C^i \gets C_1 A^i C_2$
\State $\varepsilon \gets $\Call{Norm}{$\lambda^{-1} \sum_{j,k,l} T^{i,j}_{k,l} F_L^k A_C^j F_R^l - A_L^i \sum_{k} F_L^k C F_R^k$}
\EndWhile
\State \textbf{return} $A_L,\lambda$
\EndProcedure
\end{algorithmic}
\end{algorithm}

The same comments as in the algorithm \texttt{LeftOrthonormalize} apply, i.e.\ the call to \texttt{Arnoldi} should only build up the Krylov subspace  once (no restarts), whereas the outer \texttt{while} loop can be interpreted as the restart loop. Note that we could also monitor $\lvert \lambda-\lambda'\rvert$ or differences in $\lambda^{(\prime)}$ between subsequent iterations, though these quantities typically converge as $\varepsilon^2$. Unfortunately, algorithm~\ref{alg:mpofixedpoint1} is still plagued by the inverses of $C_1$ and $C_2$, which can potentially be ill-conditioned. In a typical DMRG algorithm for finite lattices, the eigenvalue problem is always formulated for $A_C$, which is then split into an isometric tensor and a $C$ matrix which is absorbed in the subsequent site. For a uniform MPS, we cannot simply replicate this strategy without spoiling the translation invariance. However, a simple generalization can be formulated that follows our Galerkin condition closely. Note that Figure~\ref{fig:mpographs}(c) contains separate eigenvalue equations for $A_C$ and $C$ and thus allow for computing an updated $A_C'$ and $C'$, which should define an updated $A_L'$ or $A_R'$ via $A_C^{\prime i} = A_L^{\prime i} C' = C' A_R^{\prime i}$. One might consider computing $A_L^{\prime i} = A_C^{\prime i} (C')^{-1}$. This, however, requires again to invert $C$ (actually, now, $C'$), which is exactly the stap we are trying to avoid. Furthermore, aside from the numerical inaccuracies related to this inverse, the resulting $A_L'$ would violate the left orthonormality condition at second order in the change $\Delta A_C= A_C'-A_C$. Therefore, instead, we determine $A_L'$ as the right orthonormal tensor that minimizes $\sum_{i} \lVert A_C' - A_L' C' \rVert^2$, where we now interpret $A_C'$ and $A_L'$ as rectangular matrices by merging the physical index and the virtual row index. Henceforth dropping the indices, the exact solution of this minimization problem is obtained by computing the singular value decomposition $ A_C C^\dagger = U S V^\dagger$ and setting $A_L = U V^\dagger$. However, if $A_C$ is indeed equal to $A_L C$, then this approach amounts to first computing $A_CC^\dagger   = A_L C^\dagger C = U S V^\dagger$. Thus, the singular values contained in $S$ are the square of the singular values in $C$ and many of the smallest can be below machine precision $\epsilon$. Correspondingly, the singular vectors in $U$ and $V$ will not be accurately computed. Therefore, we resort to a different solution, namely by identifying the elements in the QR-decomposition of both terms. If $A_C = Q_1 R_1$ and $C = Q_2 R_2$, then an exact equality $A_C = A_L C$ requires that $A_L = Q_1 Q_2^\dagger $ and $R_1 = R_2$. As approximate solution, we still use $A_L =  Q_2^\dagger Q_1$. The error is then given by $\delta = \lVert R_1 - R_2\rVert$. It turns out that if we obtain the updated $A_C$ and $C$ from the eigenvalue problems in Figure~\ref{fig:mpographs}(c), the resulting $\delta$ is roughly proportional to the convergence measure $\varepsilon$ can then thus be used instead. Using a similar strategy for determining an updated $A_R$, this gives rise to Algorithm~\ref{alg:mpofixedpoint2}.

\begin{algorithm}
\caption{Find the optimal MPS approximation for the fixed point of the MPO $\hat{O}(T)$ \label{alg:mpofixedpoint2}}
\begin{algorithmic}[1]
\Procedure{MPOFixedPoint2}{$T,A_R,\eta$}\Comment{Initial guess $A_R$ and a tolerance $\eta$}
\State $A_L,C \gets$ \Call{RightOrthonormalize}{$A_R$}
\Repeat
\State $\lambda, F_L \gets $ \Call{Arnoldi}{$F^l \to \sum_{i,j,k} T^{i,j}_{k,l} (A_L^i)^\dagger F^k A_L^j$, $F_L$, $\delta/10$}
\State $\sim, F_R \gets $ \Call{Arnoldi}{$F^k \to \sum_{i,j,l} T^{i,j}_{k,l} A_R^j F^l (A_R^i)^\dagger$, $F_R$, $\delta/10$}
\State $F_L \gets F_L / (\sum_{k} \mathrm{tr}[F_L^k C F_R^k C^\dagger])$
\State $A_C^i \gets C A_R^i$
\State $\lambda',A_C \gets $ \Call{Arnoldi}{$A^i_C \to \sum_{j,k,l} T^{i,j}_{k,l} F_L^k A_C^j F_R^l$, $A_C$, $\delta/10$}
\State $\mu',C \gets $ \Call{Arnoldi}{$C \to \sum_{k} F_L^k C F_R^k$, $C$, $\delta/10$}
\State $Q_A, R_A \gets $ \Call{QRPos}{$A_C$}
\State $Q_C, R_C \gets $ \Call{QRPos}{$C$}
\State $A_L \gets Q_A Q_C^\dagger$
\State $L_A, Q_A \gets $ \Call{LQPos}{$A_C$}
\State $L_C, Q_C \gets $ \Call{LQPos}{$C$}
\State $A_R \gets Q_C^\dagger Q_A$
\State $\delta \gets$ \Call{Max}{$\lVert R_A - R_C\rVert$,$\lVert L_A - L_C\rVert$}
\Until{$\delta < \eta$}
\State \textbf{return} $\lambda, A_R$
\EndProcedure
\end{algorithmic}
\end{algorithm}

Let us now compare our approach to the CTM method. More specifically, we start by comparing Algorithm~\ref{alg:mpofixedpoint1} to the so-called full one-directional CTM method described in Appendix~A of Ref.~\cite{orus2012exploring}. We can make the following identification
\begin{align}
C_1 &= C_1,&C_2&=C_2,&T_1 &= A,& T_2 &= F_L,&T_3&= F_R,\\
U&= A_L,&V^\dagger &= A_R,&P&= C_2 F_R C_2^{-1},&P^{(-1)}&= C_1^{-1} F_L C_1.
\end{align}
and notice that we replace the step $C_1\to \tilde{C}_1\to C_1'$ and the determination of $U=A_L$ in the so-called $x$ move [FIG.~25 of Ref.~\cite{orus2012exploring}] with the routine \texttt{LeftOrthonormalize}. In particular, the determination of $U$ in Ref.~\cite{orus2012exploring} is based on the full density matrix containing the square of the singular values, whereas \texttt{LeftOrthonormalize} works with the singular values directly in order to obtain a higher accuracy. The same applies for $C_2$ and $V=A_R$. The update of $T_i\to \tilde{T}_i \to T_i'$ for $i=1,2,3$ [FIG.~25 and 26 of Ref.~\cite{orus2012exploring}] can be interpreted as a single step of the power method in the CTM method, where we apply a full Arnoldi step to obtain faster convergence. Note that in our identification, $P^{(-1)}$ might not be the exact pseudo inverse (unlike in the CTM method), because $P^{(-1)} P = \sum_{k} C_1^{-1} F_L^k C_1 C_2 F_R^k C_2^{-1}$ will only approximate the identity as Algorithm~\ref{alg:mpofixedpoint1} converges. However, we anyway prefer to avoid the inverses all together by using Algorithm~\ref{alg:mpofixedpoint2} instead, which has no direct analogue in the CTM language.

A particularly interesting case that deserves further consideration is when $T^{i,j} = (T^{i,j})^\dagger$ (i.e.\ $T^{i,j}_{k,l} = \bar{T}^{i,j}_{l,k}$). This happens typically in the case of classical partition functions where $T$ is real anyway and we do not need the complex conjugation. Then there is alternative reformulation of Algorithm~\ref{alg:mpofixedpoint1} which avoids inverses. In this case, we can choose $(A^i)^\dagger = A^i$, which leads to $C_1 = C_2^\dagger$ and $A_L^i = (A_R^i)^\dagger$, and therefore also $F_L^k = (F_R^k)^\dagger$. Let us henceforth call $B^k = F_L^k$. The eigenvalue problem for $C = C_1 C_1^\dagger$ than amounts to $\sum_{k} B^k C_1 C_1^\dagger (B_k)^\dagger = C_1 C_1^\dagger$. This implies that $C_1$ is not only the gauge transform to bring $A$ into left orthonormal form, but also to bring $B$ into the right orthonormal form  $B_R^k = C_1^{-1} B_k C_1$. Hence, we can use \texttt{RightOrthonormalize} instead of using the explicit inversion of $C_1$. This enables an alternative reformulation of Algorithm~\ref{alg:mpofixedpoint1} that simplifies down into Algorithm~\ref{alg:mpofixedpoint3}.

\begin{algorithm}
\caption{Find the optimal MPS approximation for the fixed point of the MPO $\hat{O}(T)$ \label{alg:mpofixedpoint3}}
\begin{algorithmic}[1]
\Procedure{MPOFixedPoint3}{$T,A,\eta$}\Comment{Initial guess $A$ and a tolerance $\eta$}
\Repeat
\State $A_L,C_1 \gets$ \Call{LeftOrthonormalize}{$A$,$C_1'$}
\State $\lambda, B \gets $ \Call{Arnoldi}{$B^l \to \sum_{i,j,k} T^{i,j}_{k,l} (A_L^i)^\dagger B^k A_L^j$}
\State $B_R,C_1' \gets$ \Call{RightOrthonormalize}{$B$,$C_1$}
\State $\lambda',A \gets $ \Call{Arnoldi}{$A^i \to  \sum_{j,k,l} T^{i,j}_{k,l} B_R^k A^j (B_R^l)^\dagger$}
\State $\delta = \lVert C_1 - C_1'\rVert$
\Until{$\delta < \eta$}
\State \textbf{return} $A_L,\lambda$
\EndProcedure
\end{algorithmic}
\end{algorithm}
Note that, because $C_1$ is upper triangular and $C_1'$ is lower triangular, their convergence implies that the final $S=C_1=C_1'$ will be diagonal and can be identified with the eigenvalues of Baxter's corner transfer matrix, which are actually the square root of the singular values $C=C_1 C_1^\dagger = S^2$ of the state $\ket{\Psi(A)}$ and thus the fourth root of the spectrum of the half chain density matrix. Furthermore, this convergence implies that
\begin{align}
\sum_{j,k} T^{i,j}_{k,l} B^k C_1 A^j = \lambda C_1 A^i C_1^{-1} B^l C_1 + V^i X W^l = \lambda A_L^i C_1 B_R^l + V^i X W^l
\end{align}
where the $D \times (d-1)D$ matrix $V^i$ satisfies $\sum_i (A_L^i)^\dagger V^i = 0$ and $\sum_i (V^i)^\dagger V^i = 1$, i.e.\ it is an orthonormal basis for the subspace orthogonal to $A_L$. Similarly, we have $\sum_i W^l (B_R^l)^\dagger = 0 $ and $\sum_l W^l (W_l)^\dagger = 1$. By absorbing the unitary factors of the $(d-1)D \times (d-1)D$ matrix $X$ into the definition of $V$ and $W$, we obtain the singular value decomposition of the left hand side, where the $D$  largest singular values are contained in $\lambda C_1$ and the $(d-1)D$ smallest singular values are contained in $X$. Turning this observation into an algorithm exactly yields the CTM renormalization group of Baxter \cite{baxter1978variational}, later generalized to the nonsymmetric setting by Nishino and Okunishi \cite{nishino1996corner,nishino1997corner} and also called the two-directional CTM method in Ref.~\cite{orus2012exploring}. This approach allows to increase the bond dimension $D$ (similar to the two-site DMRG algorithm) by including more singular values in the renormalization step. As a final note, we remark that when $X=0$, we have obtained exact solution which can be identified (up to details such as reinterpreting the MPO diagonally) with the algebra constructed to solve the ASEP in Section~\ref{ss:exact:asep}.

Clearly, many generalizations are possible, to finite systems or systems with a periodically repeated unit cell. Two-site versions of the reported algorithms with a dynamically increasing bond dimension can be formulated and compared to the aforementioned CTM renormalization group. Furthermore, oblique projection methods can be formulated which compute the fixed points of the linear transfer matrices in the four directions simultaneously, similar to the fully general CTM renormalization group method \cite{nishino1996corner,nishino1997corner,orus2009simulation}.

\subsection{Algorithms for the excited states of a matrix product operator}
Finally, we discuss how to obtain information about excited states of a matrix product operator. For Hamiltonians, it has recently proven useful to construct approximations to low-lying excited states within the tangent space of the MPS ground state $\ket{\Psi(A)}$ \cite{haegeman2012variational}. Strictly speaking, the tangent space $T_{\ket{\Psi(A)}}\mathcal{M}$ to the manifold of uniform MPS only contains translation invariant states (momentum zero), but it is easy to generalise the definition of tangent vectors to have a well-defined non-zero momentum $k$. We thereto use the states
\begin{equation}
\ket{\Phi_k(B;A)} = \sum_{n=-\infty}^{+\infty} \mathrm{e}^{\mathrm{i} k n} B^{i_n}_{\alpha,\beta} \ket{\Psi_{\alpha}^{[-\infty,n-1]}}\otimes \ket{i_n}\otimes \ket{\Psi_{\beta}^{[n+1,+\infty]}}.\label{eq:mps:deftangent2}
\end{equation}
which span a space $T^{[k]}_{\ket{\Psi(A)}}\mathcal{M}$, with $k\in[-\pi,+\pi)$. Note the parameterisation redundancy $B^i \to B^i + \mathrm{e}^{\mathrm{i} k } A^i_L X - X A_R^i$, $\forall X \in \mathbb{C}^{D\times D}$. This can be used to choose e.g.\ $\sum_{i} (A_L^i)^\dagger B^i = 0$, in which case the overlap between two such states evaluates to
\begin{equation}
\braket{\Phi_{k}(B;A)|\Phi_{k'}(B';A)} = 2\pi\delta(k-k') \sum_{\alpha,i,\beta} \bar{B}^{i}_{\alpha,\beta} B^{\prime i}_{\alpha\beta}.
\end{equation}

For low-lying excited states of Hamiltonians, this tangent space approach has been generalised to continuous MPS \cite{draxler2013particles} and PEPS \cite{vanderstraeten2015excitations}, and can be further extended to define and compute scattering states \cite{vanderstraeten2014s,vanderstraeten2015scattering}. Furthermore, for locally interacting Hamiltonians, it can be theoretically argued that these states do indeed capture the eigenvectors associated with isolated dispersion curves in the energy-momentum spectrum. For MPOs on the other hand, any such intuition or theoretical argument about the nature of excited states is missing. Nevertheless, using the subspace $T^{[k]}_{\ket{\Psi(A)}}\mathcal{M}$ still seems to work well in practice \cite{haegeman2014shadows} and gives rise to eigenvalues of the form $\mu(k)\lambda^N = \exp(-N f - e(k))$ with $f$ the (complex) free energy density of the fixed point and $e(k)$ a complex excitation energy.

Finding approximate excited states of the MPO thus amounts to an ordinary orthogonal projection method in the subspace of tangent vectors. We thus have to compute the matrix element of the MPO $\hat{O}(T)$ between two MPS tangent vectors. For simplicity, we rescale the MPO tensor $T$ with $\lambda^{-1}$, such that the corresponding free energy of the new MPO is zero. Equivalently, the largest magnitude eigenvalue of $\mathbb{T}$ is one. We also introduce the more general notation
\begin{equation}
\mathbb{T}^{A}_{B} = \sum_{i,j} \bar{B}^j \otimes T^{i,j} \otimes A^j
\end{equation}
such that $\mathbb{T}_L = \mathbb{T}^{A_L}_{A_L}$ and $\mathbb{T}_R = \mathbb{T}^{A_R}_{A_R}$.

Expressing that $\ket{\Phi_{k}(B;A)}$ is an approximate eigenvector with eigenvalue $\theta$ then gives rise to the Galerkin condition that, $\forall B'$ (satisfying $\sum_{i} (B^{\prime i})^\dagger A_L^i=0$)
\begin{align}
0=&\braket{\Phi_{k'}(B';A)|\hat{O}(T)- \theta |\Phi_{k}(B;A)} \\
=& 2\pi \delta(k-k') \bigg[ \sum_{n=0}^{+\infty} \mathrm{e}^{\mathrm{i} k (n+1)} \braket{F_L| \mathbb{T}^{A_L}_{B'} (\mathbb{T}^{A_L}_{A_R})^{n-1} \mathbb{T}^{B}_{A_R} | F_R} + \braket{F_L|\mathbb{T}^{B}_{B'}|F_R}\\
&\qquad\qquad\qquad+ \sum_{n=0}^{+\infty} \mathrm{e}^{-\mathrm{i} k (n+1)} \braket{F_L| \mathbb{T}^{B}_{A_L} (\mathbb{T}^{A_R}_{A_L})^{n-1} \mathbb{T}^{A_L}_{B'} | F_R}  - \theta \sum_{i} \mathrm{tr}[B^i (B^{\prime i})^\dagger] \bigg]
\end{align}
The first term contains a geometric series of the operator $\mathrm{e}^{i k} \mathbb{T}^{A_L}_{A_R}$, which has an eigenvalue $\mathrm{e}^{i k}$ of unit magnitude, with corresponding right eigenvector $\ket{G}$ given as 3-index tensor by $G_{\alpha,\beta,\gamma} = \sum_{\alpha'} C_{\alpha,\alpha'} (F_R)_{\alpha',\beta,\gamma}$. This follows from $A_L^i C = C A_R^i$. We can however check that $\braket{F_L|\mathbb{T}^{A_L}_{B}|G}=\braket{F_L|\mathbb{T}^{A_C}_{B}|F_R} = \sum_{i} \mathrm{tr}( (B^i)^\dagger A_C^i) = \sum_{i} \mathrm{tr}( (B^i)^\dagger A_L^i C) = 0$, as follows from the gauge choice of $B$ and the fact that $A_C$ satisfies Eq.~\eqref{eq:mps:eigAC} with $\lambda=1$. Thus, the eigenspace corresponding to the eigenvalue of unit magnitude does not contribute to the geometric series and all other eigenvalues of $\mathrm{e}^{i k} \mathbb{T}^{A_L}_{A_R}$ have a magnitude strictly smaller than one\footnote{Note that, if we would have obtained the MPS fixed point $\ket{\Psi(A)}$ using an oblique projection method, e.g. $P_{T_{\ket{\Psi(A')}}\mathcal{M}} (\hat{O}(T)-\theta)\ket{\Psi(A)}=0$ with $\ket{\Psi(A')}$ the MPS approximation of the left fixed point of $\hat{O}(T)$, then we also need to compute the
excited states in this oblique setting.}. A similar argument can be used for the geometric series in the third term, so that we obtain
\begin{align}
0 = & \mathrm{e}^{\mathrm{i} k} \braket{F_L| \mathbb{T}^{A_L}_{B'} (\openone - \mathrm{e}^{\mathrm{i} k}\mathbb{T}^{A_L}_{A_R})^{-1} \mathbb{T}^{B}_{A_R} | F_R} + \braket{F_L|\mathbb{T}^{B}_{B'}|F_R}\\
&+ \mathrm{e}^{-\mathrm{i} k } \braket{F_L| \mathbb{T}^{B}_{A_L} (\openone - \mathrm{e}^{-\mathrm{i} k}\mathbb{T}^{A_R}_{A_L})^{-1} \mathbb{T}^{A_L}_{B'} | F_R} - \theta \sum_{i} \mathrm{tr}[B^i (B^{\prime i})^\dagger] \bigg]
\end{align}
which has to be satisfied for any $B'$ satisfying $\sum_{i} (B^{\prime i})^\dagger A_L^i=0$.
This gives rise to an eigenvalue equation for $B$ with an effective operator that can be applied to $B$ efficiently, provided we use an iterative linear solver [e.g.\ the generalized minimal residual or stabilized biconjugate gradient method \cite{barrett1994templates}] to compute the inverses appearing in the above expression.

\subsection{Examples}
We now discuss a number of examples where we utilize both the fixed point algorithms and the algorithm for finding excited states of matrix product operators.

\subsubsection{Hard squares}
As discussed in the section \ref{hardsq}, counting the number of hard square configurations is equivalent to calculating the leading eigenvalue of the transfer matrix

\[A^{ij}_{\alpha\beta}=\mat{cc}{1 & 0\\0 & p}_{i\alpha}\cdot \mat{cc}{1 & 1\\1 &0}_{ij}\cdot\mat{cc}{1 & 1\\1 &0}_{\alpha\beta} \]

\begin{figure}
\begin{subfigure}{.5\textwidth}
  \centering
  \includegraphics[width=\linewidth]{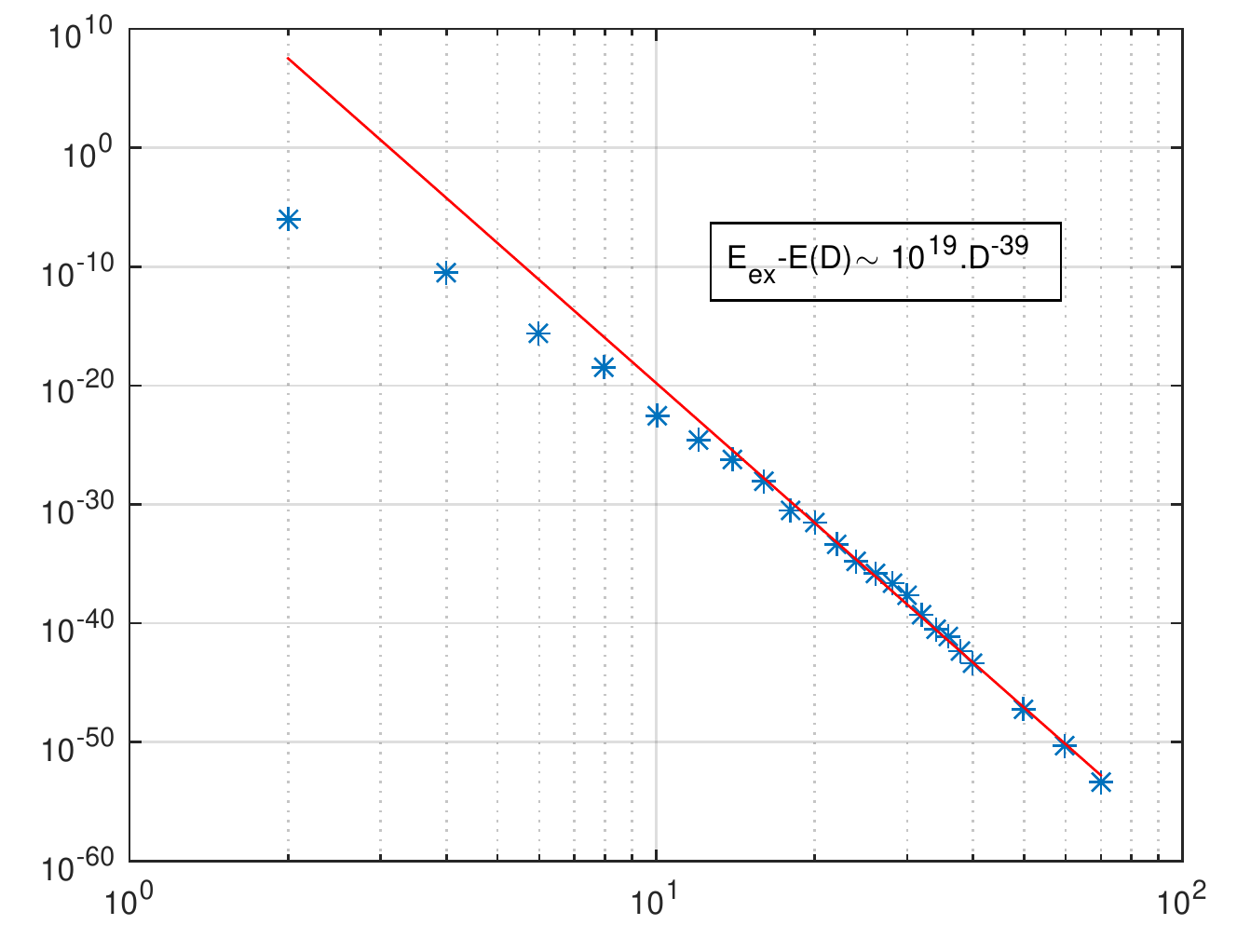}
  \label{fig:sub1}
\end{subfigure}%
\begin{subfigure}{.5\textwidth}
  \centering
  \includegraphics[width=\linewidth]{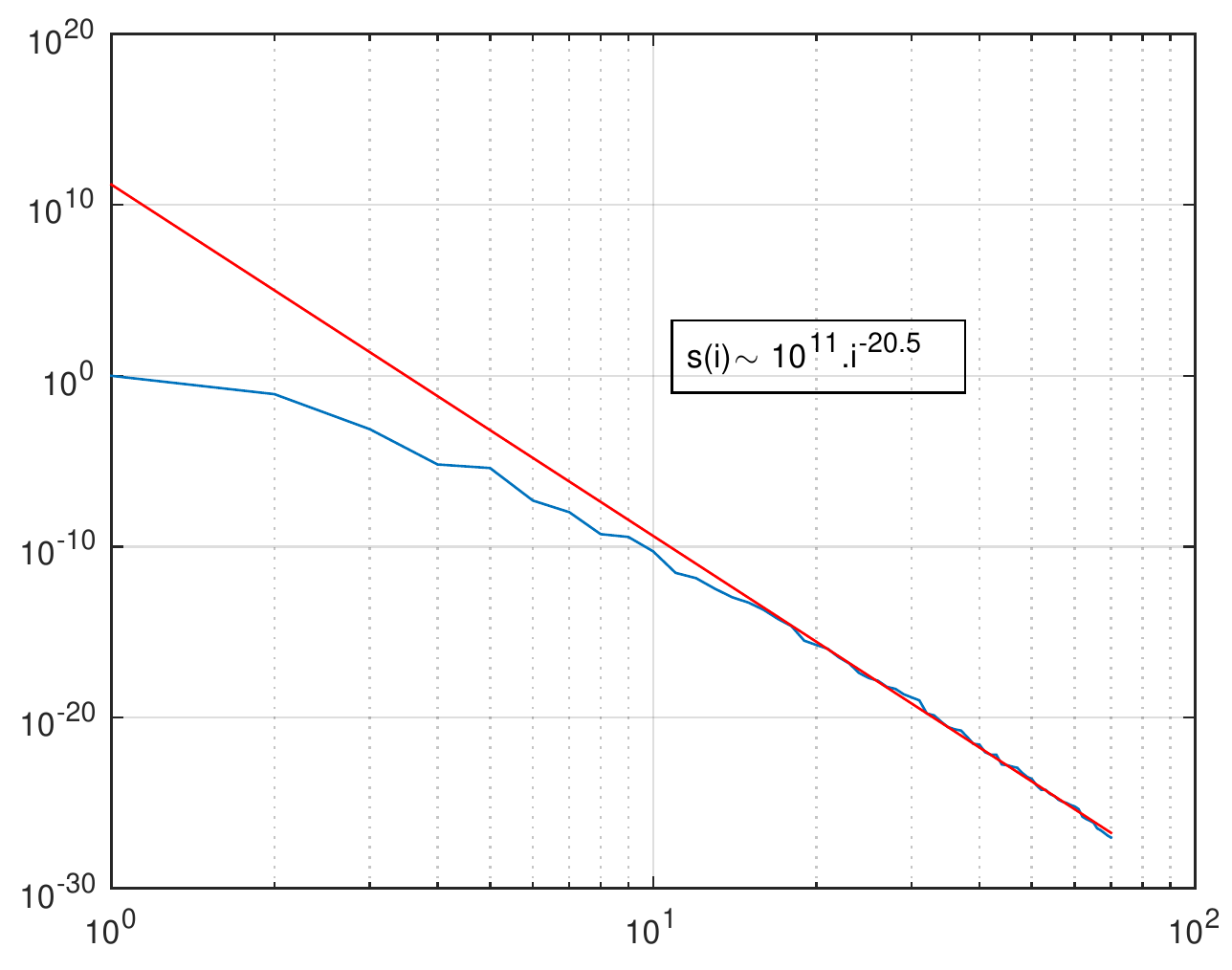}
  \label{fig:sub2}
\end{subfigure}
\caption{\label{hardsq1} (a) Convergence of free energy versus bond dimension for the hard squares model and (b) decay of Schmidt coefficients for $D=70$.}
\end{figure}

For the case where the chemical potential is zero ($p=1$), Baxter wrote a remarkable paper \cite{baxter1999planar} in which he used series expansions to calculate the corresponding free energy to 43 digits of precision. This turned out to be possible because for this value of $p$ the gap of the transfer matrix is large. We used the algorithms above with a multiprecision toolbox, and were able to get a precision of 58 digits with modest effort; see table \ref{tablehs}. In figure \ref{hardsq1}, we plot the precision of the free energy as a function of the bond dimension, and clearly observe a power law behaviour. We also illustrate the decay of Schmidt coefficients for the case of $D=70$, whose smallest ones can be fitted to a polynomial. This polynomial as opposed to exponential decay for the Schmidt coefficients can be understood for integrable models, where the exponential decay of Schmidt coefficients goes hand in hand with an exponential increase in their degeneracy in the form $\exp(\pi\sqrt{n/3})/n^{3/4}$.

\begin{table}
\begin{center}
\begin{tabular}{|l|l|}
\hline
2  &      1.5030477\\
4  &      1.50304808246\\
6  &      1.50304808247533218\\
8  &      1.5030480824753322642\\
10 &       1.503048082475332264322058\\
20 &       1.50304808247533226432206632947554\\
30 &       1.503048082475332264322066329475553689377\\
{\rm Baxter}\cite{baxter1999planar} & 1.503048082475332264322066329475553689385781\\
40 &       1.50304808247533226432206632947555368938578102\\
50 &       1.503048082475332264322066329475553689385781038609\\
60 &       1.503048082475332264322066329475553689385781038610303\\
70 &       1.503048082475332264322066329475553689385781038610305061\\
80 &       1.503048082475332264322066329475553689385781038610305062026556\\
\hline
\end{tabular}
\end{center}
\caption{\label{tablehs} Free energy of the hard squares model; with bond dimension $D=80$, we get 58 digits of precision.}
\end{table}

\begin{figure}
\begin{subfigure}{.5\textwidth}
  \centering
  \includegraphics[width=\linewidth]{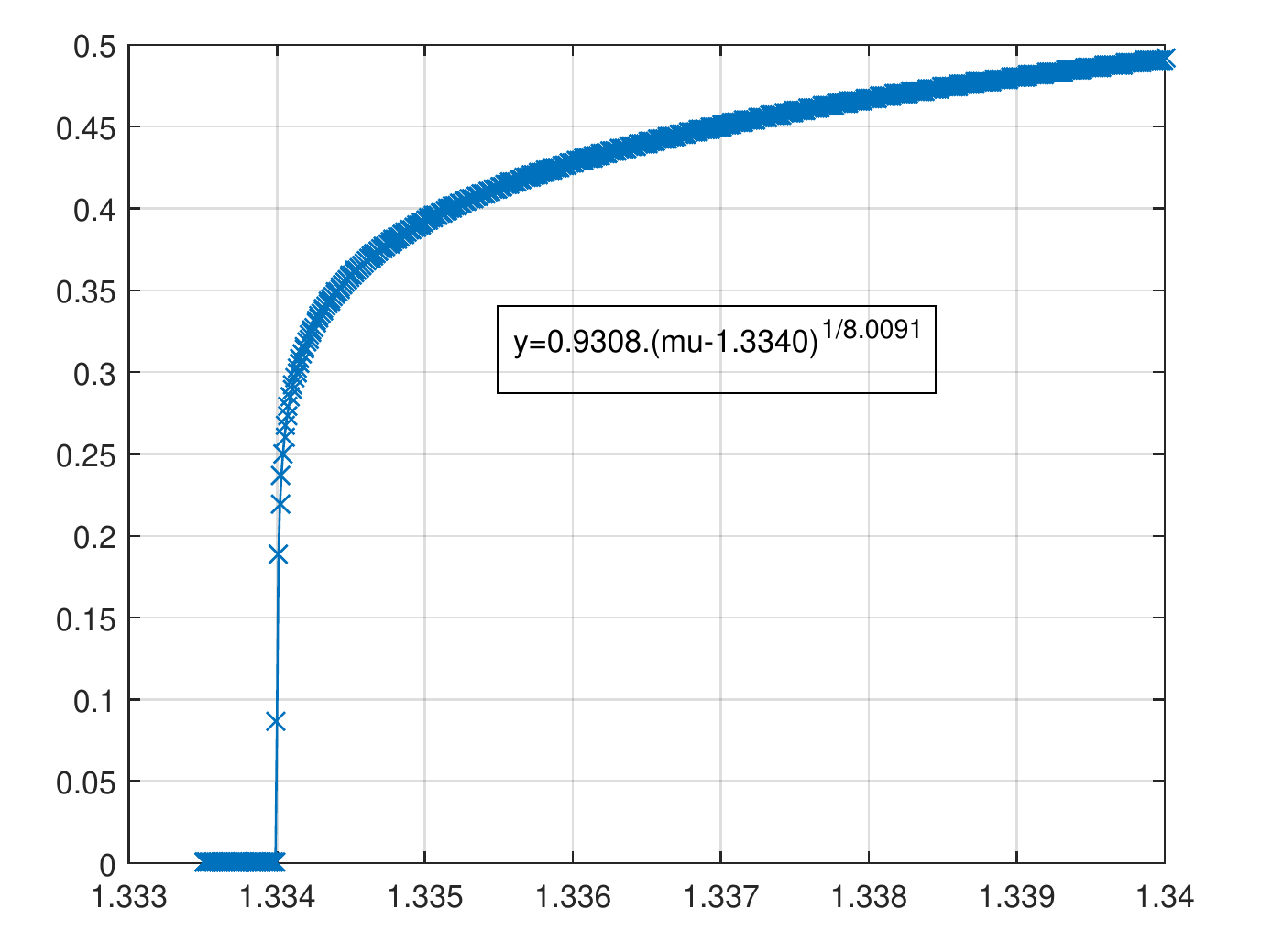}
  \label{fig:sub1}
\end{subfigure}%
\begin{subfigure}{.5\textwidth}
  \centering
  \includegraphics[width=\linewidth]{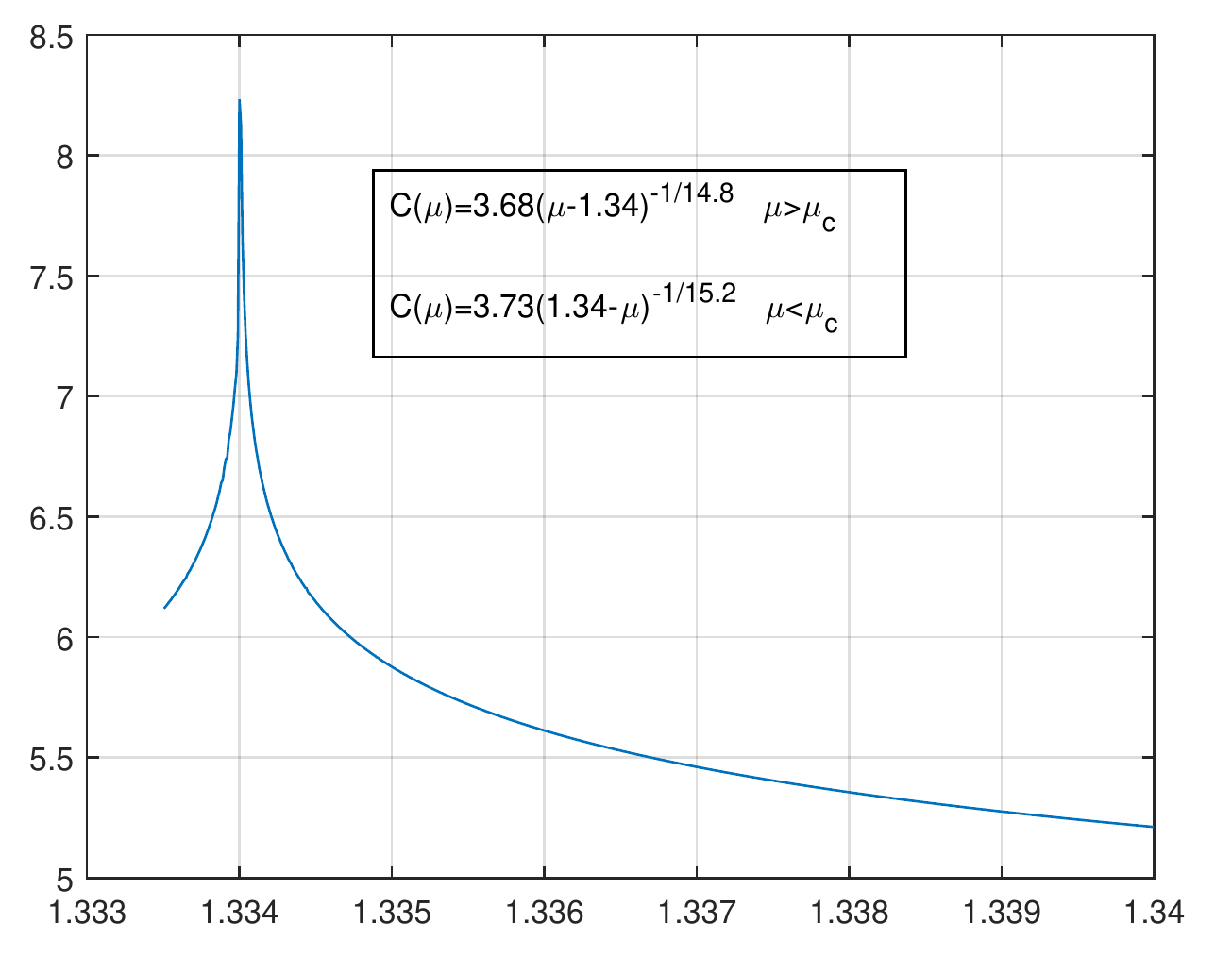}
  \label{fig:sub2}
\end{subfigure}
\caption{\label{hardsq2} Magnetization and compressibility of the hard square model as a function of the chemical potential as calculated with the variational MPS method with bond dimension $D=128$.}
\end{figure}

This model exhibits an Ising phase transition at  $p=\exp(\mu)$ with the chemical potential $\mu\simeq 1.334$. To simulate this model, it turns out to be beneficial to work with the "dual" matrix product operator
\[A^{ij}_{\alpha\beta}=\left(|0000\rangle+\exp(\mu/2)\left(|0001\rangle+|0010\rangle+|0100\rangle+|1001\rangle\right)+\exp(\mu)\left(|0011\rangle+|1100\rangle\right)\right)_{\alpha\beta}\]
as this MPO still yields a translational invariant fixed point in the symmetry broken phase. In figure \ref{hardsq2}, the results for the magnetization and the compressibility  are presented as a function of the chemical potential as calculated with the variational MPS algorithm with bond dimension $D=128$. A better accuracy of the critical exponents could have been obtained by doing a finite entanglement scaling of the results, which has not been done here.

\subsubsection{Transfer matrix spectrum of the $q$-level Potts model}
Next, we study the transfer matrix of the two-dimensional classical Potts model described by the Hamiltonian
\begin{equation}
	H = -J \sum_{\langle i,j\rangle} \delta_{s_i,s_j}
\end{equation}
where every site spin $s$ can take values $1,2,\ldots,q$. Like the classical Ising model (which is equivalent to the case $q=2$), this model transitions from a disordered (symmetric) phase at high temperatures to an ordered (symmetry broken) phase at low temperatures. The phase transition point is known exactly to be $(\beta J)_c = \log(\sqrt{q}+1)$. However, the nature of the phase transition changes from second order for $q=2,3,4$ to first order for $q >4$.

\begin{figure}[h]
  \centering
  \includegraphics[width=0.9\textwidth]{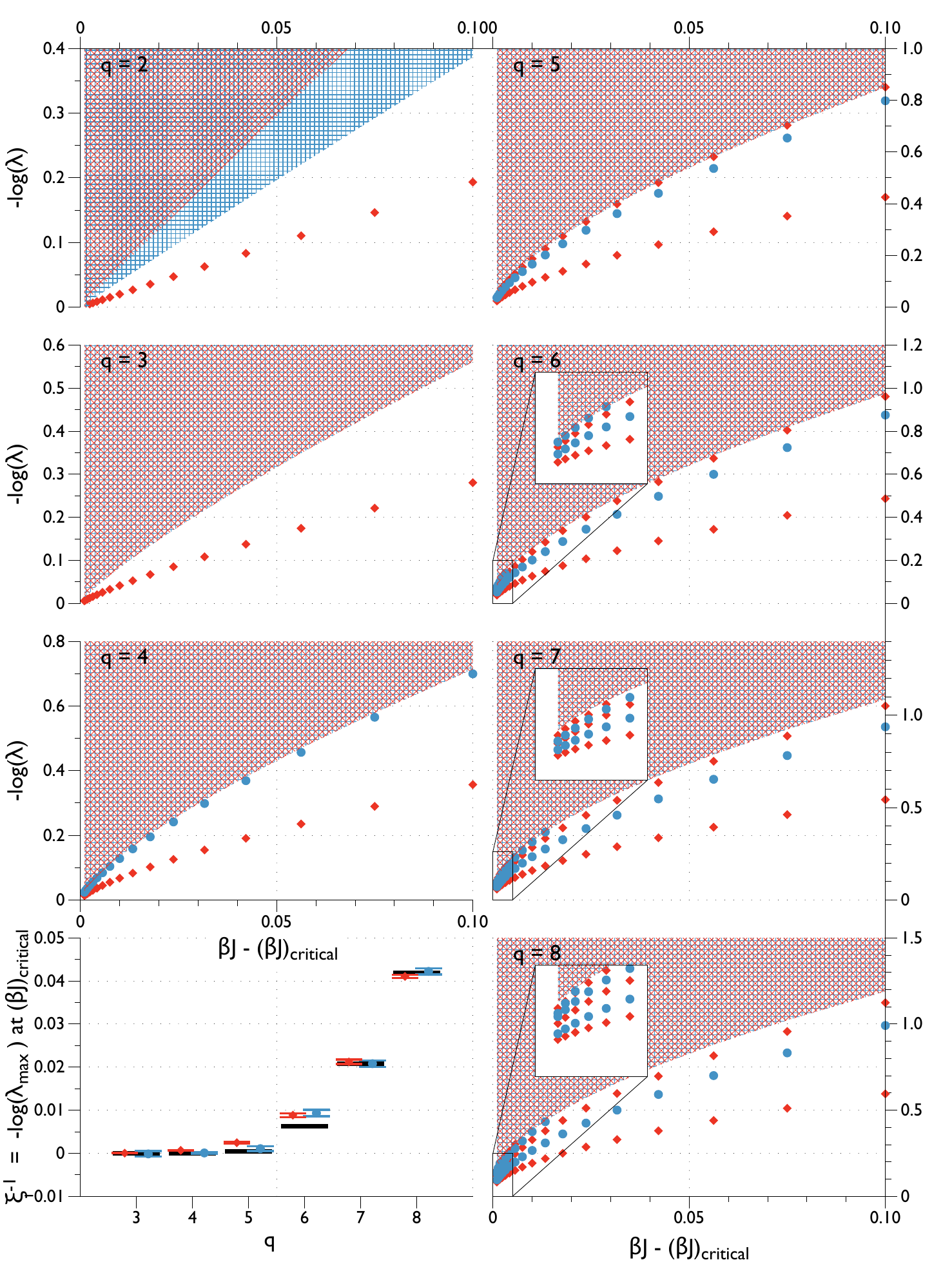}
\caption{\label{fig:pottsTm}Transfer matrix spectra at zero momentum of the $q$-level Potts model for $q=2,3,\ldots,8$ in the ordered (symmetry-broken) phase $\beta J > (\beta J)_c = \log(\sqrt{q}+1)$, plotted as function of $\beta J$. Blue dots indicate topologically trivial excitations, whereas red diamonds are obtained with the topologically non-trivial ansatz. Blue arced region (horizontal and vertical) indicate continuous bands in the trivial sector, whereas red diagonally arced regions indicate bands in the topological sector. Lower left panel shows the extrapolated largest eigenvalue at $(\beta J)_c$, with the exact values in black. See text for additional discussion.
}
\end{figure}

We focus our attention on the ordered regime $\beta J > (\beta J)_c$ in the vicinity of the phase transition. Rather than looking at conventional physical observables like the order parameter or the susceptibility, we analyse the eigenvalue spectrum of the (row-to-row) transfer matrix using the methods discussed in this section and plot the results in Figure~\ref{fig:pottsTm} below. The symmetry breaking is also manifest in the fixed point structure of the transfer matrix. Consequently, the infinite transfer matrix has topologically distinct excitation sectors corresponding to domain walls that interpolate between different fixed points. The difference in ground state ordering at $+\infty$ and $-\infty$ gives rise to a topological quantum number $s \in \mathbb{Z}_q$. Interestingly, we observe the same spectrum in all sectors with $s \neq 0$, and encode these using the red color in the plots. The trivial sector with $s=0$ is encoded using blue. In every sector, we observe one or more isolated bands, followed by clusters of eigenvalues which are indicative of the existence of continuous bands in the true spectrum. As for every sector and any isolated excitation branch, the minimum occurs at momentum $p=0$, we only plot (minus the logarithm) of the MPO eigenvalues at zero momentum. Furthermore, the results we plot are obtained by extrapolating our results obtained with MPS bond dimensions $D$ in the range $[16,256]$, based on a quadratic fit in $D^{-1}$.

For $q=2$, there is one topologically non-trivial sector which contains an isolated branch, corresponding to an `elementary' kink or domain wall excitation. The trivial sector only has continuous bands corresponding to multi-kink states with an even number of kinks, starting at the energy $2 m_K$, with $m_K$ the mass or lowest energy of the kink. The continuous band in the  topological sector starts at $3 m_K$, corresponding to a $3$-kink state. For $q=3$, not much changes, except that there are now two distinct topologically non-trivial sectors. The corresponding elementary excitations are kink and antikink and have the same energy. As a consequence, the continuous band in the topological sectors also starts at $2 m_K$, as a two-kink state is equivalent to an antikink and vice versa. Surprisingly, for all higher values of $q$, we also find that the spectra in all of the topologically non-trivial sectors are equivalent. For $q=4$, we observe an isolated excitation in the trivial sector just below the continuum, which we can interpret as a kink-antikink bound state. For every higher value of $q$, our simulation results indicate the appearance of a new isolated excitation, at least when sufficiently close to the phase transition $(\beta J)_c$. The new excitation appears alternatingly in the topological sector for odd $q$ and in the trivial sector for even $q$.

\begin{figure}[h]
  \centering
  \includegraphics[width=\textwidth]{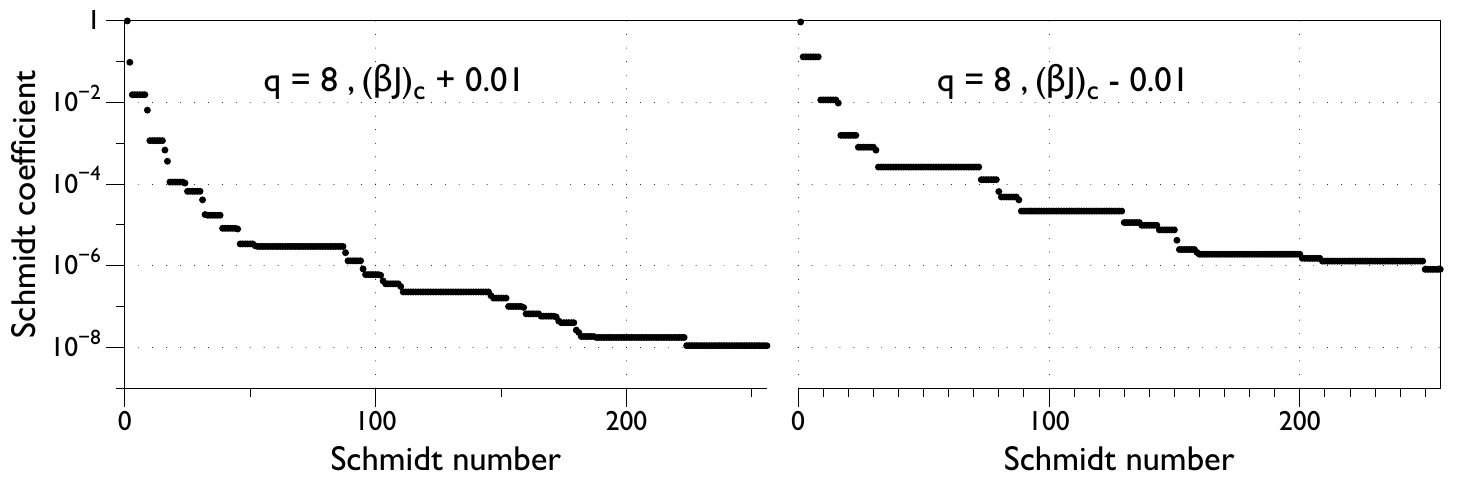}
\caption{\label{fig:pottsschmidt}Spectrum of Schmidt coefficients (singular values) of the $D=256$ MPS approximation to the fixed point of the transfer matrix of the $q=8$ Potts model, in the ordered phase (left panel) and disordered phase (right panel). The smallest degeneracies appearing in the ordered (symmetry broken) phase are $(q-2)$-fold, whereas in the disordered (symmetric) phase they are $(q-1)$-fold.
}
\end{figure}

The lowest excitation energy in the trivial sector also defines the correlation length of the system via $\xi^{-1} = - \log \lambda$. We have used our excitation energies at various values of $\beta J$ to extrapolate the value of $\xi^{-1}$ at the phase transition point $\beta J_c$, for the values $q=3,4,\ldots 8$. For $q=3,4$, the phase transition is still second order and the exact value is $\xi^{-1}=0$. An exact value of $\xi_d^{-1}$ is also known for $q < 4$ when approaching $(\beta J)_c$ from within the disordered phase $\beta J < (\beta J)_c$ by relating the transfer matrix to that of the 6-vertex model \cite{buddenoir1993correlation}; the value is given by
\begin{equation}
\xi_d^{-1} = 4 \sum_{n=0}^{\infty} \log\left(\frac{1+[\sqrt{2}\cosh((\pi^2/2v)(n+1/2))]^{-1}}{1-[\sqrt{2}\cosh((\pi^2/2v)(n+1/2))]^{-1}}\right)\label{eq:corrPottsexact}
\end{equation}
with $2 \cosh(v) = \sqrt{2+\sqrt{q}}$. Here, we approach $(\beta J)_c$ from within the ordered (symmetry broken) phase and the resulting correlation length $\xi_o$ is not necessarily equal. Indeed, it was once speculated that $\xi_o = \xi_d/2$ \cite{borgs1992explicit}. This would be understood from the behavior in the Ising model at first order away from criticality. The (trivial) excitation gap in the disordered phase at $(\beta J)_c - \epsilon$ is equal to the excitation gap in the topological sector at $(\beta J)_c + \epsilon$ (at first order in $\epsilon$). But the correlation length of two-point functions is set by the gap in the trivial sector, which is twice as large in the case of the Ising model. In the Potts model we find bound states in the trivial sector, and our extrapolation results (lower left panel of Figure~\ref{fig:pottsTm}) seem to indicate that in the limit $\beta J \stackrel{>}{\to} (\beta J)_c$, the gap in the trivial and topological sector become equal and correspond exactly to the analytical result Eq.~\eqref{eq:corrPottsexact} obtained on the other side of the phase transition in the disordered phase, in correspondence with more recent Monte Carlo results \cite{janke19952d}. Note that $\xi_d=2512.2468\ldots$ for $q=5$ and still $\xi_d=23.8782\ldots$ for $q=8$. Indeed, for values of $q \gtrsim 5$, the phase transition in the Potts model is said to be weak first order. Together with the huge degeneracies in the spectrum of Schmidt coefficients (see Figure~\ref{fig:pottsschmidt}), this explains the large bond dimensions required to study the Potts model in this regime.

\subsubsection{Schmidt spectrum of the Heisenberg model}
As a further illustration, we now compute the fixed point of the transfer matrix of the 6-vertex model $T^{ij}_{kl}=\delta_{i,j}\delta_{k,l}-\frac{1}{2}\delta_{i,k}\delta_{j,l}$ numerically. We have already discussed the exact solution in terms of the Bethe ansatz
 The leading eigenvector of the transfer matrix is the ground state of the Heisenberg antiferromagnet.  Without using symmetries or other techniques, Algorithm~\ref{alg:mpofixedpoint2} or \ref{alg:mpofixedpoint3} allows us to readily find an MPS approximation with $D=512$ or larger. We present the corresponding Schmidt spectrum, which shows an exact two-fold degeneracy, in Figure~\ref{fig:spectra}(a).

\subsubsection{Spectrum of the corner transfer matrix for the filtered Fibonacci string-net model}
\begin{figure}
\includegraphics[width=\textwidth]{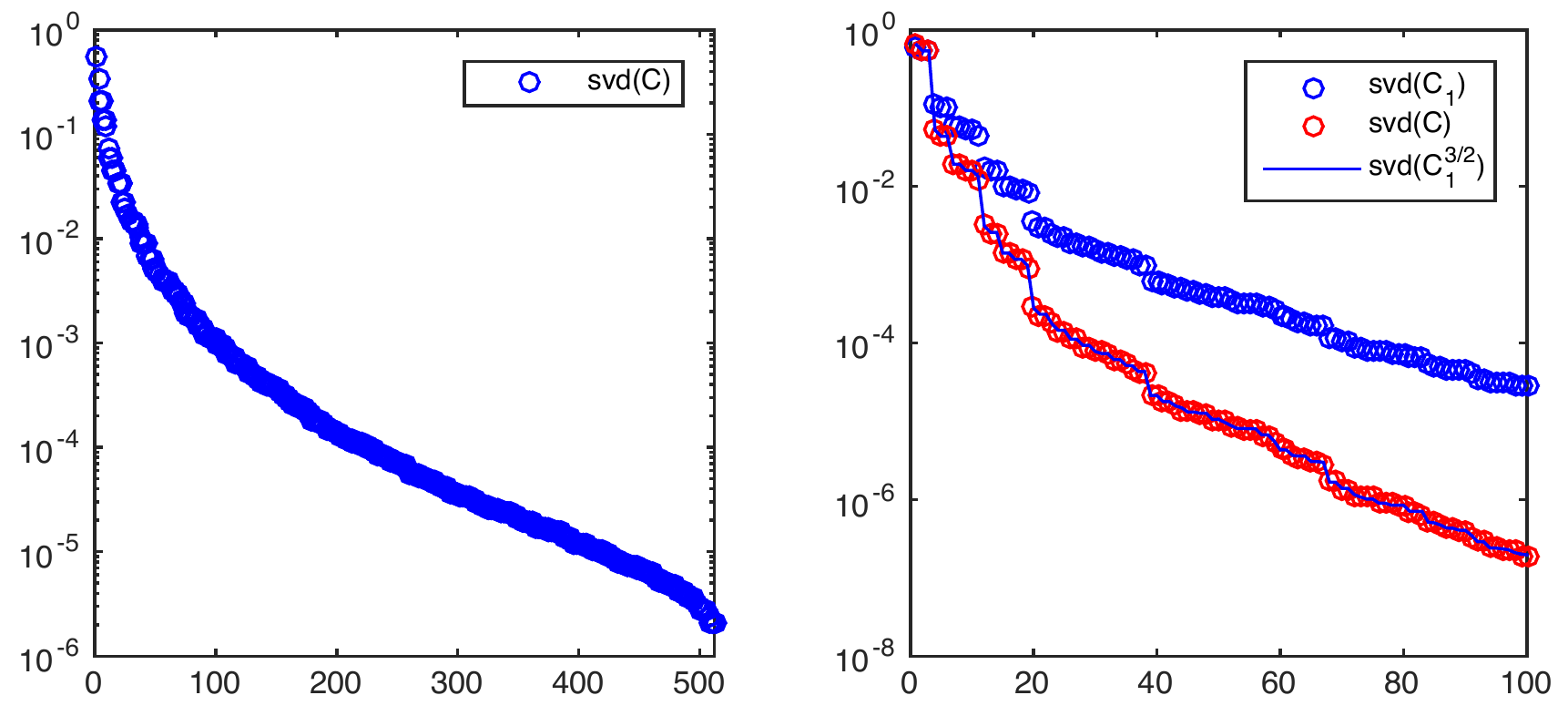}
\caption{\label{fig:spectra} Left (a): Schmidt spectrum of the MPS approximation with $D=512$ to the ground state of the Heisenberg model, or equivalently, the fixed point of the transfer matrix of the 6-vertex model. Right (b): Schmidt spectrum of the fixed point of the PEPS transfer matrix of the filtered Fibonacci string net model, as well as spectrum of the CTM. These two spectra are equal up to a power $3/2$, related to the geometric interpretation where $C$ represents a corner of $180^\circ$ and $C_1$ represents a corner of $120^\circ$.}
\end{figure}

Here, we start from the PEPS description of the Fibonacci string-net model \cite{levin2005string,buerschaper2009explicit}, to which we add a string tension by applying the filtering operation $\exp(\beta (\sigma^z-1)/2)$. This model undergoes a topological phase transition by increasing $\beta$, but we choose a value of $\beta$ that is still in topological phase \cite{Marien:2016aa}. We block two sites of the hexagonal lattice into a single square PEPS tensor. The resulting PEPS transfer matrix forms an MPO with bond dimension $D'=25$ and has a two-fold degenerate fixed point subspace. Because of the underlying hexagonal structure, the corresponding MPO has no reflection symmetry. Nevertheless, we can still apply Algorithm~\ref{alg:mpofixedpoint3}, even though the relation $C = C_1 C_1^\dagger$ breaks down (i.e.\ the singular values $S$ of the CTM $C_1$ are no longer the square root of the Schmidt spectrum of the fixed point). Geometrically, however, this is to be expected, as the corner now represents a corner over $120^\circ$. If instead we try to compare the spectrum $S^(3/2)$ with the Schmidt spectrum of the fixed point, we find perfect agreement (after matching the normalization), as illustrated in Figure~\ref{fig:spectra}(b).

\subsubsection{Entanglement spectrum of a two-dimensional chiral spin liquid}
\begin{figure}
\includegraphics[width=0.5\textwidth]{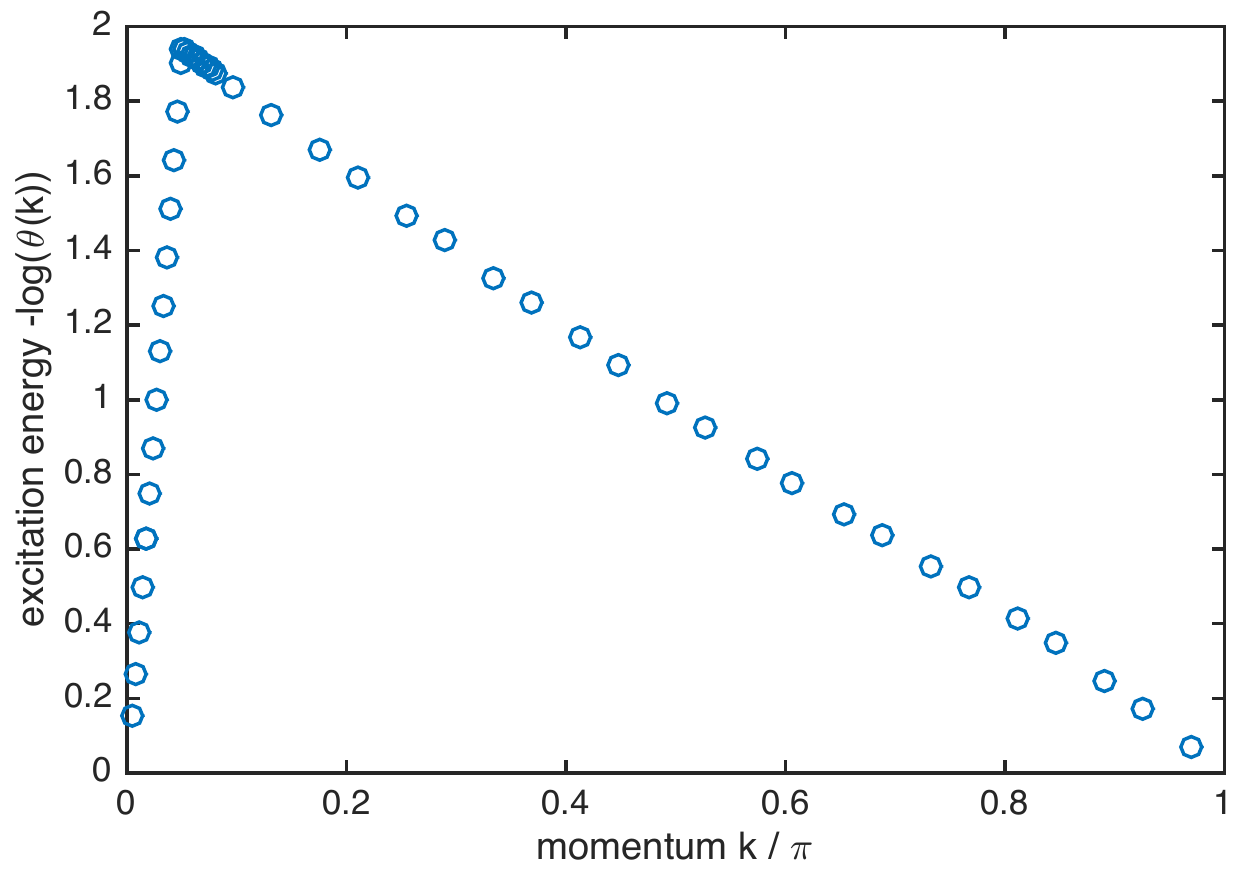}
\caption{\label{fig:spectra2d} Energy momentum spectrum of the entanglement Hamiltonian of the chiral spin liquid PEPS of Ref.~\cite{PhysRevB.91.224431}, computed directly in the thermodynamic limit using the MPS methods presented in this section.}
\end{figure}

As a final illustration of the excitation ansatz, we here compute the lowest excited state of the entanglement Hamiltonian of a PEPS that was recently shown to capture a chiral spin liquid \cite{PhysRevB.91.224431}. Unlike in Ref.~\cite{PhysRevB.91.224431}, we work directly in the thermodynamic limit. Firstly, we compute the left and right fixed points $\sigma_L$ and $\sigma_R$ of the infinite PEPS transfer matrix. We then construct the new MPO $\sigma_L \sigma_R$, which has the same spectrum as the reduced density matrix of the half infinite plane. Of this MPO, we again construct the leading eigenvector as uniform MPS (momentum zero) and then use the excitation ansatz of Eq.~\eqref{eq:mps:deftangent2} to compute the lowest lying excitated states for all $k\in [0,\pi]$. As explained, we are interested in the logarithm of the eigenvalues, which are variational approximations to the excitation energies of the entanglement Hamiltonian. The results are illustrated in Figure~\ref{fig:spectra2d} and clearly show the signature of the chiral edge modes.

\section{Outlook and conclusion}
Transfer matrices and matrix product operators pop up everywhere in the field of many body physics. On the one hand, this follows from the fact that the diagonalization of such operators has advanced the understanding of quantum and classical many body phases and the associated phase transitions enormously. On the other hand, MPOs exhibit a fascinating algebraic structure, as witnessed by the structure of the algebraic Bethe ansatz and of MPO representations of tensor fusion categories. As we have illustrated, this algebraic structure can also be used as a guideline to develop new algorithms for determining the eigenstructure numerically.

The real challenge is to construct an algebraic and computational framework for dealing with the 2-dimensional version of MPOs, namely PEPOs. This would allow a much better understanding of 3D classical statistical mechanical models and of 2D quantum phases. First steps in this direction have certainly be set. Zamolodchikov introduced a framework generalizing the algebraic Bethe ansatz to 3 dimensions involving the tetrahedral equations \cite{zamolodchikov1980tetrahedra}. It has however proven difficult to find nontrivial solutions of it. Similarly, Walker and Wang introduced a class of nontrivial 3D quantum theories exhibiting topological quantum order \cite{walker20123+}; those theories can be understood in terms of algebras of PEPOs. As far as we are aware of, no exact diagonalizations have been reported for a family of PEPOs exhibiting phase transitions (such as the 3D Ising model). It therefore looks like the numerical approach will be the only recourse, and an algebraic framework should be the guiding factor in constructing fast and reliable computational methods.

\appendix

\section{Explicit representation of the matrix product state steady state for the asymmetric exclusion process with parallel updates}

Here we represent the explicit solution of the matrix product state for the ASEP with parallel updates. Although the tensors $A^i$ are exactly as obtained in \cite{evans1999exact}, the way of proving the correctness of this solution by introducing tensors $B^i$ which satisfy the zipper conditions seems to be new and simpler than the original proofs.  We are looking for operators $A^i,B^i$ satisfying
\bea
A^1 B^0&=&(1-p)B^1 A^0 + B^2 A^0\\
A^0 B^2&=&p B^1 A^0\\
A^0 B^0 &=& B^0 A^0\\
A^0 B^1&=& B^0 A^1\\
A^1 B^1&=& B^1 A^1 + B^2 A^1
\eea
and vectors $|L\rangle,|R\rangle$ satisfying
\bea
\langle L|\left(B^2-\alpha A^0\right)&=&0\\
\langle L|\left(B^0-(1-\alpha) A^0\right)&=&0\\
\langle L|\left(B^1- A^1\right)&=&0\\
\left(A^1-B^2-(1-\beta)B^1\right)|R\rangle&=&0\\
\left(A^0-B^0-\beta B^1\right)|R\rangle&=&0\\
\eea
By some guess work, an explicit solution in terms of infinite dimensional matrices can be found:
\bea
\langle L|&=& \mat{cccccccc}{1 & 0 & 0 & \cdots & \frac{\alpha}{1-\alpha}\sqrt{\frac{1-p}{p}} & 0 & 0 & \cdots}\\
\langle R|&=& \mat{cccccccc}{1 & 0 & 0 & \cdots & \frac{\beta}{1-\beta}\sqrt{\frac{1-p}{p}} & 0 & 0 & \cdots}\\
\eea
\bea
A^0 &=& \mat{cc}{\sqrt{1-p}K_p & \sqrt{p}K_m \\ 0 & 0} \hspace{3cm} A^1  =  \mat{cc}{\sqrt{1-p}K_m & 0\\ \sqrt{p}K_p & 0}\\
B^0 &=& \mat{cc}{(1-p)^2 X & (1-p)\sqrt{p(1-p)}Y\\ 0 & 0}\hspace{1cm}  B^1 = \mat{cc}{ Z & 0\\0 & 0}\hspace{1cm}B^2 = \mat{cc}{ 0 & 0\\(1-p)\sqrt{p(1-p)}X & p(1-p)Y}
\eea
where
\[
K_p=\mat{ccccc}{\frac{p(1-\alpha)}{\alpha\sqrt{1-p}} & 0 & 0 & \cdots & \cdots\\ \frac{b}{\sqrt{1-p}} & \sqrt{1-p} & 0 & 0 & \cdots\\
0 & 1 & \sqrt{1-p} & 0 & \ddots\\
\vdots & 0 & 1 & \sqrt{1-p} & \ddots\\
\vdots & \ddots & \ddots & \ddots & \ddots}
\hspace{1cm}
K_m=\mat{ccccc}{\frac{p(1-\beta)}{\beta\sqrt{1-p}} & \frac{b}{\sqrt{1-p}} & 0 & 0 & \cdots\\
0 & \sqrt{1-p} & 1 & 0 & \ddots \\ \vdots & \ddots & \sqrt{1-p} & 1 & \ddots \\ \vdots & \ddots & \ddots & \ddots & \ddots}
\]

\[X= \mat{ccccc}{\frac{p(1-\alpha)^2}{\alpha(1-p)^2} & 0 & 0 & \cdots & \cdots\\
                 \frac{b(1-\alpha p)}{(1-p)^2} & 1 & 0 & 0 & \ddots\\
                 -\frac{\alpha p b}{(1-p)^3}\left(-\sqrt{1-p}\right)^3 & \frac{1+p}{\sqrt{1-p}}                              & 1 & 0 & \ddots\\
                 -\frac{\alpha p b}{(1-p)^3}\left(-\sqrt{1-p}\right)^4 &  -\frac{p^2}{(1-p)^{5/2}}\left(-\sqrt{1-p}\right)^3 & \frac{1+p}{\sqrt{1-p}}  & 1 & \ddots\\
                 -\frac{\alpha p b}{(1-p)^3}\left(-\sqrt{1-p}\right)^5 &  -\frac{p^2}{(1-p)^{5/2}}\left(-\sqrt{1-p}\right)^4 & -\frac{p^2}{(1-p)^{5/2}}\left(-\sqrt{1-p}\right)^3 & \ddots & \ddots\\
                 \vdots                                                &     \vdots    & \ddots & \ddots &\ddots}\]

\[Y= \mat{ccccccc}{\frac{ p(1-\alpha)(1-\beta)}{\beta (1-p)^2}                          & \frac{b(1-\alpha)}{(1-p)^2} & 0 & 0 & \cdots &\cdots &\cdots\\
                \frac{\alpha p (1-\beta) b}{\beta (1-p)^3}\left(-\sqrt{1-p}\right)^2  & \frac{p(\alpha -p)(1-\beta)}{\beta (1-p)^3}\left(-\sqrt{1-p}\right)^2 & \frac{1}{\sqrt{1-p}}   & 0 & \ddots & \ddots &\ddots\\
                \frac{\alpha p (1-\beta) b}{\beta (1-p)^3}\left(-\sqrt{1-p}\right)^3  & \frac{p(\alpha -p)(1-\beta)}{\beta (1-p)^3}\left(-\sqrt{1-p}\right)^3 &
                 \frac{1}{1-p} &     \frac{1}{\sqrt{1-p}}   & 0  & \ddots &\ddots\\
                 \frac{\alpha p (1-\beta) b}{\beta (1-p)^3}\left(-\sqrt{1-p}\right)^4 & \frac{p(\alpha -p)(1-\beta)}{\beta (1-p)^3}\left(-\sqrt{1-p}\right)^4 &
                  0 &     \frac{1}{1-p} &     \frac{1}{\sqrt{1-p}}   & \ddots &\ddots\\
                 \frac{\alpha p (1-\beta) b}{\beta (1-p)^3}\left(-\sqrt{1-p}\right)^5 & \frac{p(\alpha -p)(1-\beta)}{\beta (1-p)^3}\left(-\sqrt{1-p}\right)^5 &
                  0 &  0 &    \frac{1}{1-p} &     \frac{1}{\sqrt{1-p}}  &\ddots\\
                  \vdots      & \vdots   & \vdots & \vdots & \ddots & \ddots & \ddots}\]
\[Z= \mat{cccccc}{\frac{p}{\beta} & b & 0 & 0 & \cdots & \cdots\\
                 0 & 1 & \sqrt{1-p} & 0 & \ddots & \ddots\\
                 0 & 0 & 1 & \sqrt{1-p} & 0 & \ddots\\
                 \vdots & \ddots &\ddots &\ddots & \ddots & \ddots}
\]
and $b=\sqrt{\frac{p}{\alpha\beta}(1-p-(1-\alpha)(1-\beta))}$.\\

\section*{DISCLOSURE STATEMENT}
The authors are not aware of any affiliations, memberships, funding, or financial holdings that might be perceived as affecting the objectivity of this review.

\section*{ACKNOWLEDGMENTS}
We are grateful to Paul Fendley, Julien Vidal, Laurens Vanderstraeten, Micha\"el Mari\"en, Matthew Fishman and Valentin Stauber-Zauner for valuable discussions. We acknowledge funding from the Research Foundation Flanders, the European projects QUTE and SIQS, and Austrian Science Fund (FWF) project F40 (Foqus) and F41 (Vicom).

\bibliographystyle{ar-style4}
\bibliography{citations.bib}

\begin{thebibliography}{136}
\expandafter\ifx\csname natexlab\endcsname\relax\def\natexlab#1{#1}\fi

\bibitem{Feynman}
Feynman R. 1987.
Proceedings of the international workshop on variational calculations in
  quantum field theory held in wangarooge, west germany.
World Scientific Publishing, Singapore

\bibitem{DMFT}
Georges A, Kotliar G, Krauth W, Rozenberg MJ. 1996.
\textit{Reviews of Modern Physics} 68:13

\bibitem{Horodecki4}
Horodecki R, Horodecki P, Horodecki M, Horodecki K. 2009.
\textit{Reviews of modern physics} 81:865

\bibitem{faithful}
Verstraete F, Cirac JI. 2006.
\textit{Physical Review B} 73:094423

\bibitem{arealaw1}
Wolf MM, Verstraete F, Hastings MB, Cirac JI. 2008.
\textit{Physical review letters} 100:070502

\bibitem{hastings2007area}
Hastings MB. 2007.
\textit{Journal of Statistical Mechanics: Theory and Experiment} 2007:P08024

\bibitem{eisert2010colloquium}
Eisert J, Cramer M, Plenio MB. 2010.
\textit{Reviews of Modern Physics} 82:277

\bibitem{verstraete2004renormalization}
Verstraete F, Cirac JI. 2004.
\textit{arXiv preprint cond-mat/0407066}

\bibitem{baxter1978variational}
Baxter R. 1978.
\textit{Journal of Statistical Physics} 19:461--478

\bibitem{Affleck2004}
Affleck I, Kennedy T, Lieb EH, Tasaki H. 2004.
Condensed matter physics and exactly soluble models: Selecta of elliott h.
  lieb, chap. Valence Bond Ground States in Isotropic Quantum Antiferromagnets.
Berlin, Heidelberg: Springer Berlin Heidelberg,  253--304

\bibitem{fannes1992finitely}
Fannes M, Nachtergaele B, Werner RF. 1992.
\textit{Communications in mathematical physics} 144:443--490

\bibitem{white1992density}
White SR. 1992.
\textit{Physical Review Letters} 69:2863

\bibitem{klumper1993matrix}
Kl{\"u}mper A, Schadschneider A, Zittartz J. 1993.
\textit{EPL (Europhysics Letters)} 24:293

\bibitem{ostlund1995thermodynamic}
{\"O}stlund S, Rommer S. 1995.
\textit{Physical review letters} 75:3537

\bibitem{nishino1995density}
Nishino T. 1995.
\textit{Journal of the Physical Society of Japan} 64:3598--3601

\bibitem{vidal2003efficient}
Vidal G. 2003.
\textit{Physical Review Letters} 91:147902

\bibitem{verstraete2004density}
Verstraete F, Porras D, Cirac JI. 2004.
\textit{Physical review letters} 93:227205

\bibitem{perez2006matrix}
Perez-Garcia D, Verstraete F, Wolf MM, Cirac JI. 2006.
\textit{arXiv preprint quant-ph/0608197}

\bibitem{schollwock2005density}
Schollw{\"o}ck U. 2005.
\textit{Reviews of modern physics} 77:259

\bibitem{schollwock2011density}
Schollw{\"o}ck U. 2011.
\textit{Annals of Physics} 326:96--192

\bibitem{shi2006classical}
Shi YY, Duan LM, Vidal G. 2006.
\textit{Physical Review A} 74:022320

\bibitem{PhysRevB.82.205105}
Murg V, Verstraete F, Legeza O, Noack RM. 2010.
\textit{Phys. Rev. B} 82:205105

\bibitem{verstraete2008matrix}
Verstraete F, Murg V, Cirac JI. 2008.
\textit{Advances in Physics} 57:143--224

\bibitem{dukelsky1998equivalence}
Dukelsky J, Mart{\'\i}n-Delgado MA, Nishino T, Sierra G. 1998.
\textit{EPL (Europhysics Letters)} 43:457

\bibitem{nishino2001two}
Nishino T, Hieida Y, Okunishi K, Maeshima N, Akutsu Y, Gendiar A. 2001.
\textit{Progress of Theoretical Physics} 105:409--417

\bibitem{verstraete2006criticality}
Verstraete F, Wolf MM, Perez-Garcia D, Cirac JI. 2006.
\textit{Physical review letters} 96:220601

\bibitem{schuch2007computational}
Schuch N, Wolf MM, Verstraete F, Cirac JI. 2007.
\textit{Physical review letters} 98:140506

\bibitem{murg2007variational}
Murg V, Verstraete F, Cirac JI. 2007.
\textit{Physical Review A} 75:033605

\bibitem{jordan2008classical}
Jordan J, Or{\'u}s R, Vidal G, Verstraete F, Cirac JI. 2008.
\textit{Physical review letters} 101:250602

\bibitem{orus2009simulation}
Or{\'u}s R, Vidal G. 2009.
\textit{Physical Review B} 80:094403

\bibitem{bauer2011implementing}
Bauer B, Corboz P, Or{\'u}s R, Troyer M. 2011.
\textit{Physical Review B} 83:125106

\bibitem{corboz2011stripes}
Corboz P, White SR, Vidal G, Troyer M. 2011.
\textit{Physical Review B} 84:041108

\bibitem{vidal2007entanglement}
Vidal G. 2007{\natexlab{a}}.
\textit{Physical review letters} 99:220405

\bibitem{evenbly2009algorithms}
Evenbly G, Vidal G. 2009.
\textit{Physical Review B} 79:144108

\bibitem{evenbly2015tensor}
Evenbly G, Vidal G. 2015{\natexlab{a}}.
\textit{Physical Review Letters} 115:200401

\bibitem{bal2015matrix}
Bal M, Rams MM, Zauner V, Haegeman J, Verstraete F. 2015.
\textit{arXiv preprint arXiv:1509.01522}

\bibitem{verstraete2004matrix}
Verstraete F, Garcia-Ripoll JJ, Cirac JI. 2004.
\textit{Physical review letters} 93:207204

\bibitem{zwolak2004mixed}
Zwolak M, Vidal G. 2004.
\textit{Physical review letters} 93:207205

\bibitem{pirvu2010matrix}
Pirvu B, Murg V, Cirac JI, Verstraete F. 2010.
\textit{New Journal of Physics} 12:025012

\bibitem{mcculloch2007density}
McCulloch IP. 2007.
\textit{Journal of Statistical Mechanics: Theory and Experiment} 2007:P10014

\bibitem{onsager1944crystal}
Onsager L. 1944.
\textit{Physical Review} 65:117

\bibitem{kramers1941statistics}
Kramers HA, Wannier GH. 1941.
\textit{Physical Review} 60:252

\bibitem{lieb1967residual}
Lieb EH. 1967{\natexlab{a}}.
\textit{Physical Review} 162:162

\bibitem{faddeev1980quantum}
Faddeev LD, Sklyanin EK, Takhtajan LA. 1980.
\textit{Theoretical and Mathematical Physics} 40:688

\bibitem{korepin1997quantum}
Korepin VE, Bogoliubov NM, Izergin AG. 1997.
Quantum inverse scattering method and correlation functions.
Cambridge university press

\bibitem{levin2005string}
Levin MA, Wen XG. 2005.
\textit{Physical Review B} 71:045110

\bibitem{chen2013symmetry}
Chen X, Gu ZC, Liu ZX, Wen XG. 2013.
\textit{Physical Review B} 87:155114

\bibitem{csahinouglu2014characterizing}
{\c{S}}ahino{\u{g}}lu MB, Williamson D, Bultinck N, Mari{\"e}n M, Haegeman J,
  et~al. 2014.
\textit{arXiv preprint arXiv:1409.2150}

\bibitem{williamson2014matrix}
Williamson DJ, Bultinck N, Mari{\"e}n M, Sahinoglu MB, Haegeman J, Verstraete
  F. 2014.
\textit{arXiv preprint arXiv:1412.5604}

\bibitem{bultinck2015anyons}
Bultinck N, Mari{\"e}n M, Williamson DJ, {\c{S}}ahino{\u{g}}lu MB, Haegeman J,
  Verstraete F. 2015.
\textit{arXiv preprint arXiv:1511.08090}

\bibitem{derrida1993exact}
Derrida B, Evans M, Hakim V, Pasquier V. 1993.
\textit{Journal of Physics A: Mathematical and General} 26:1493

\bibitem{blythe2007nonequilibrium}
Blythe RA, Evans MR. 2007.
\textit{Journal of Physics A: Mathematical and Theoretical} 40:R333

\bibitem{domany1984equivalence}
Domany E, Kinzel W. 1984.
\textit{Physical review letters} 53:311

\bibitem{evans1978spectral}
Evans DE, H{\o}egh-Krohn R. 1978.
\textit{Journal of the London Mathematical Society} 2:345--355

\bibitem{perez2008string}
P{\'e}rez-Garc{\'\i}a D, Wolf MM, Sanz M, Verstraete F, Cirac JI. 2008.
\textit{Physical review letters} 100:167202

\bibitem{newmpo2016}
Cirac J, Perez-Garcia D, Schuch N, Verstraete F. 2016.
\textit{arXiv preprint arXiv:1606.00608}

\bibitem{metropolis1953equation}
Metropolis N, Rosenbluth AW, Rosenbluth MN, Teller AH, Teller E. 1953.
\textit{The journal of chemical physics} 21:1087--1092

\bibitem{baxter1980hard}
Baxter RJ. 1980.
\textit{Journal of Physics A: Mathematical and General} 13:L61

\bibitem{kasteleyn1963dimer}
Kasteleyn PW. 1963.
\textit{Journal of Mathematical Physics} 4:287--293

\bibitem{lieb1967solution}
Lieb EH. 1967{\natexlab{b}}.
\textit{Journal of Mathematical Physics} 8:2339--2341

\bibitem{pauling1935structure}
Pauling L. 1935.
\textit{Journal of the American Chemical Society} 57:2680--2684

\bibitem{foulkes2001quantum}
Foulkes W, Mitas L, Needs R, Rajagopal G. 2001.
\textit{Reviews of Modern Physics} 73:33

\bibitem{evertz2003loop}
Evertz HG. 2003.
\textit{Advances in Physics} 52:1--66

\bibitem{troyer2005computational}
Troyer M, Wiese UJ. 2005.
\textit{Physical review letters} 94:170201

\bibitem{perez2008peps}
Perez-Garcia D, Verstraete F, Wolf MM, Cirac JI. 2008.
\textit{Quantum Information \& Computation} 8:650--663

\bibitem{li2008entanglement}
Li H, Haldane FDM. 2008.
\textit{Physical review letters} 101:010504

\bibitem{cirac2011entanglement}
Cirac JI, Poilblanc D, Schuch N, Verstraete F. 2011.
\textit{Physical Review B} 83:245134

\bibitem{schuch2010peps}
Schuch N, Cirac I, P{\'e}rez-Garc{\'\i}a D. 2010.
\textit{Annals of Physics} 325:2153--2192

\bibitem{buerschaper2014twisted}
Buerschaper O. 2014.
\textit{Annals of Physics} 351:447--476

\bibitem{kitaev2003fault}
Kitaev AY. 2003.
\textit{Annals of Physics} 303:2--30

\bibitem{kitaev2006topological}
Kitaev A, Preskill J. 2006.
\textit{Physical review letters} 96:110404

\bibitem{levin2006detecting}
Levin M, Wen XG. 2006.
\textit{Physical review letters} 96:110405

\bibitem{schuch2013topological}
Schuch N, Poilblanc D, Cirac JI, Perez-Garcia D. 2013.
\textit{Physical review letters} 111:090501

\bibitem{haegeman2014shadows}
Haegeman J, Zauner V, Schuch N, Verstraete F. 2015.
\textit{Nature Communications} 6:8284

\bibitem{gu2014symmetry}
Gu ZC, Wen XG. 2014.
\textit{Physical Review B} 90:115141

\bibitem{barkeshli2014symmetry}
Barkeshli M, Bonderson P, Cheng M, Wang Z. 2014.
\textit{arXiv preprint arXiv:1410.4540}

\bibitem{derrida1998exactly}
Derrida B. 1998.
\textit{Physics Reports} 301:65--83

\bibitem{evans1999exact}
Evans M, Rajewsky N, Speer E. 1999.
\textit{Journal of statistical physics} 95:45--96

\bibitem{de1999exact}
de~Gier J, Nienhuis B. 1999.
\textit{Physical Review E} 59:4899

\bibitem{katsura2010derivation}
Katsura H, Maruyama I. 2010.
\textit{Journal of Physics A: Mathematical and Theoretical} 43:175003

\bibitem{verstraete2009quantum}
Verstraete F, Cirac JI, Latorre JI. 2009.
\textit{Physical Review A} 79:032316

\bibitem{schultz1964two}
Schultz TD, Mattis DC, Lieb EH. 1964.
\textit{Reviews of Modern Physics} 36:856

\bibitem{yang1952spontaneous}
Yang CN. 1952.
\textit{Physical Review} 85:808

\bibitem{bethe1935statistical}
Bethe HA. 1935.
Statistical theory of superlattices. In \textit{Proc. Roy. Soc. London A}, vol.
  150

\bibitem{yang1968s}
Yang C. 1968.
\textit{Physical Review} 168:1920

\bibitem{baxter1971eight}
Baxter RJ. 1971.
\textit{Physical Review Letters} 26:832

\bibitem{murg2012algebraic}
Murg V, Korepin VE, Verstraete F. 2012.
\textit{Physical Review B} 86:045125

\bibitem{lieb1968absence}
Lieb EH, Wu FY. 1968.
\textit{Physical Review Letters} 20:1445

\bibitem{zamolodchikov1979z}
Zamolodchikov AB. 1979.
\textit{Communications in Mathematical Physics} 69:165--178

\bibitem{alcaraz2003bethe}
Alcaraz FC, Lazo MJ. 2003.
\textit{Journal of Physics A: Mathematical and General} 37:L1

\bibitem{alcaraz2006generalization}
Alcaraz FC, Lazo MJ. 2006.
\textit{Journal of Physics A: Mathematical and General} 39:11335

\bibitem{etingof2015tensor}
Etingof P, Gelaki S, Nikshych D, Ostrik V. 2015.
Tensor categories.
vol. 205.
American Mathematical Soc.

\bibitem{nayak2008non}
Nayak C, Simon SH, Stern A, Freedman M, Sarma SD. 2008.
\textit{Reviews of Modern Physics} 80:1083

\bibitem{wang2010topological}
Wang Z. 2010.
Topological quantum computation.
No. 112. American Mathematical Soc.

\bibitem{temme2010stochastic}
Temme K, Verstraete F. 2010.
\textit{Physical review letters} 104:210502

\bibitem{essler1996representations}
Essler FH, Rittenberg V. 1996.
\textit{Journal of Physics A: Mathematical and General} 29:3375

\bibitem{de2005bethe}
de~Gier J, Essler FH. 2005.
\textit{Physical review letters} 95:240601

\bibitem{vidal2004efficient}
Vidal G. 2004.
\textit{Physical review letters} 93:040502

\bibitem{vidal2007classical}
Vidal G. 2007{\natexlab{b}}.
\textit{Physical review letters} 98:070201

\bibitem{nishino1996corner}
Nishino T, Okunishi K. 1996.
\textit{Journal of the Physical Society of Japan} 65:891--894

\bibitem{bursill1996density}
Bursill R, Xiang T, Gehring G. 1996.
\textit{Journal of Physics: Condensed Matter} 8:L583

\bibitem{wang1997transfer}
Wang X, Xiang T. 1997.
\textit{Physical Review B} 56:5061

\bibitem{shibata1997thermodynamics}
Shibata N. 1997.
\textit{Journal of the Physical Society of Japan} 66:2221--2223

\bibitem{sirker2005real}
Sirker J, Kl{\"u}mper A. 2005.
\textit{Physical Review B} 71:241101

\bibitem{murg2005efficient}
Murg V, Verstraete F, Cirac JI. 2005.
\textit{Physical review letters} 95:057206

\bibitem{carlon1999density}
Carlon E, Henkel M, Schollw{\"o}ck U. 1999.
\textit{The European Physical Journal B-Condensed Matter and Complex Systems}
  12:99--114

\bibitem{carlon2001critical}
Carlon E, Henkel M, Schollw{\"o}ck U. 2001.
\textit{Physical Review E} 63:036101

\bibitem{haegeman2014geometry}
Haegeman J, Mari{\"e}n M, Osborne TJ, Verstraete F. 2014{\natexlab{a}}.
\textit{Journal of Mathematical Physics} 55:021902

\bibitem{haegeman2011time}
Haegeman J, Cirac JI, Osborne TJ, Pi{\v{z}}orn I, Verschelde H, Verstraete F.
  2011.
\textit{Physical review letters} 107:070601

\bibitem{haegeman2014unifying}
Haegeman J, Lubich C, Oseledets I, Vandereycken B, Verstraete F.
  2014{\natexlab{b}}.
\textit{arXiv preprint arXiv:1408.5056}

\bibitem{orus2008infinite}
Orus R, Vidal G. 2008.
\textit{Physical Review B} 78:155117

\bibitem{saad2011numerical}
Saad Y. 2011.
Numerical methods for large eigenvalue problems: Revised edition.
vol.~66.
Siam

\bibitem{lehoucq1998arpack}
Lehoucq RB, Sorensen DC, Yang C. 1998.
Arpack users' guide: solution of large-scale eigenvalue problems with
  implicitly restarted arnoldi methods.
vol.~6.
Siam

\bibitem{morgan1996restarting}
Morgan R. 1996.
\textit{Mathematics of Computation of the American Mathematical Society}
  65:1213--1230

\bibitem{anderson1967infrared}
Anderson PW. 1967.
\textit{Physical Review Letters} 18:1049

\bibitem{nishino1997corner}
Nishino T, Okunishi K. 1997.
\textit{Journal of the Physical Society of Japan} 66:3040--3047

\bibitem{orus2012exploring}
Or{\'u}s R. 2012.
\textit{Physical Review B} 85:205117

\bibitem{levin2007tensor}
Levin M, Nave CP. 2007.
\textit{Physical review letters} 99:120601

\bibitem{gu2008tensor}
Gu ZC, Levin M, Wen XG. 2008.
\textit{Physical Review B} 78:205116

\bibitem{Evenbly:2015aa}
Evenbly G, Vidal G. 2015{\natexlab{b}}.
\textit{Physical Review Letters} 115:180405

\bibitem{yang2015loop}
Yang S, Gu ZC, Wen XG. 2015.
\textit{arXiv preprint arXiv:1512.04938}

\bibitem{haegeman2012variational}
Haegeman J, Pirvu B, Weir DJ, Cirac JI, Osborne TJ, et~al. 2012.
\textit{Physical Review B} 85:100408

\bibitem{draxler2013particles}
Draxler D, Haegeman J, Osborne TJ, Stojevic V, Vanderstraeten L, Verstraete F.
  2013.
\textit{Physical review letters} 111:020402

\bibitem{vanderstraeten2015excitations}
Vanderstraeten L, Mari{\"e}n M, Verstraete F, Haegeman J. 2015.
\textit{Physical Review B} 92:201111

\bibitem{vanderstraeten2014s}
Vanderstraeten L, Haegeman J, Osborne TJ, Verstraete F. 2014.
\textit{Physical review letters} 112:257202

\bibitem{vanderstraeten2015scattering}
Vanderstraeten L, Verstraete F, Haegeman J. 2015.
\textit{Physical Review B} 92:125136

\bibitem{barrett1994templates}
Barrett R, Berry MW, Chan TF, Demmel J, Donato J, et~al. 1994.
Templates for the solution of linear systems: building blocks for iterative
  methods.
vol.~43.
Siam

\bibitem{baxter1999planar}
Baxter R. 1999.
\textit{Annals of Combinatorics} 3:191--203

\bibitem{buddenoir1993correlation}
Buddenoir E, Wallon S. 1993.
\textit{Journal of Physics A: Mathematical and General} 26:3045

\bibitem{borgs1992explicit}
Borgs C, Janke W. 1992.
\textit{Journal de Physique I} 2:2011--2018

\bibitem{janke19952d}
Janke W, Kappler S. 1995.
\textit{EPL (Europhysics Letters)} 31:345

\bibitem{buerschaper2009explicit}
Buerschaper O, Aguado M, Vidal G. 2009.
\textit{Physical Review B} 79:085119

\bibitem{Marien:2016aa}
Mari\"en M. 2016.
In prepration

\bibitem{PhysRevB.91.224431}
Poilblanc D, Cirac JI, Schuch N. 2015.
\textit{Phys. Rev. B} 91:224431

\bibitem{zamolodchikov1980tetrahedra}
Zamolodchikov A. 1980.
\textit{Soviet Physics JETP} 52:325--336

\bibitem{walker20123+}
Walker K, Wang Z. 2012.
\textit{Frontiers of Physics} 7:150--159

\end{thebibliography}

\end{document}